\documentclass{article}
\usepackage[utf8]{inputenc}

\usepackage{amsmath}
\usepackage{graphicx}
\usepackage{subcaption}
\usepackage{caption}
\usepackage{array}
\usepackage{float}
\usepackage{multirow}
\usepackage{amsfonts}
\usepackage{authblk}  
\usepackage{color}

\usepackage{tikz}

\newcommand{\change}[1]{ #1 }

\title{Numerical study of the effect of the relative mobilities of chemical components on the Non solvent induced phase separation process for membrane elaboration}
\small\author[1,*]{A. Bounjad}
\author[2]{A. Wu}

\author[2]{C. Chevarin}
\author[3]{P. Guenoun}
\author[3]{F. Malloggi}
\author[2]{J. Mericq}
\author[3]{C. Merzougui}

\author[2]{D. Bouyer}
\author[1]{H. Henry}

\affil[1]{ Laboratoire de Physique de la Matière Condensée,
CNRS,
École Polytechnique,
Institut Polytechnique de Paris,
Palaiseau 91120, France}
\affil[2]{
CEA Saclay,
NIMBE,
UMR 3685, LIONS, Gif-Sur-Yvette  91191, France
NIMBE}
\affil[3]{IEM (Institut Européen des Membranes)
Université Montpellier,
UMR5635, CNRS, ENSCM, Montpellier 34090, France }
\affil[*]{\small \texttt{abderraouf.bounjad@polytechnique.edu}}

\begin{document}

\maketitle

\textbf{Keywords :} Phase separation, NIPS, Spinodal decomposition, Cahn-Hilliard.
\begin{abstract}
 The filtration membranes are often elaborated through a phase separation process where a polymer rich phase and a  polymer poor phase spontaneously form through spinodal decomposition. One process that is still not well  understood from a theoretical point of view is the Non-Solvent induced phase separation where a thermodynamically stable film of a a polymer mixture is put in contact with a bad solvent of the polymer. The invasion of the film by this non-solvent drives the film out of stability and leads to spinodal decomposition. During this phase separation polymer poor and polymer rich regions form.   In this article we present a numerical study of the effect of kinetic coefficients: namely the relative mobilities of polymer and solvent/non-solvent on the observed patterns. Using 2D numerical simulations of the ternary Cahn-Hilliard model  we show that  for a given thermodynamic landscape, this parameter has dramatic effects: depending on its value phase separation can be observed or not. We also show that it can affect the nature of the observed pattern.  In addition analysing 3D simulations we analyse the final pattern using a quantitative indicator of its connectivity and show that for a wide range of  initial composition of the film the final pattern is bicontinuous. We also quantify the transport properties of both polymer rich and polymer poor domains. 
\end{abstract}

\section{Introduction}

A widely used type of water  filtration membranes\cite{Mulder2012} consist of a nanometer thick polymer layer attached to a thicker porous polymeric layer. The former is the actual filter. The latter must let the flow pass through the membrane and ensures its  mechanical stability. This supporting layer  is a bicontinuous structure in order to meet both requirement. Voids must be continuous to let the flow pass  and the polymeric phase must be continuous to ensure mechanical cohesion. Obviously the performance of the membrane (its flow throughput) will depend on the spatial organisation of the void phase while its mechanical stability will depend on  the spatial organisation of the polymeric phase. \\

A process to create the  supporting  relies on the phase separation or spinodal decomposition or phase inversion that is known to lead to the formation of such bicontinuous patterns\cite{Cahn1958,Cahn1963}. This phenomenon occurs when an  initially thermodynamically stable homogeneous mixture of polymer and solvent is driven in a state where the homogeneous mixture is no longer stable. Then,   it  separates spontaneously (through a linear instability and not nucleation) into polymer rich and polymer poor domains. In the case of membrane formation the loss of stability can be induced by temperature change (TIPS)\cite{LI2006}, solvent evaporation (VIPS)\cite{Elford1937,VIPS2,Venault2013} or non solvent invasion (NIPS)\cite{Koenhen1977,Guillen2011}. After the initial phase separation, a purely anti-diffusive process, that takes place at very small lengthscales, a coarsening process takes place: the total interface area between the phases decreases leading to an increase of the characteristic length of the microstructure. This process is driven by the reduction of the interfacial energy and can be either due to diffusion\cite{Binder1987,PGG1980}  or hydrodynamic~\cite{Siggia1979,Kendon2001,Charaff}. This later process is self similar in most usual cases.\\

Since the properties of the porous material  strongly depend on the organisation of phases, it is highly desirable to optimize this microstructure to improve the membrane performance.
To this purpose a better understanding of the phase separation process and of the subsequent coarsening is necessary. Among the tools available, numerical simulations of the process describe the evolution of the whole microstructure and  provide information on the actual organisation of the phases and of their global properties.   In the case of TIPS large scale simulations can now be performed on a routine basis for homogeneous systems with or without flow\cite{Tanaka,Henry,henry2018,Kendon2001} and with realistic rheological properties\cite{Tanaka2}.\\

Here we consider the Non-solvent induced phase separation process (NIPS). It is a widely used industrial technique for fabricating polymer membranes supporting layers. In this process, a polymer-solvent solution film is immersed in a non-solvent bath, leading to an exchange of solvent and non-solvent. The solvent migrates from the polymer solution to the bath, while the non-solvent follows the opposite path. This compositional change leads the film in a thermodynamically unstable state, leading to phase separation in an isothermal manner. The film transforms from a homogeneous state to a heterogeneous one, characterized by polymer-rich and polymer-depleted phases. As the concentration in the polymer-rich phase increases, it solidifies, forming the membrane's microstructure\cite{Guillen2011}.  The outcome of the membrane formation mechanism was found to depend on the preparation process\cite{cohen1979}. First modeling efforts\cite{cohen1979,StrathmannKock1977} aimed at describing the solvent-nonsolvent exchange between different layers constitutive of the membrane. The phase separation process was then considered as enslaved to this transport phenomena and the diffusion layer was seen as at quasi-equilibrium. Later models\cite{radanovic1992} used less crude approximation of the mass transport process through the polymer film layer. However the use of  unidimensional model did not allow to capture the phase separation process properly. Recently modeling relying on the use of computer simulations\cite{Tree2017} have been introduced and applied  on the problem of NIPS. They use well established models\cite{Lowengrub,Toth2015} that describe the diffusive transport driven by chemical potential derived from a realistic thermodynamic functional: the Flory-Huggins free energy. Hence they reproduce both normal diffusion and phase separation.  In addition such model incorporate the description of the fluid flow driven by the heterogeneities of composition, including surface tension.  \\

The use of such models\cite{Tree2017,Tree2018,Tree2019,Garcia2020,Manzanares} has lead to significant progress. It was shown that putting  a mixture of polymer and solvent  into contact with a non solvent bath would lead to the formation of oriented patterns close to the interface as in surface directed spinodal decomposition\cite{Puri2017,Plapp1999,Ball1990}  and to the formation of more isotropic structure deeper in the polymer film. However most simulations were performed with an initially unstable film. Therefore they are  missing one key aspect of the NIPS process: the initial stability of the polymeric layer.  In another work, few simulations  show the initially at  equilibrium film driven out of equilibrium by diffusion of the non solvent are presented that. In this later case it must be noted that the initial film composition was close to the critical point~\cite{Tree2018}. \\

Here, using a similar setup and model we study the influence of the respective mobilities of the chemical species on the pattern formation process. This choice is driven  by the fact that  kinetic aspects of the transport process have been shown to affect dramatically the coarsening process. In the case of diffusion, it has been shown that faster diffusive transport in one phase could lead to the breakup of a bicontinuous pattern\cite{beck2020}. In the same spirit in the case of hydrodynamic coarsening it has been shown\cite{Henry} that the thinner phase  loses  its connectivity at a higher volume fraction than the more viscous one during the coarsening process.\\

Here  we show through numerical simulations that the respective mobilities of polymer, solvent and non solvent affect dramatically the process and that depending on their values very different patterns appear during the early stages. The article is organised as follows. We first present  model and the model parameters. We  then discuss the numerical results in 2D system that show the effect of the mobility on the film: if the polymer mobility is high enough, the film composition for which a phase separation is observed is limited to the vicinity of the critical point while if is low compared to solvent and non-solvent this domain is much larger. In a second results section we discuss our results for 3D system. Finally we draw conclusions and offer some perspectives.

\begin{figure}
    \centering
    \includegraphics[width=0.65\linewidth]{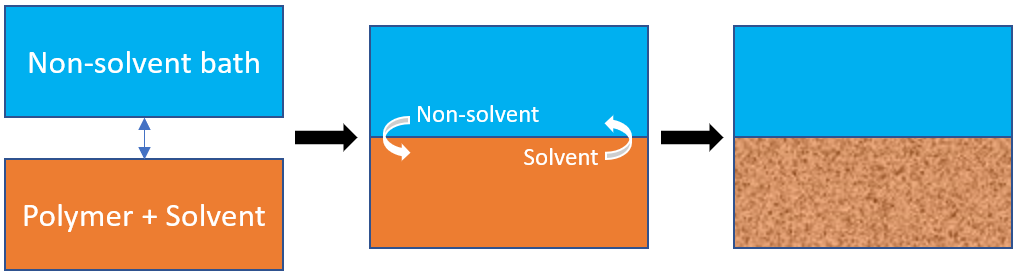}
    \caption{A schematic showing the NIPS membrane manufacturing process}
    \label{fig:NIPS}
\end{figure}

\section{Model and methods}
 The aim of the model used here is to describe the transport and phase separation that occur in a film composed of a polymer and a solvent put into contact with a bad solvent of the polymer. Since we consider the spatio-temporal evolution of the system, we need to resolve the composition of any point of space at any time in the laboratory frame of reference.  For the sake of simplicity we make the following simplifying assumptions: the molar volume of all chemical species considered here is independent of the local composition. There is no fluid flow and we neglect the effects of gravity.

\subsection{Model}
We describe the film and bath with a ternary Cahn-Hilliard model where the molar volume and mass density of constituents are assumed to be constant.  The three constituent that are considered are  the polymer (p), the solvent (s), and the non-solvent (n). Each of these components is characterized by its volume fraction $\phi_i$, where $i=\{\mathrm{p},\mathrm{s},\mathrm{n}\}$ and the $\phi_i$ obey:
\begin{equation}
    \phi_p+\phi_n+\phi_s=1\label{eq:incompressibility}
\end{equation}
 The total free energy of the system writes:
\begin{equation}
    \mathcal{F}=\int f(\{\phi_i\}) +\Sigma_i \kappa_i (\nabla \phi_i)^2
\end{equation}
Where the $\kappa_i$ are model parameters and $f(\{\phi_i\}) $ is the Flory-Huggins free energy density\cite{flory1942,huggins1942theory}. It writes:
\begin{equation}
f(\{\phi_i\}) = A\times  \left(\sum\limits_{i}^{p,n,s} \frac{\phi_i}{N_i} \ln(\phi_i) + \frac{1}{2} \sum\limits_{i \neq j}^{p,n,s} \chi_{ij} \phi_i \phi_j\right)
\label{eq:flory-huggins}
\end{equation}

Where $N_i$ denote the degrees of polymerization, and $\chi_{ij}$ represents the Flory interaction parameters. $A$ is a prefactor that is equal to $k_B T/v_0  $ with  $v_0$ the monomer volume, $K_b$ the Boltzmann constant and  $T$  the absolute temperature and $V_0$ the monomer volume. It must be noted that  the interface thickness between phase separated domains is proportional to $\sqrt{A/\kappa_i}$ while the interfacial energy between phase separated domains is proportional to $\sqrt{A \kappa_i}$\cite{Mirantsoa2022}.\\

The chemical potential for each component writes:
\begin{equation}
\mu_i = \frac{\partial f}{\partial \phi_i}-\kappa_i\nabla^2\phi_i
\label{eq:potentiel}
\end{equation}

The evolution of the system  obeys  the   non linear diffusion equations that  write:

\begin{eqnarray}
\frac{\partial \phi_p}{\partial t} &=& \nabla \!\cdot\! \left[ M_{pn} \nabla \mu_{pn}+M_{ps} \nabla \mu_{ps}\right]
\label{eq:eq1}\\
\frac{\partial \phi_n}{\partial t} &=& \nabla \!\cdot\! \left[ M_{np} \nabla \mu_{np}+M_{ns} \nabla \mu_{ns}\right]
\label{eq:eq2}\\
\frac{\partial \phi_s}{\partial t} &=& \nabla \!\cdot\! \left[ M_{sp} \nabla \mu_{sp}+M_{sn} \nabla \mu_{sn}\right]
\label{eq:eq3}
\end{eqnarray}

The  $M_{ij}$ are the  exchange mobilities that are such that $M_{ij}=M_{ji}$ and the $\mu_{ij}$ are the   exchange chemical potentials, defined as:
\begin{equation}
    \mu_{ij}=\mu_i-\mu_j
\end{equation}
Eq \ref{eq:eq3} can be deduced from the incompressibility condition (eq. \ref{eq:incompressibility}) and eqs. \ref{eq:eq1}, \ref{eq:eq2}.  As a result  evolution of the system is fully described by \ref{eq:eq1}, \ref{eq:eq2} and the incompressibility condition. For $M_{ij}$ taken equal to a constant value $M$, these equations correspond exactly to the system discussed in \cite{Tree2017} in the absence of flow. Here we discuss the influence of mobilities and more specifically the lower polymer mobility. As a result we assume that $M_{sp}=M_{np}$ and denote it  $M_p$ while $M_{ns}$ is denoted $M$.\\

\change{Before turning to the description of the numerical method we give the numerical values of model parameters. The gradient coefficients \(\kappa_i\) are chosen to match the experimental interfacial width ($\varepsilon$) and tension ($\gamma$). This leads to  (\(\kappa \propto \gamma\,\varepsilon\)). The mobility \(M\) is chosen to yield a realistic diffusion coefficient (see Appendix \ref{Appendice_Mob}). The table \ref{tab:parameters} summarizes the numerical values of the model parameters}

\begin{table}[H]
    \centering
    \small
    \begin{tabular}{ccc}
        \hline
        \textbf{Parameter} & \textbf{Value} & \textbf{Description} \\
        \hline
        \(\kappa_i\) & 8 & Gradient coefficients\\\\
        \(\mathrm{N}_x\) & 288 & Number of mesh points in the x-direction\\\\
        \(\mathrm{N}_y\) & 128 &Number of mesh points in the y-direction\\\\
        \(\mathrm{N}_z\) & 288 &Number of mesh points in the z-direction\\\\
        \(\mathrm{N}_f\) & 128 & Number of mesh points in the z-direction of the film\\\\
        \(\mathrm{N}_b\) & 16 & Number of mesh points in the z-direction of the bath\\\\
        \(M\) & 0.25 & Non-solvent/Solvent exchange mobility\\\\
        \(\chi_{pn}\) & 1.6 & Interaction parameter of Flory-Huggins Polymer/Non-solvent\\\\
        \(\chi_{ps}\) & 0 & Interaction parameter of Flory-Huggins Polymer/Solvent\\\\
        \(\chi_{ns}\) & 0 & Interaction parameter of Flory-Huggins Non-solvent/Solvent\\\\
        \(\mathrm{N}_p\) & 5 & Degree of polymerization\\
        
        \hline
    \end{tabular}
	\caption{Numerical Values of Model Parameters}
		\label{tab:parameters}
\end{table}

To simplify the model and reduce the number of free parameters, the polymer–solvent ($\chi_{ps}$) and nonsolvent–solvent ($\chi_{ns}$) interaction parameters are assumed to be zero; the same approach is used in Refs.~\cite{Tree2017,Tree2019,Garcia2020}. This assumption is physically justified by the good miscibility between the solvent and nonsolvent, as well as the good solubility of the polymer in the solvent.

\subsection{Numerical Method and setup}
 To numerically simulate the  Cahn-Hilliard equation we  use a pseudo-spectral method as in \cite{Tree2017} and \cite{henry2018}. Here we have used the same numerical method as in \cite{henry2018} without taking into account any flow. This method was implemented in an in-house code that relies on the use of the Fast Fourier Transform \cite{Eilbeck1983}, has been used in various contexts and tested quantitatively against analytical prediction\cite{zanella2020,Mirantsoa2024}.

\begin{figure}[H]
    \centering
    \includegraphics[width=0.4\linewidth]{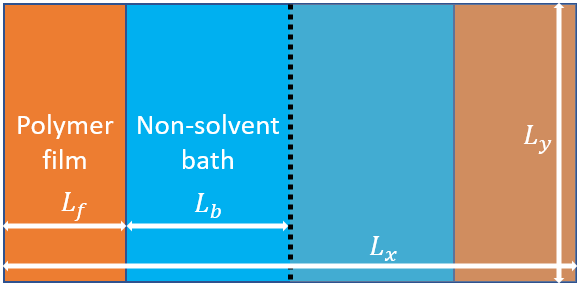}
    \caption{2D simulation geometry: Polymer film ($L_f$) and Non-solvent bath ($L_b$) with periodic boundaries}
    \label{fig:geometrie}
\end{figure}

The geometry of the simulated  system is shown in Figure \ref{fig:geometrie}. It consists of two distinct regions: a polymer film of a thickness \( L_f = N_f \times \Delta \mathrm{x} \) and a non-solvent bath with a depth \( L_b = N_b \times \Delta \mathrm{x} \), resulting in a total size \( L_x = 2\times(L_f + L_b) = N_\mathrm{x} \times \Delta \mathrm{x} \), where $\Delta \mathrm{x} > 0$ is the mesh size and $N_\mathrm{x} \in \mathbb{N}$ is the total number of mesh points in the x-direction. \change{ In experiments, the film is initially a mixture of polymer and solvent and  the initial composition of the bath is almost  pure non-solvent. However,from a numerical point of view this would lead to a singular behaviour due to the logarithmic term in the  Flory-Huggins free energy (Eq. \ref{eq:flory-huggins}). To avoid this, in our simulations,  a small amount of polymer and solvent ($\phi_p$ and $\phi_s$ are respectively 0.02 and 0.08) is added to the composition of the bath and the composition of the film is chosen inside the Gibbs triangle with varying proportions of solvent, non solvent and polymer in the stable part of the phase iagram\footnote{This allows also  to check the robustness of our simulations with respect to this changes. For low polymer mbility we found, as will be seen later that this has little effect (for a given initial polymer volume fraction). On the oposite for higher values of polymer mobilities, the behaviour of the system is affected } 
}.  Due to the use of the pseudo-spectral method, the boundary conditions are periodic, which implies that twice the domain of interest is simulated, as can be seen in Figure \ref{fig:geometrie}. Finally to mimic an infinite bath after each time step there is an exchange of solvent and non solvent between the central layers of the simulation domain (the dashed line in fig.\ref{fig:geometrie}) and a \textit{chemostat} (not represented in figure \ref{fig:geometrie}) with the initial bath composition. \change{ As a result at the central dashed line the bath composition is always such that $\mu_{ns}=\mu_{ns}^{\text{bath}}$ where $\mu_{ns}^{\text{bath}}$ is the solvent/non-solvent exchange chemical potential for the initial bath composition.}

Finally, to introduce the infinite bath in the model, after each iteration we update the bath composition by adding a constant non-solvent source to the central layers of the simulation domain (the dashed line in Fig.\ref{fig:geometrie}). This enforces the infinite bath condition by maintaining the composition at its initial value through coupling with a \textit{chemostat} (not represented in Fig.\ref{fig:geometrie}).

\subsection{Phase diagram}

With the chosen model parameter the phase diagram of the mixture present a stable domain where the homogenous mixture is stable, a domain where it is unstable and will spontaneously phase separate and a domain where the homogenous mixture is metastable. 
The boundary of the unstable domain is the spinodal curve it corresponds to the domain where the eigenvalues of the Hessian matrix (it is the second derivative for a binary mixture)  of $f_0$ becomes negative. The boundary between the metastable domain and the stable domain is the binodal curve. Both binodal and spinodal curves are tangent with each other at the critical point. \\

On the binodal curve, each point M can be associated to  a corresponding point N on the binodal curve on the other side of the binodal curve. N and M share the same exchange chemical potentials and obey the additional property that the tangent plane of the free energy functional in M is also tangent in N. The line MN is the tie line.  

\begin{figure}[H]
    \centering
    \includegraphics[width=0.5\linewidth]{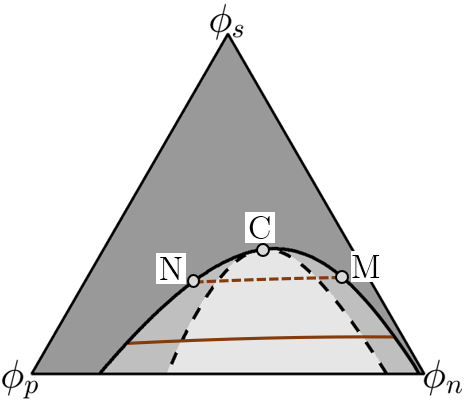}
    \caption{Ternary Gibbs phase diagram computed using an in-house simulation code. The ligh gray region corresponds to the domain of compositions for which the system is linearly unstable. The medium gray region to the metastable domain.  And the dark gray region corresponds to the homogeneous stable domain. C is the critical point, M, and N are two points that are at equilibrium with each other. The dashed line is the corresponding tie line. The solid line that  is drawn at the bottom of the binodal domain corresponds to the points for which $\mu_{ns}=\mu_{ns}^{bath}$.}
    \label{fig:diag}
\end{figure}

In Figure~\ref{fig:diag}, we present the ternary phase diagram corresponding to the model used here. The spinodal region, metastable region, and stable homogeneous domain are represented in shades of grey. The binodal curve is shown as a thick solid line, the spinodal line as a thick dashed line, and the critical point is marked as point C.\\

Additionally, the phase diagram includes the curve $\mu_{ns} = \mu_{ns}^{\text{bath}}$, which is plotted as a thin solid line. \change{ It corresponds to the set of compositions that  the film will eventually reach (if the polymer is assumed not to diffuse outside of the film) through solvent/non solvent exchange with the bath. Indeed, due to the exchange of solvent and non solvent between the simulation domain and a \textit{chemostat}, the equilibrium will be reached when in the whole simulation domain  $\mu_{ns} = \mu_{ns}^{\text{bath}}$~\footnote{a system is at equilibrium when the exchange chemical potentials are spatially homogeneous.}. Where $\mu_{ns}^{\text{bath}}$ corresponds to the exchange potential of the initial bath and of the \textit{chemostat}  that mimics an infinite bath.}

\section{Results}
  
 We now turn to the description of our numerical results. The first part consists of a two-dimensional study aimed at investigating the effect of polymer mobility and initial film composition on the pattern formation process through a systematic parametric analysis. The second part presents a three-dimensional study, focusing on the physical properties of the final structures obtained in 3D simulations.

\subsection{Equal Mobilities}
First we discuss the case that has already been discussed in \cite{Tree2017}, where all chemical species share the same mobilities. In this case, depending on the initial composition of the film, two distinct behaviors and an intermediate regime are  observed.\\ 

\begin{figure}
    \centering
    \renewcommand{\arraystretch}{1.5}
    \begin{subfigure}[b]{0.8\textwidth}
        \centering
        \begin{minipage}{0.92\textwidth}
            \centering
            \begin{tabular}{|c|c|c|c|c|}
                \hline
                t & \textbf{$t_1$} & \textbf{$t_2$} & \textbf{$t_3$} & \textbf{$t_4$} \\
                \hline
                \raisebox{0.4cm}{\textbf{(a)}} & 
                \raisebox{-0.5cm}{\includegraphics[width=0.15\textwidth]{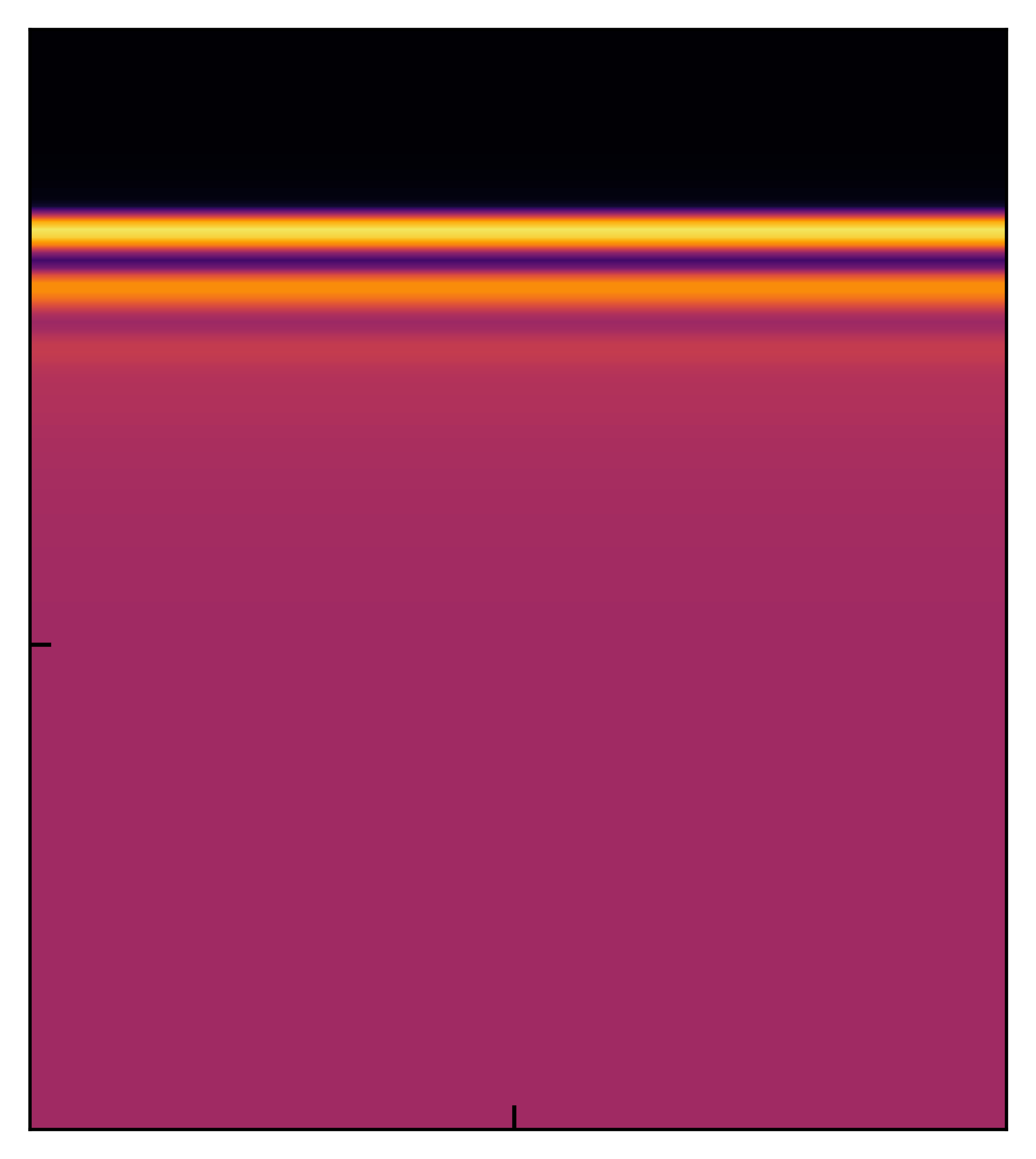}} &
                \raisebox{-0.5cm}{\includegraphics[width=0.15\textwidth]{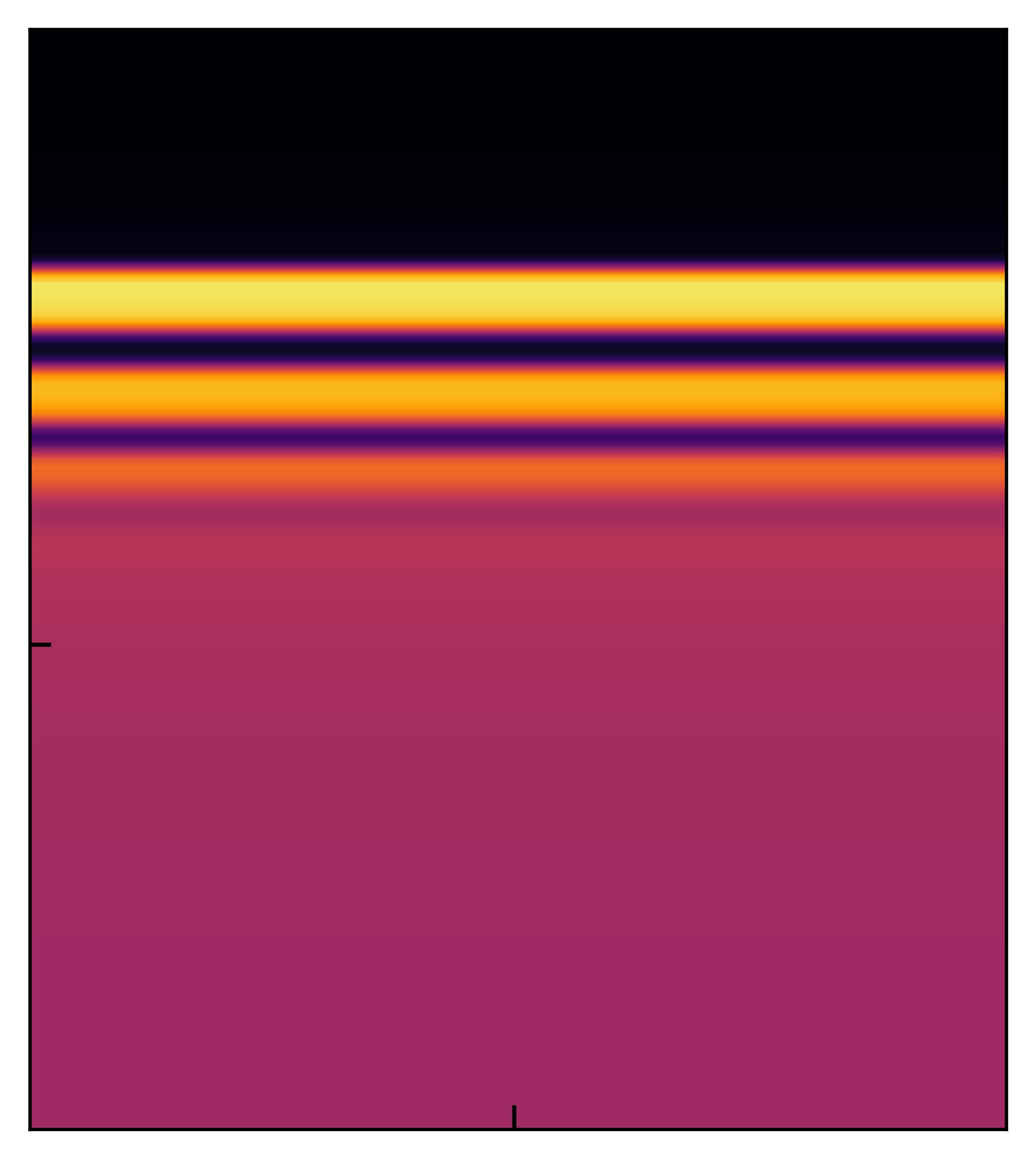}} &
                \raisebox{-0.5cm}{\includegraphics[width=0.15\textwidth]{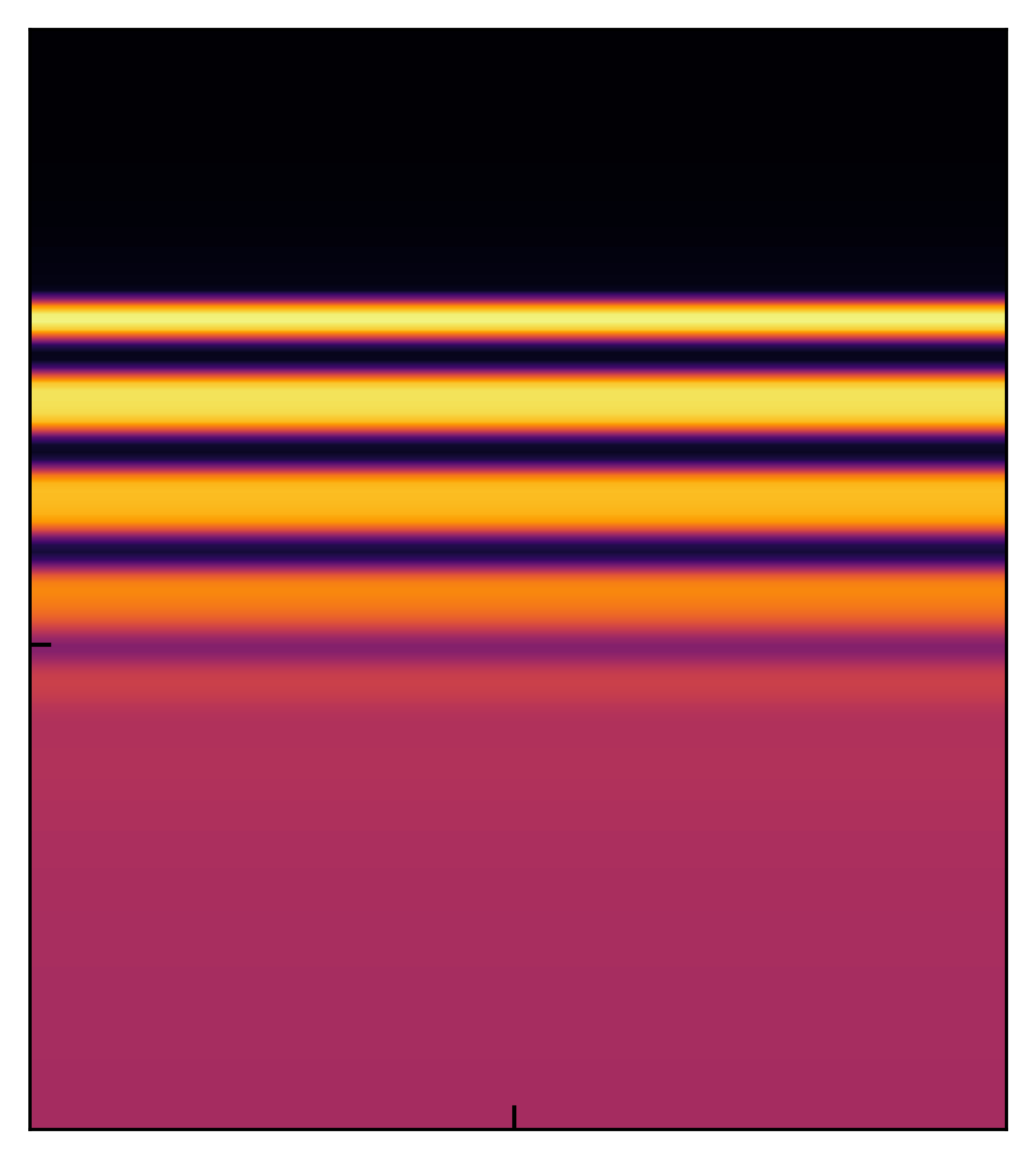}} &
                \raisebox{-0.5cm}{\includegraphics[width=0.15\textwidth]{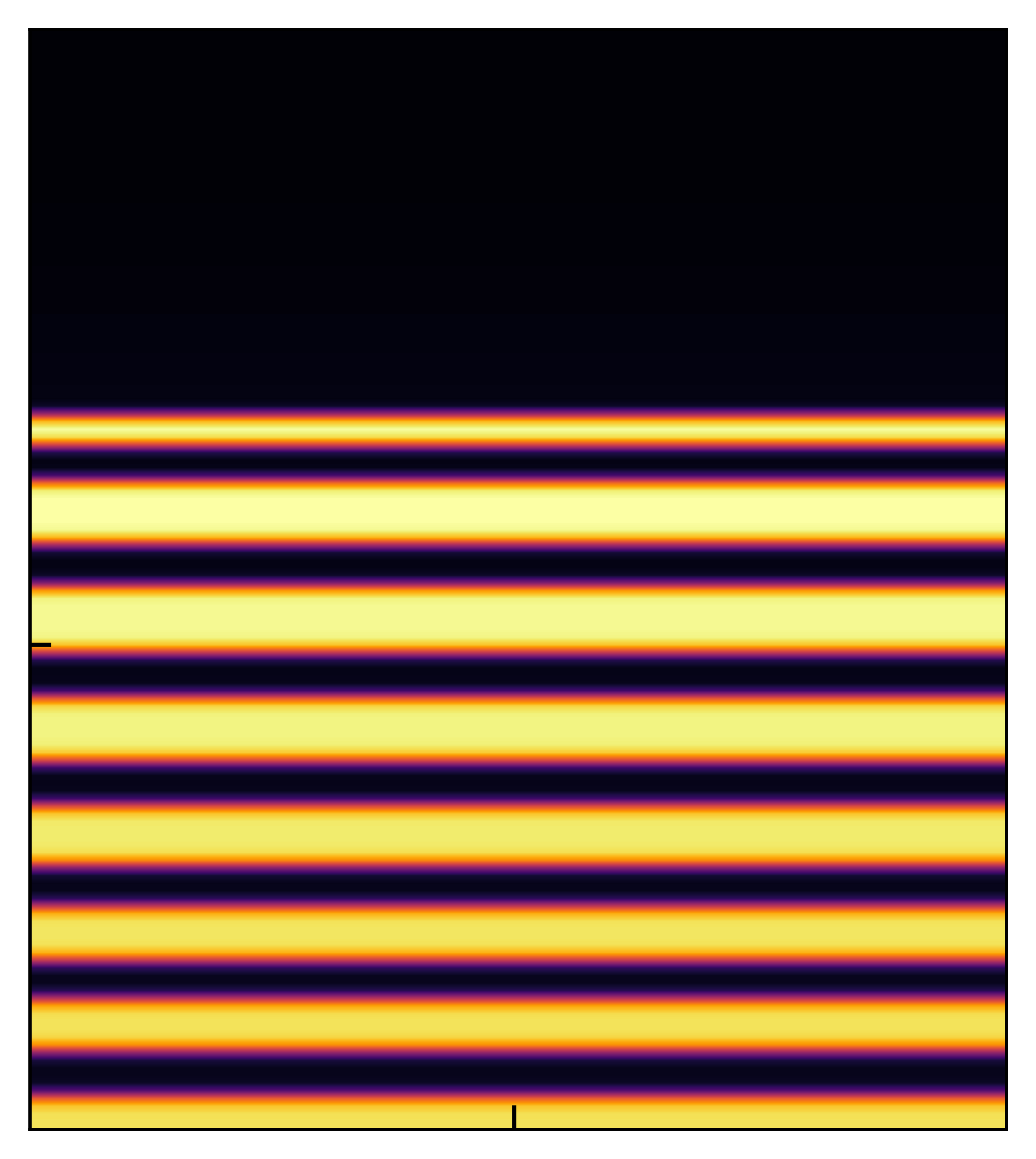}} \\
                \hline
                \raisebox{0.4cm}{\textbf{(b)}} & 
                \raisebox{-0.5cm}{\includegraphics[width=0.15\textwidth]{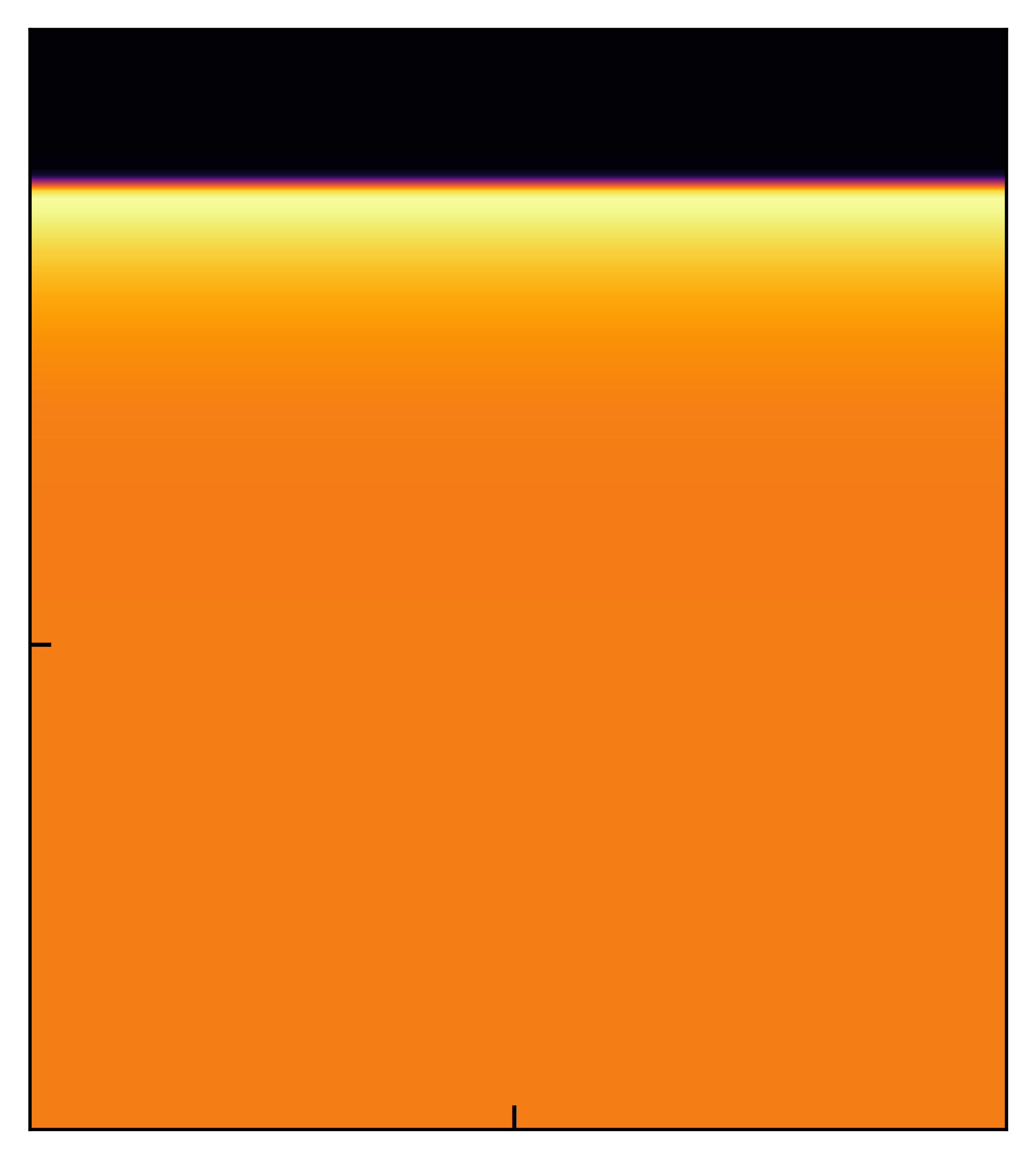}} &
                \raisebox{-0.5cm}{\includegraphics[width=0.15\textwidth]{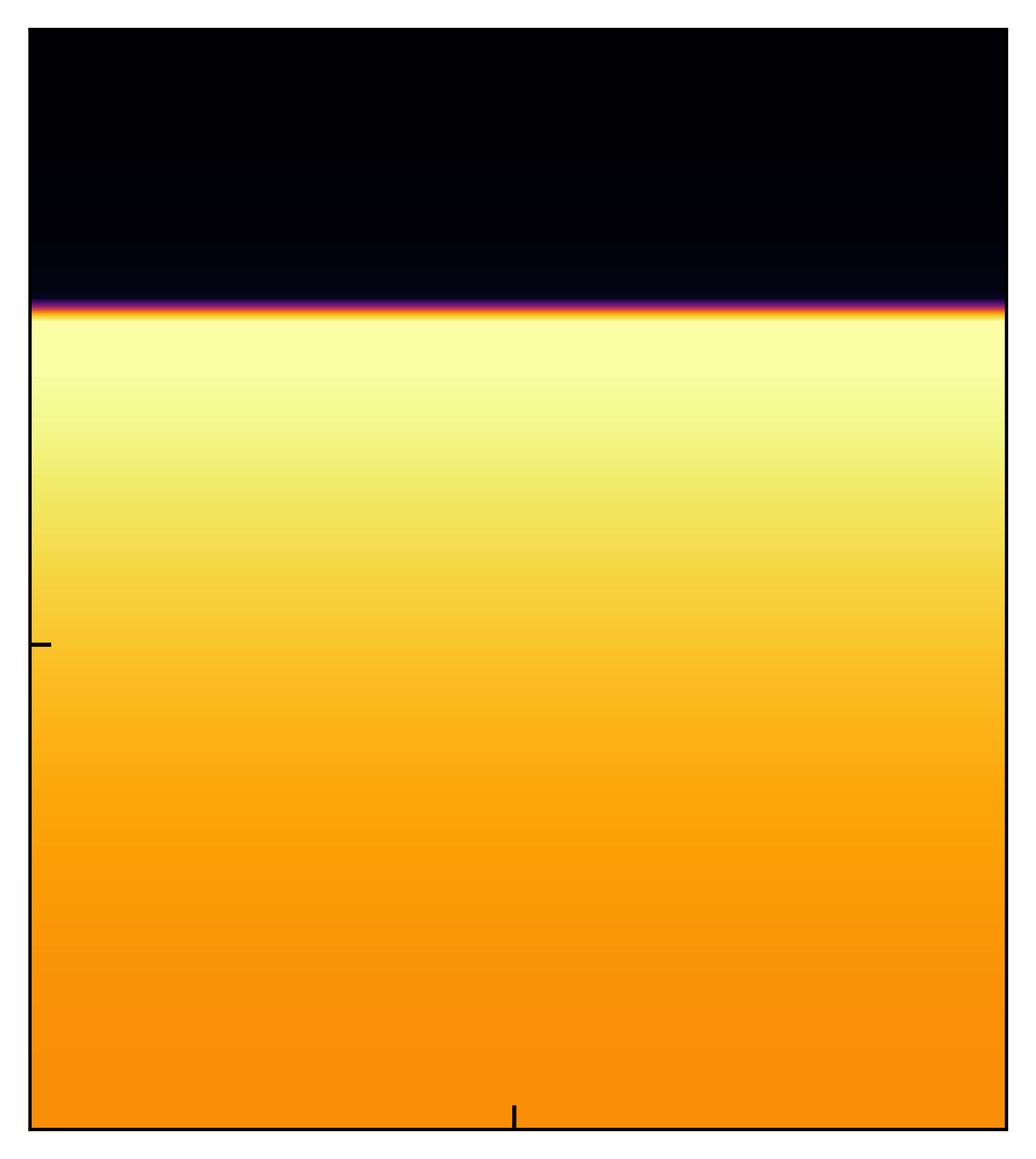}} &
                \raisebox{-0.5cm}{\includegraphics[width=0.15\textwidth]{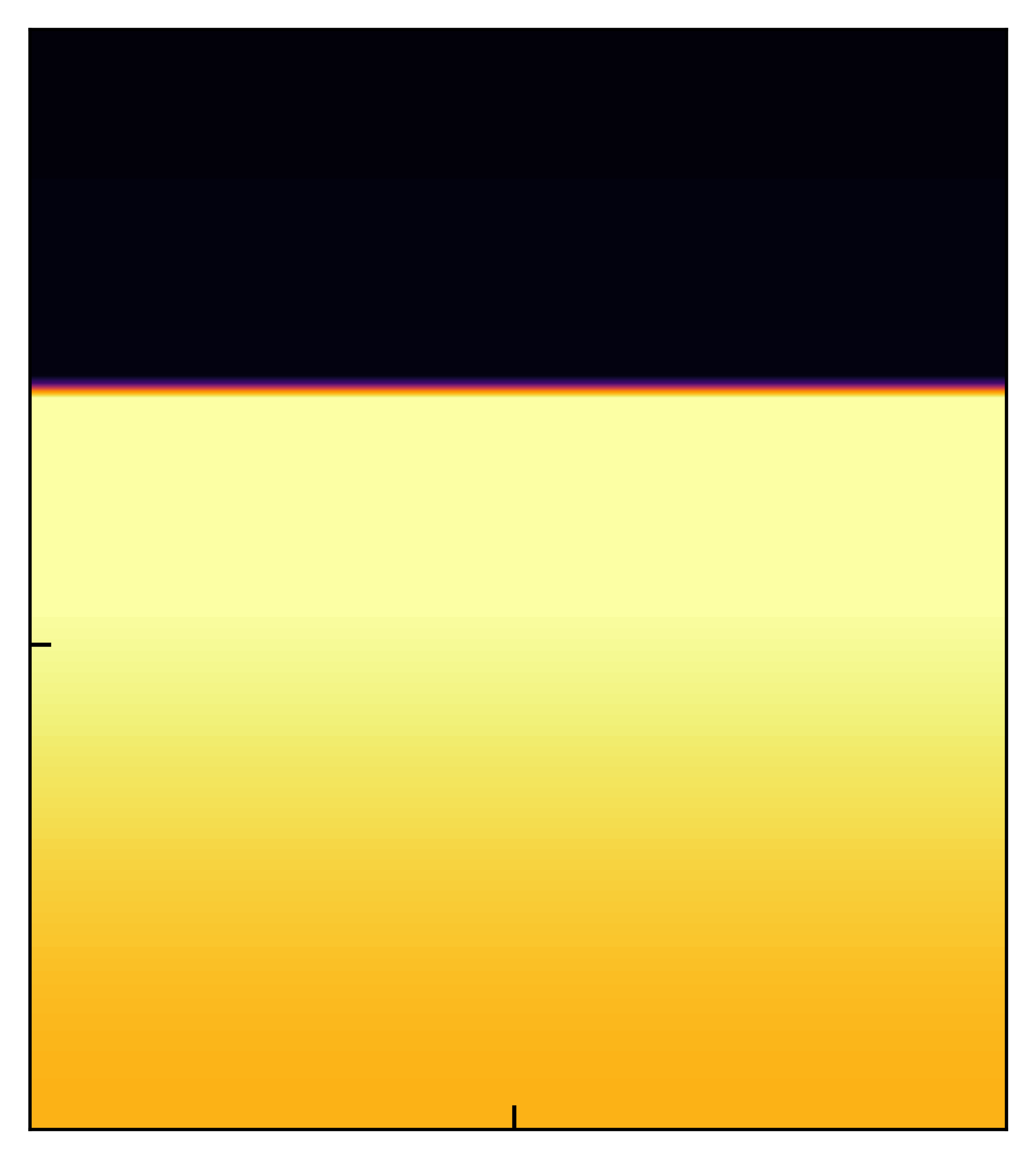}} &
                \raisebox{-0.5cm}{\includegraphics[width=0.15\textwidth]{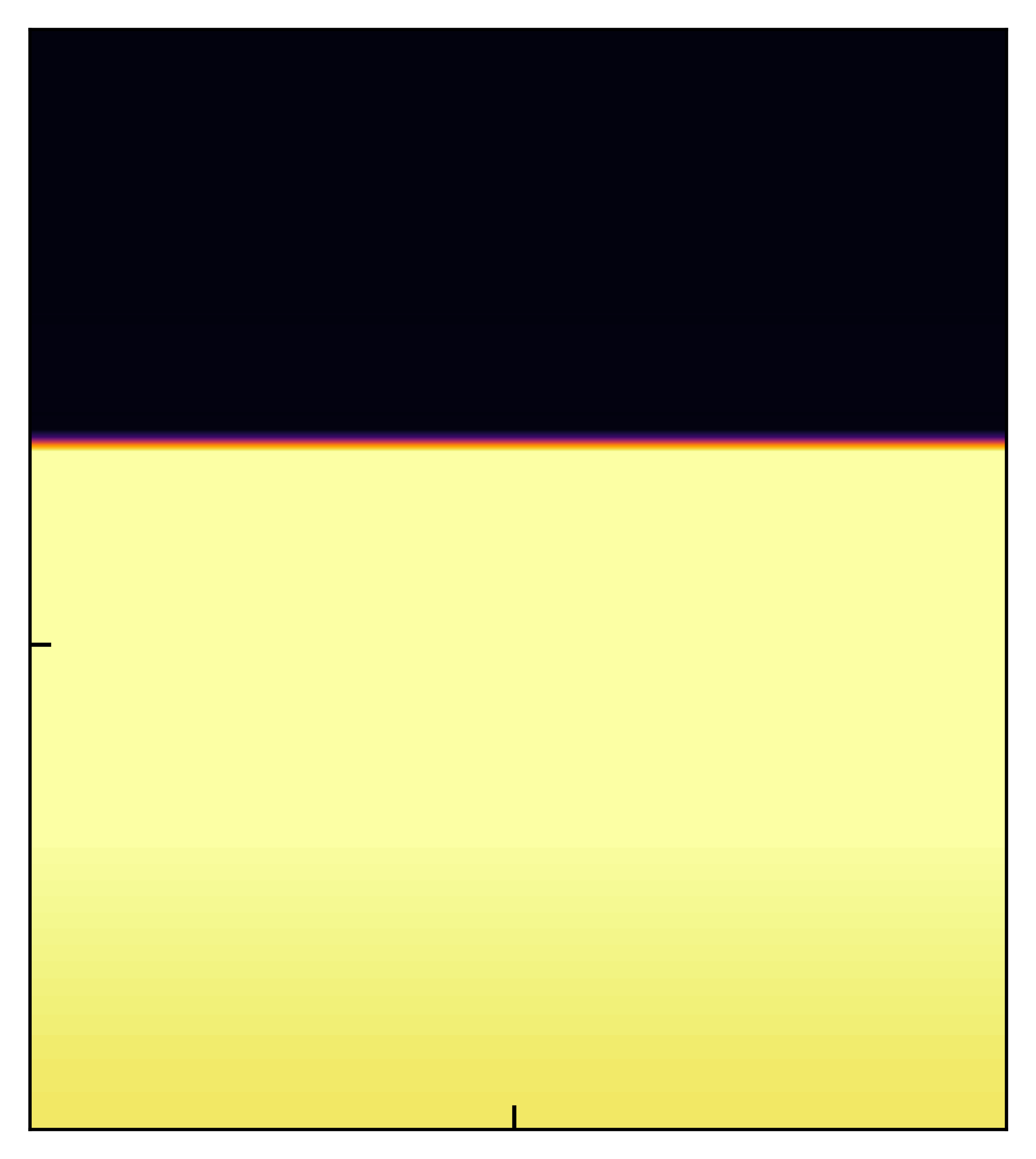}} \\
                \hline
                \raisebox{0.4cm}{\textbf{(c)}} & 
                \raisebox{-0.5cm}{\includegraphics[width=0.15\textwidth]{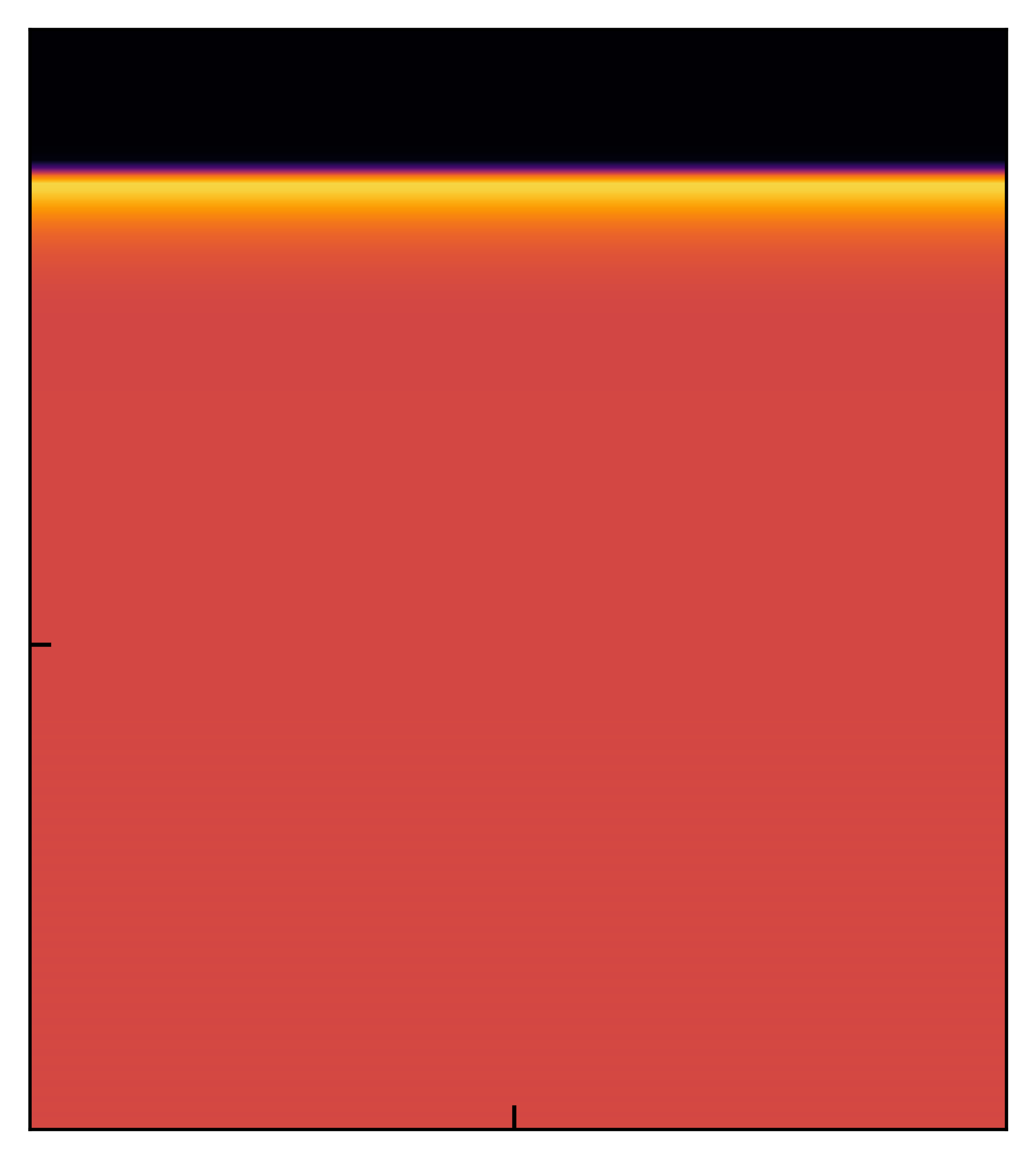}} &
                \raisebox{-0.5cm}{\includegraphics[width=0.15\textwidth]{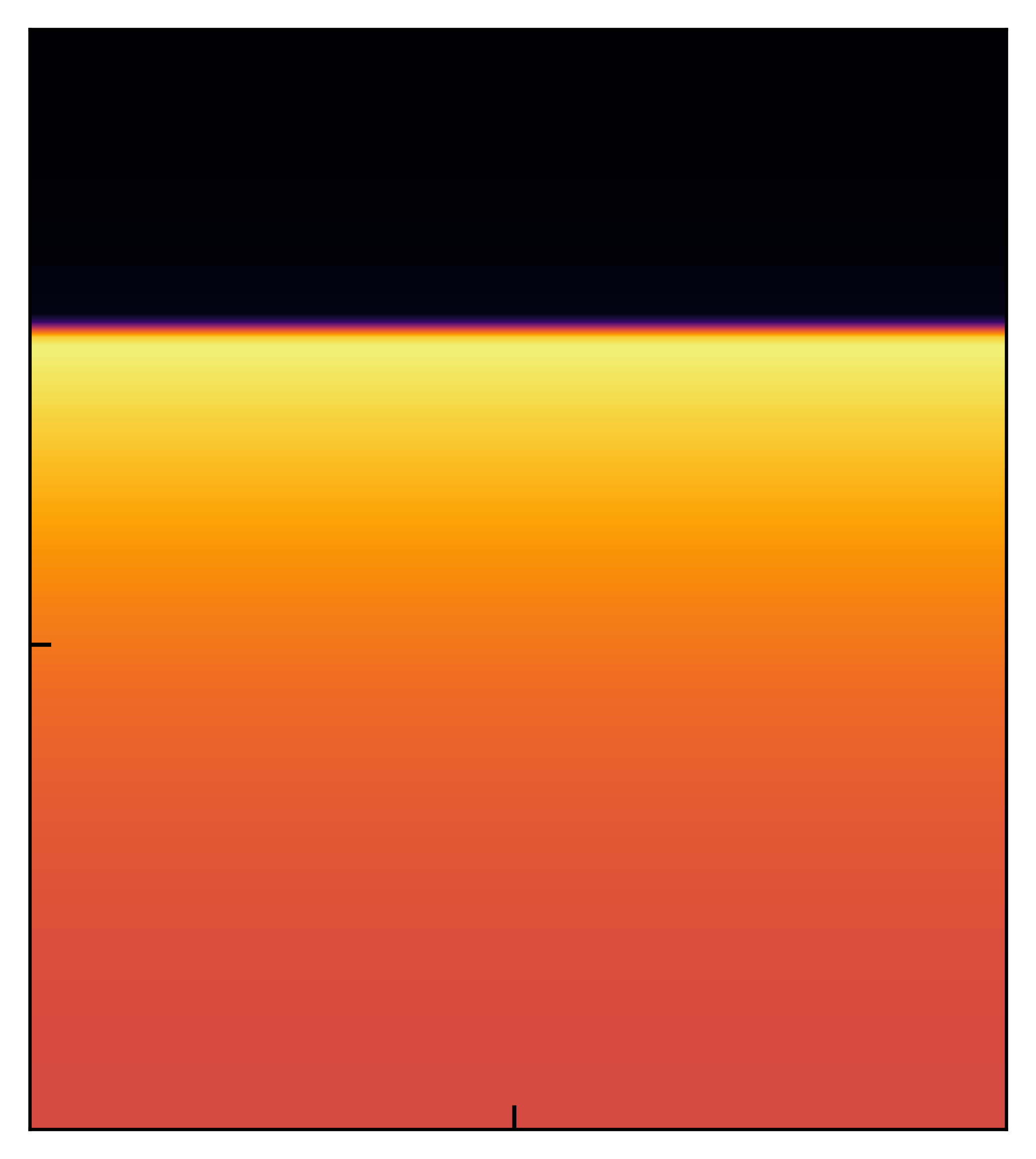}} &
                \raisebox{-0.5cm}{\includegraphics[width=0.15\textwidth]{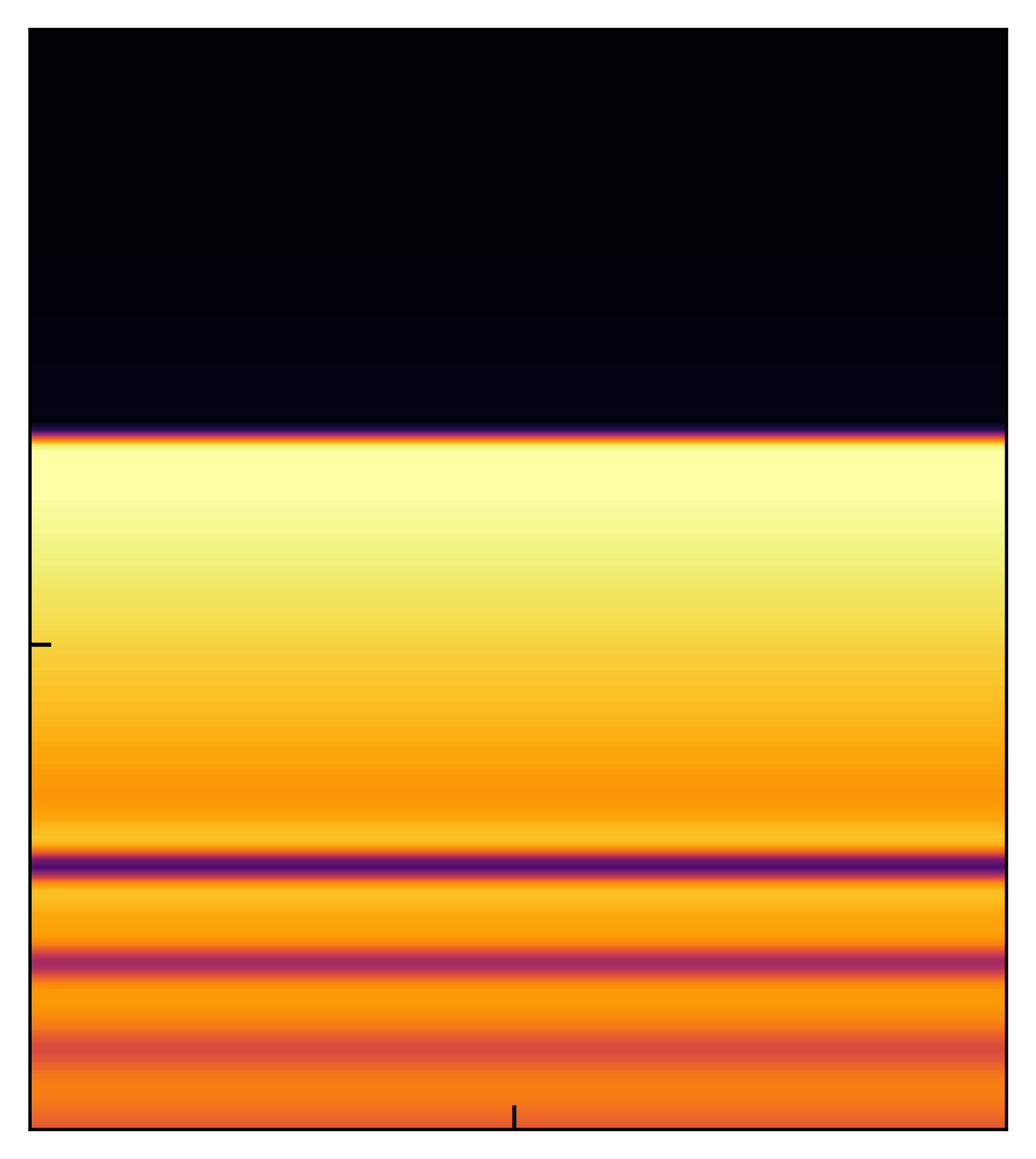}} &
                \raisebox{-0.5cm}{\includegraphics[width=0.15\textwidth]{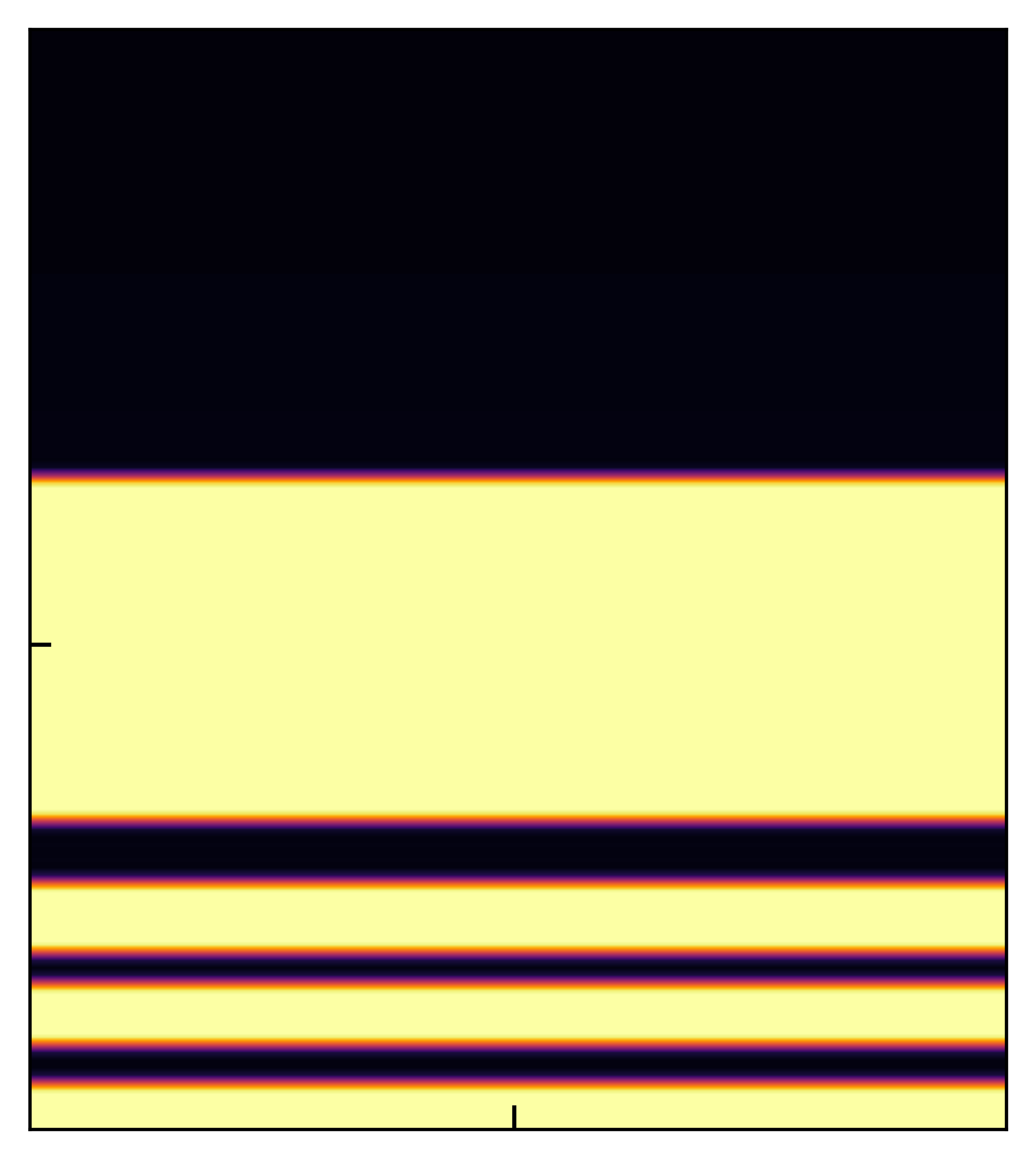}} \\
                \hline
            \end{tabular}
        \end{minipage}%
        \hfill
        \begin{minipage}{0.08\textwidth}
            \centering
            \includegraphics[width=\textwidth]{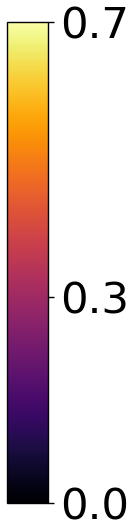}
        \end{minipage}
        
    \end{subfigure}
    \vspace{0.5em}

    \begin{subfigure}[b]{0.35\textwidth}
        \centering
        \includegraphics[width=1\textwidth,trim={1cm 0.5cm 1cm 0.5cm},clip]{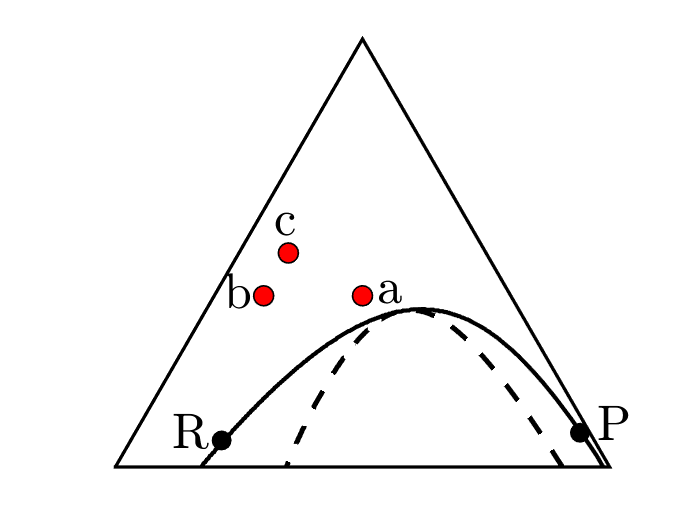}
    \end{subfigure}

    \caption{ Illustration of three behaviors observed in the NIPS process simulations, with all species having equal mobility ($M_p = M$): (a) Phase separation; (b) Delayed phase separation; (c) No phase separation. The initial composition ($\phi_p^0$, $\phi_n^0$) for each case is: (a) (0.3, 0.3); (b) (0.5, 0.1); (c) (0.4, 0.1). The evolution of the polymer volume fraction is shown at four distinct simulation times (each time is multiplied by $10^5$): (a) $[0.075, 0.22, 0.45, 1.5]$; (b) $[0.05, 1, 2.5, 5]$; (c) $[0.015, 0.75, 1.95, 3]$. The three initial compositions are indicated by points on the phase diagram shown below. \change{At final time the composition of the different phases that can be seen above  are independent of the initial film composition. They  are represented by the P (polymer poor) and R (polymer rich)  points on the phase diagram.}}
    \label{fig:Without noise}
\end{figure}

When the film composition is close to the critical point, the non-solvent is actually diffusing in the film and this process drives the film composition in the  spinodal domain. As a result, there is a phase separation that leads to formation of bands due to the symmetry breaking induced by the interface (Figure \ref{fig:Without noise}a). The  pattern is reminiscent of the so-called surface-directed spinodal decomposition (SDSD) \cite{jaiswal2012,jones1991,Ball1990}. This is shown in figure \ref{fig:Without noise} where a sequence of surface plot is presented. It must be noted that, contrarily to what is observed in classical SDSD, the band regime spans the whole simulation domain while in the case of the SDSD, its extent is limited by the competition between the spontaneous growth of the spinodal instability in the domain and the propagation of the band pattern normal to the surface~\cite{Guenoun1990}. The difference lies in the fact that the film is initially stable here which implies that there is no spontaneous spinodal decomposition in the film. \\

When the film composition is far from the critical point, there is no phase separation (Figure \ref{fig:Without noise}b). The diffusion process leads to the formation of two phases at equilibrium on the binodal curve: the bath and the film.  In this case,  the composition of the film evolves: the volume fraction of polymer increases in the film and the film thickness decreases.This is likely due to the fact that the film composition cannot enter the spinodal domain through \textit{normal} diffusion since  the spinodal domain is the domain where there is no normal diffusion. This was already discussed in\cite{cohen1979}.\\

Between these two cases, a transitional regime is observed (Figure \ref{fig:Without noise}c). In this regime, the initial \textit{equilibration} phase is followed by phase separation at the bottom of the film, resulting in the formation of a banded pattern.\\
 
These results have been obtained in an unperturbed film, which explains well the fact that the pattern is invariant along the film. In SDSD, the competition of the growth of the perturbation in an initially perturbed film and the advance of the banded pattern induces a transition between the domain close to the interface where the pattern is anisotropic and the domain far from the interface where it is isotropic. Here the mechanism leading to the banded pattern is different. Therefore it is interesting to test whether a behaviour similar to SDSD is observed in an initially perturbed film.\\

\begin{figure}
    \centering
    \renewcommand{\arraystretch}{1.5}
    \begin{subfigure}[b]{0.8\textwidth}
        \centering
        \begin{minipage}{0.92\textwidth}
            \centering
            \begin{tabular}{|c|c|c|c|c|}
                \hline
                t & \textbf{$0.3$} & \textbf{$0.9$} & \textbf{$1.5$} & \textbf{$3$} \\
                \hline
                \raisebox{0.4cm}{\textbf{(a)}} & 
                \raisebox{-0.5cm}{\includegraphics[width=0.15\textwidth]{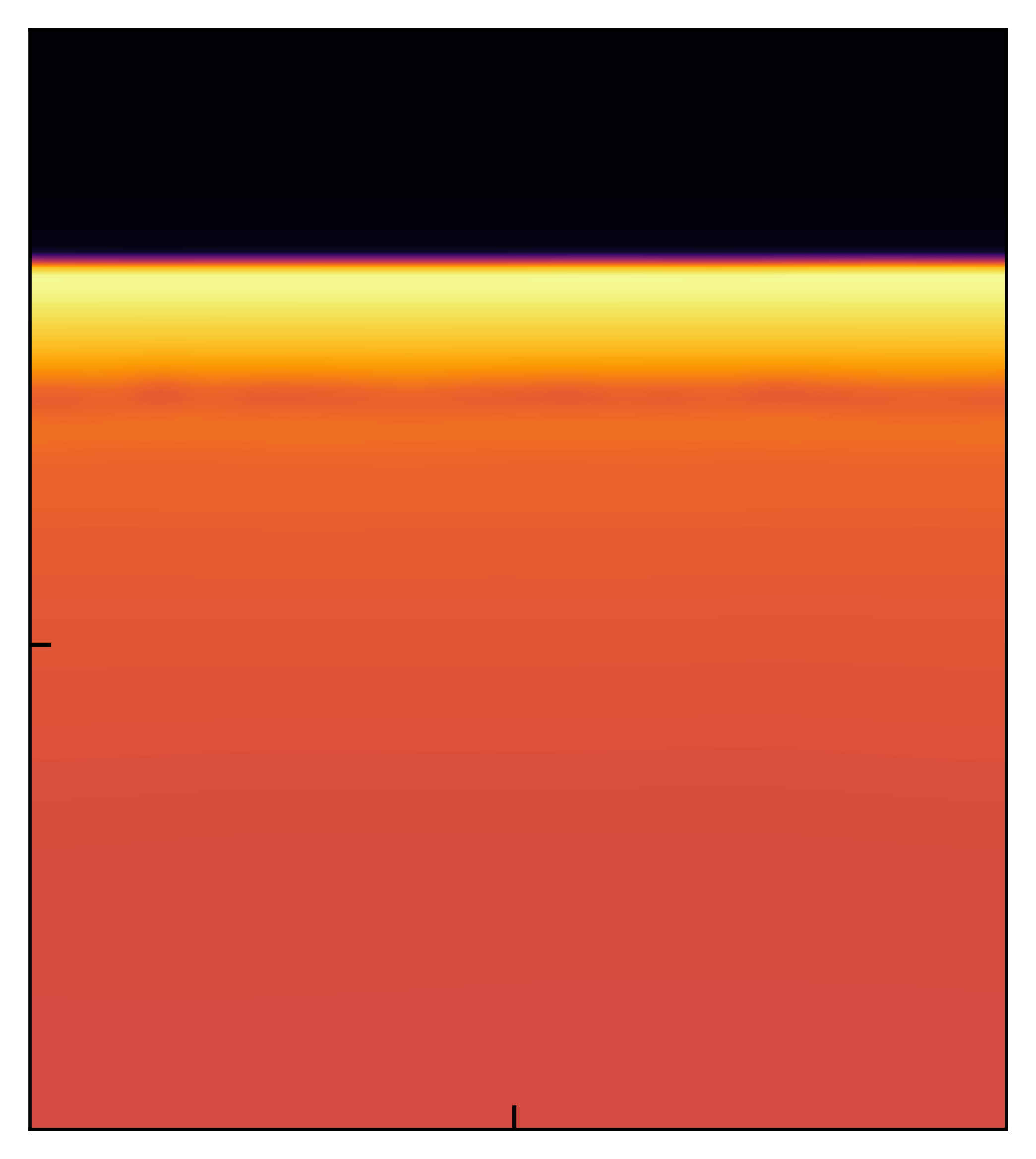}} &
                \raisebox{-0.5cm}{\includegraphics[width=0.15\textwidth]{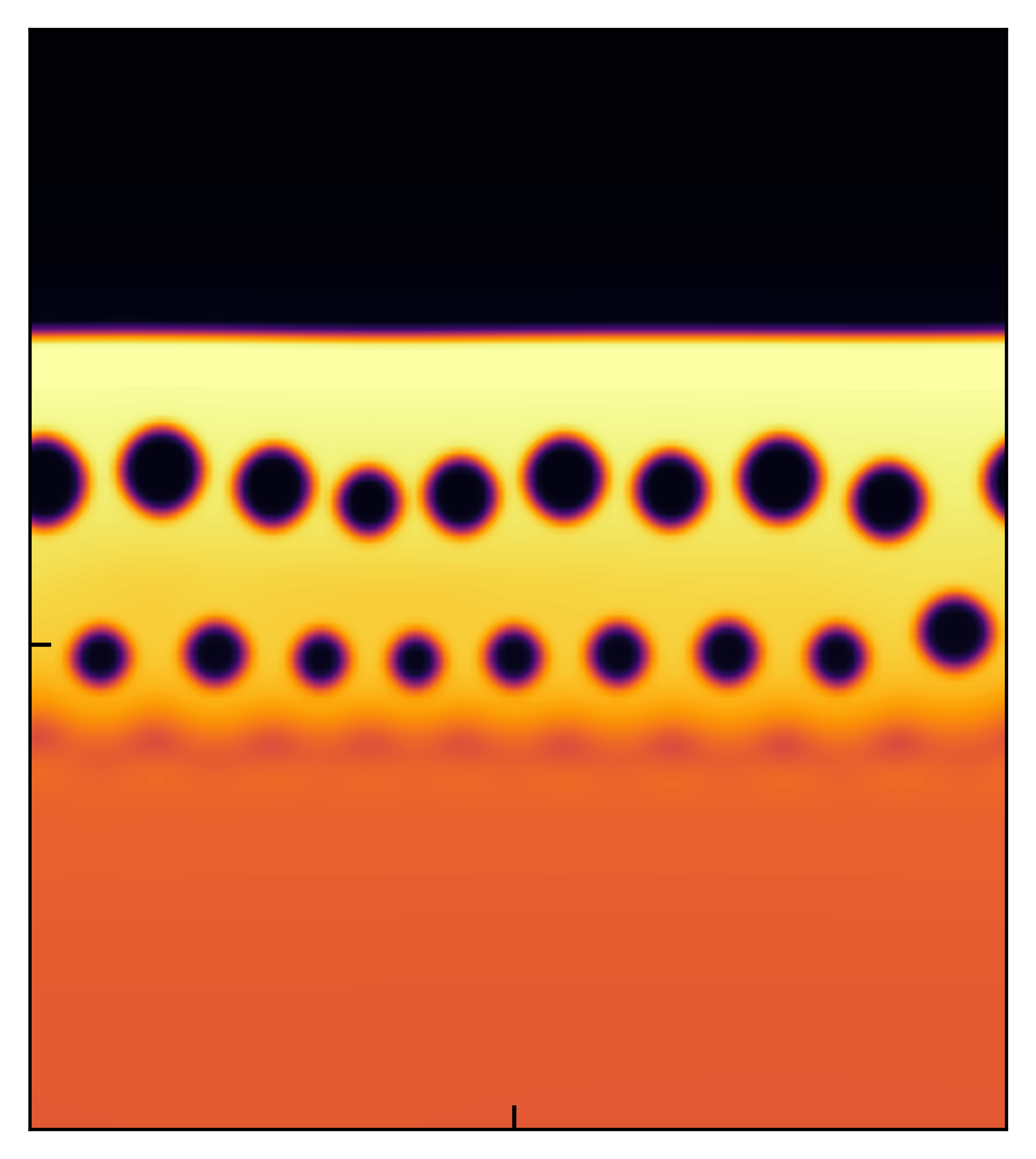}} &
                \raisebox{-0.5cm}{\includegraphics[width=0.15\textwidth]{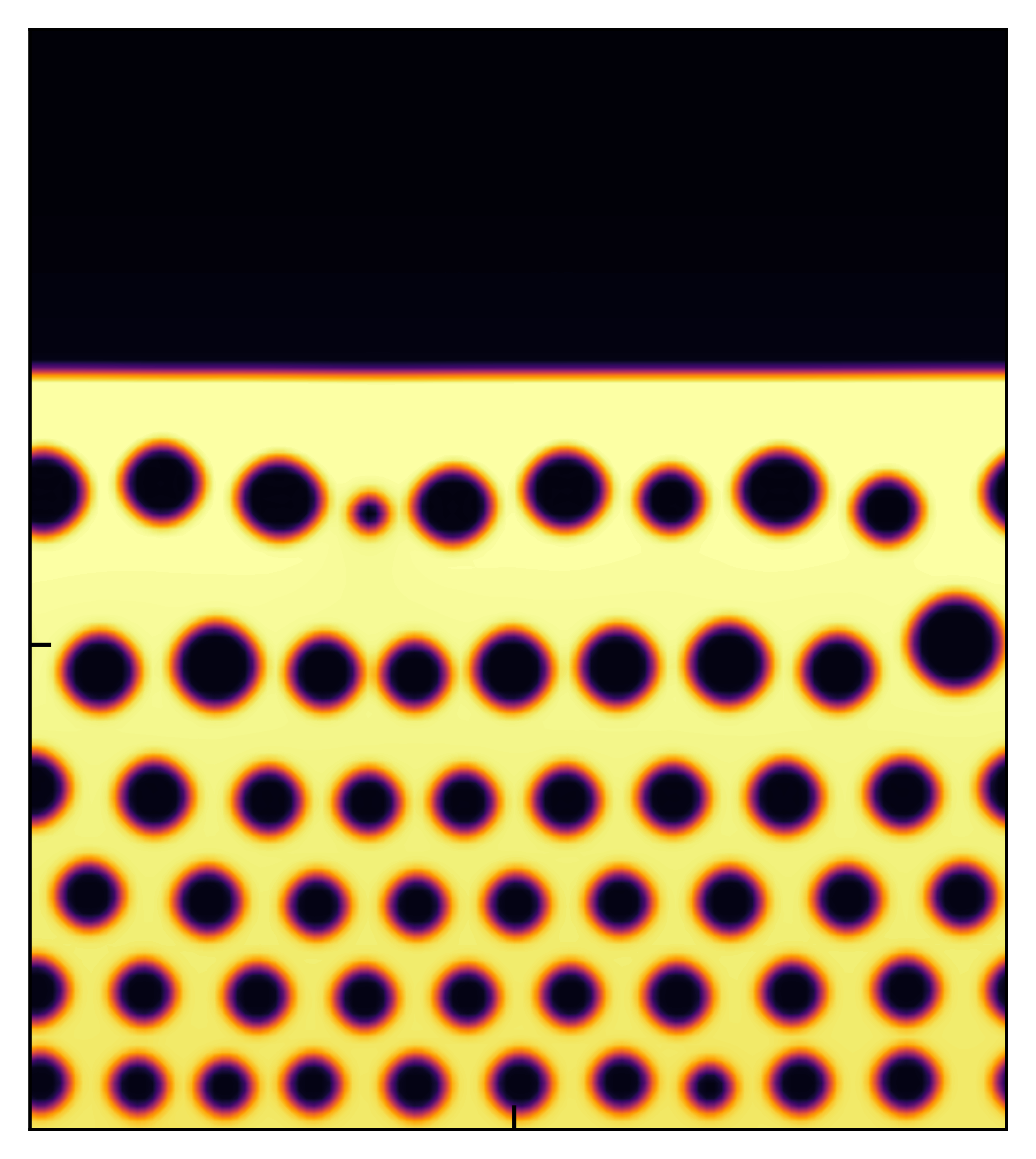}} &
                \raisebox{-0.5cm}{\includegraphics[width=0.15\textwidth]{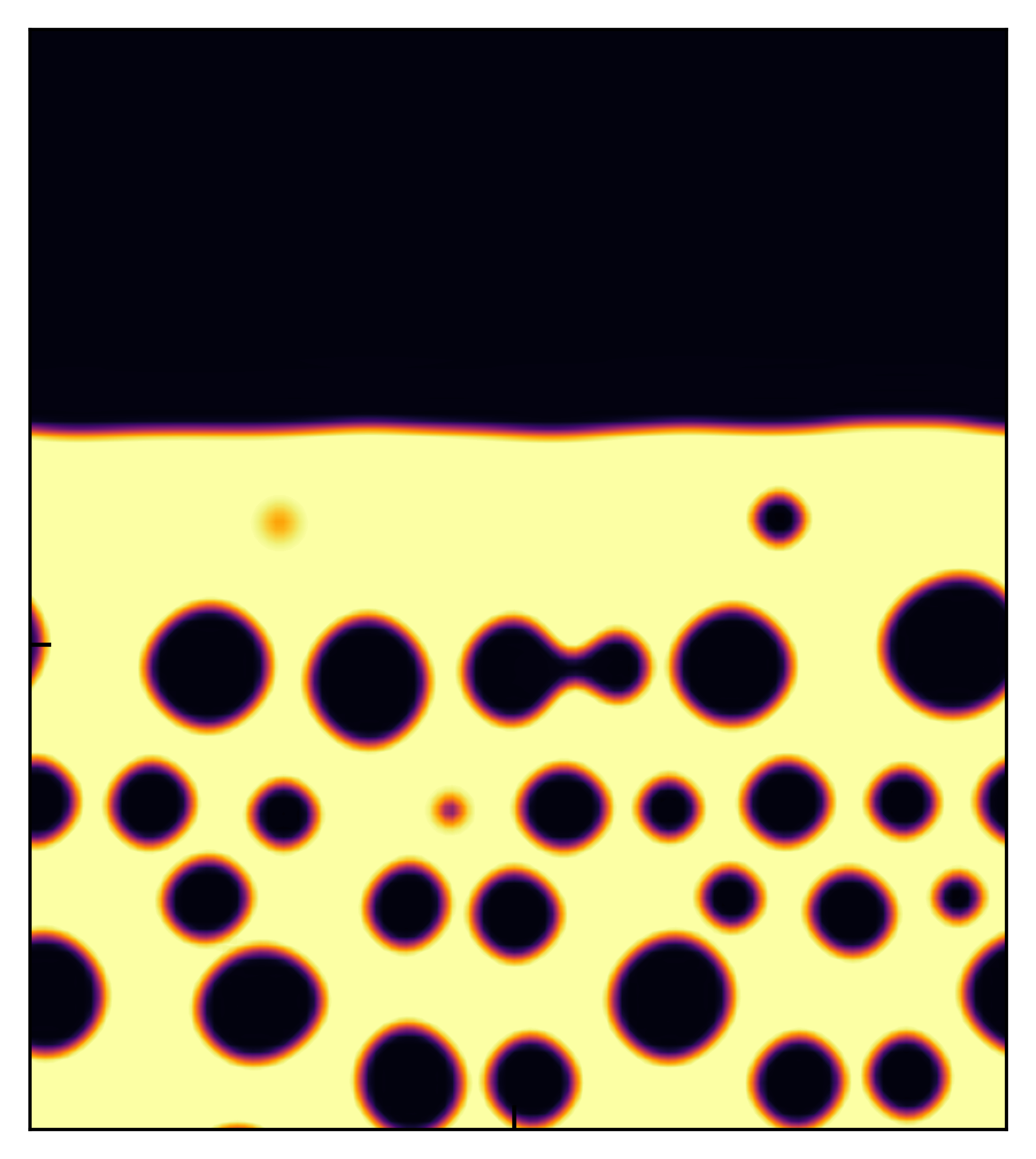}} \\
                \hline

                \raisebox{0.4cm}{\textbf{(b)}} & 
                \raisebox{-0.5cm}{\includegraphics[width=0.15\textwidth]{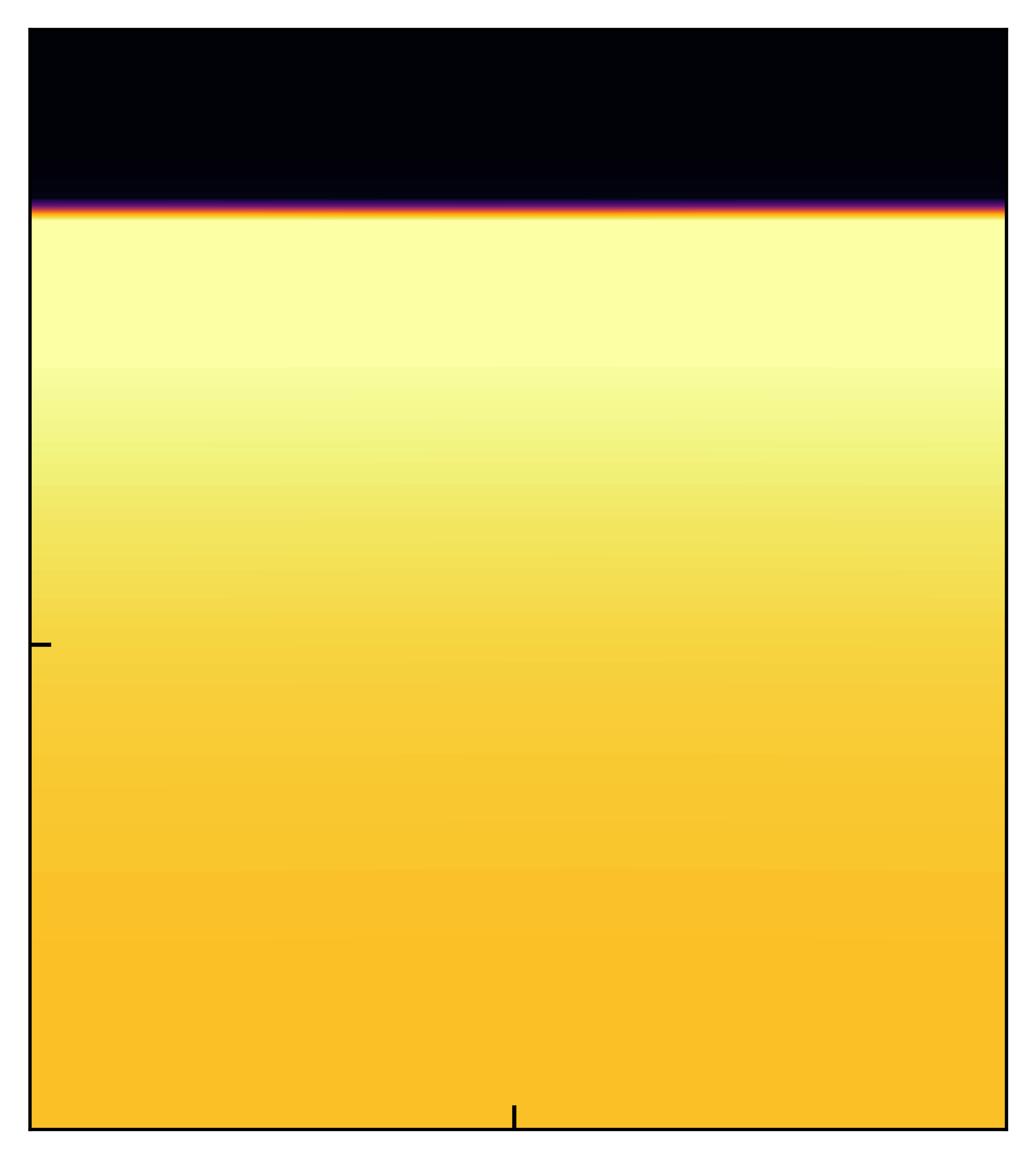}} &
                \raisebox{-0.5cm}{\includegraphics[width=0.15\textwidth]{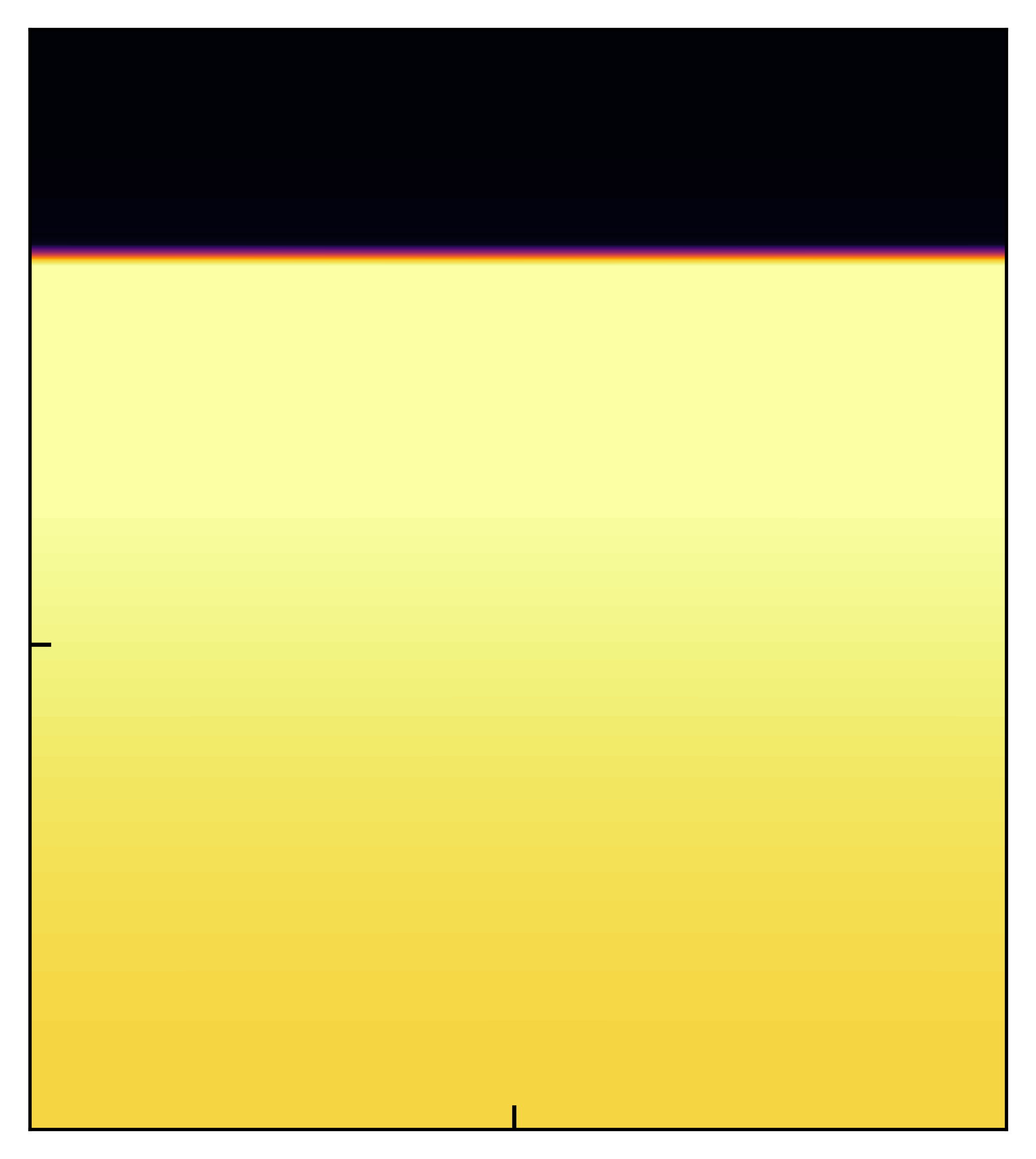}} &
                \raisebox{-0.5cm}{\includegraphics[width=0.15\textwidth]{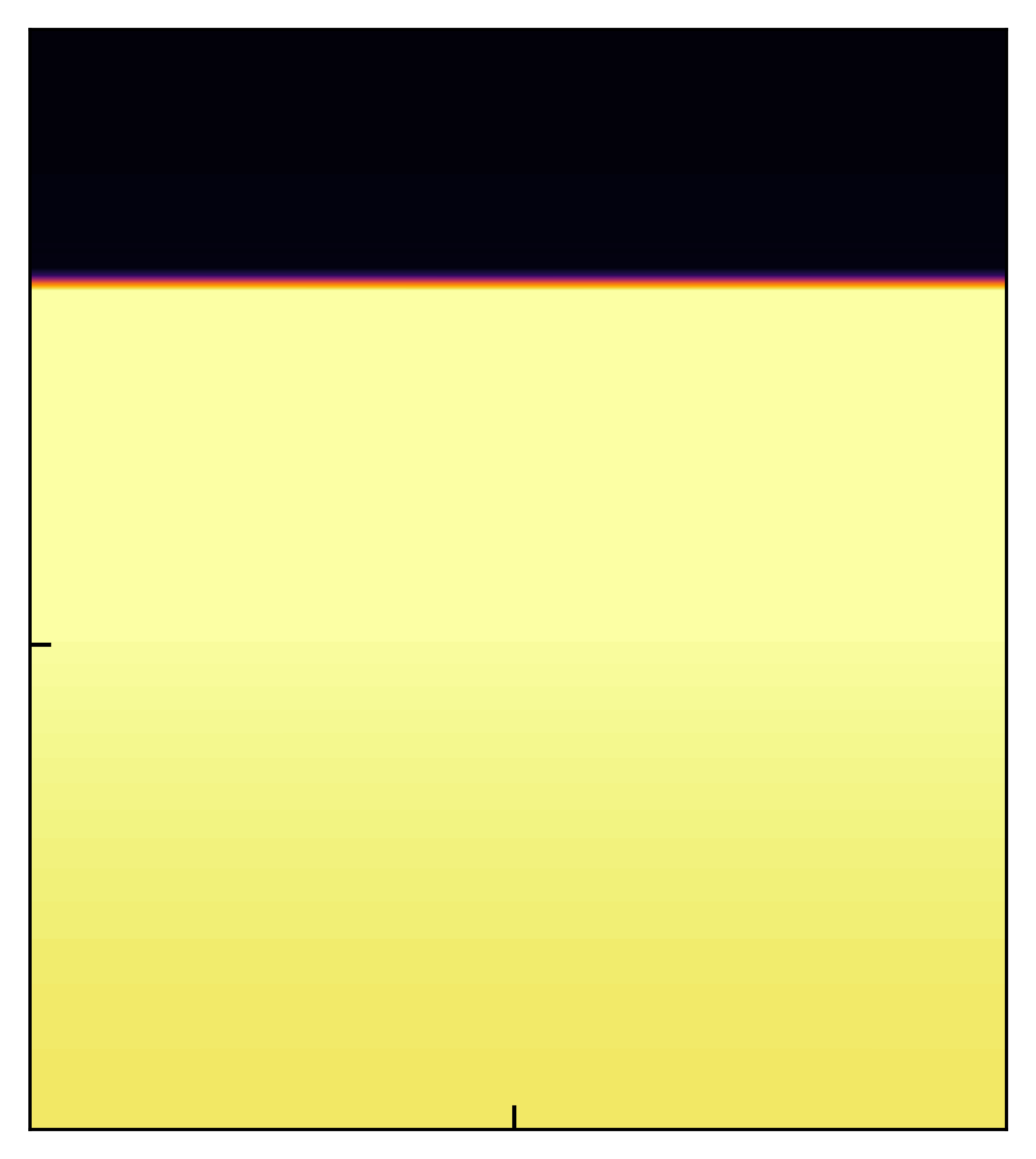}} &
                \raisebox{-0.5cm}{\includegraphics[width=0.15\textwidth]{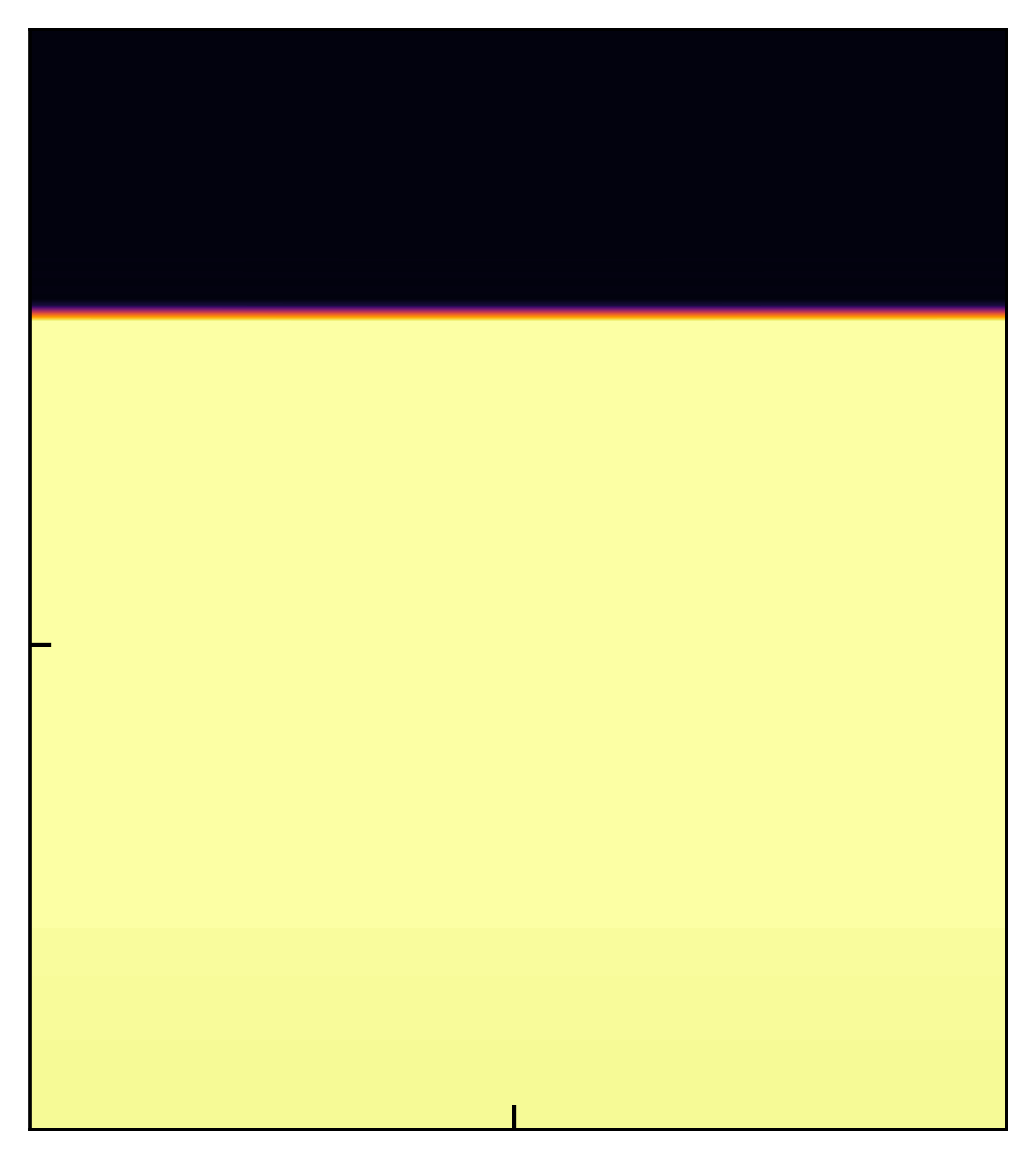}} \\
                \hline
                
                \raisebox{0.4cm}{\textbf{(c)}} & 
                \raisebox{-0.5cm}{\includegraphics[width=0.15\textwidth]{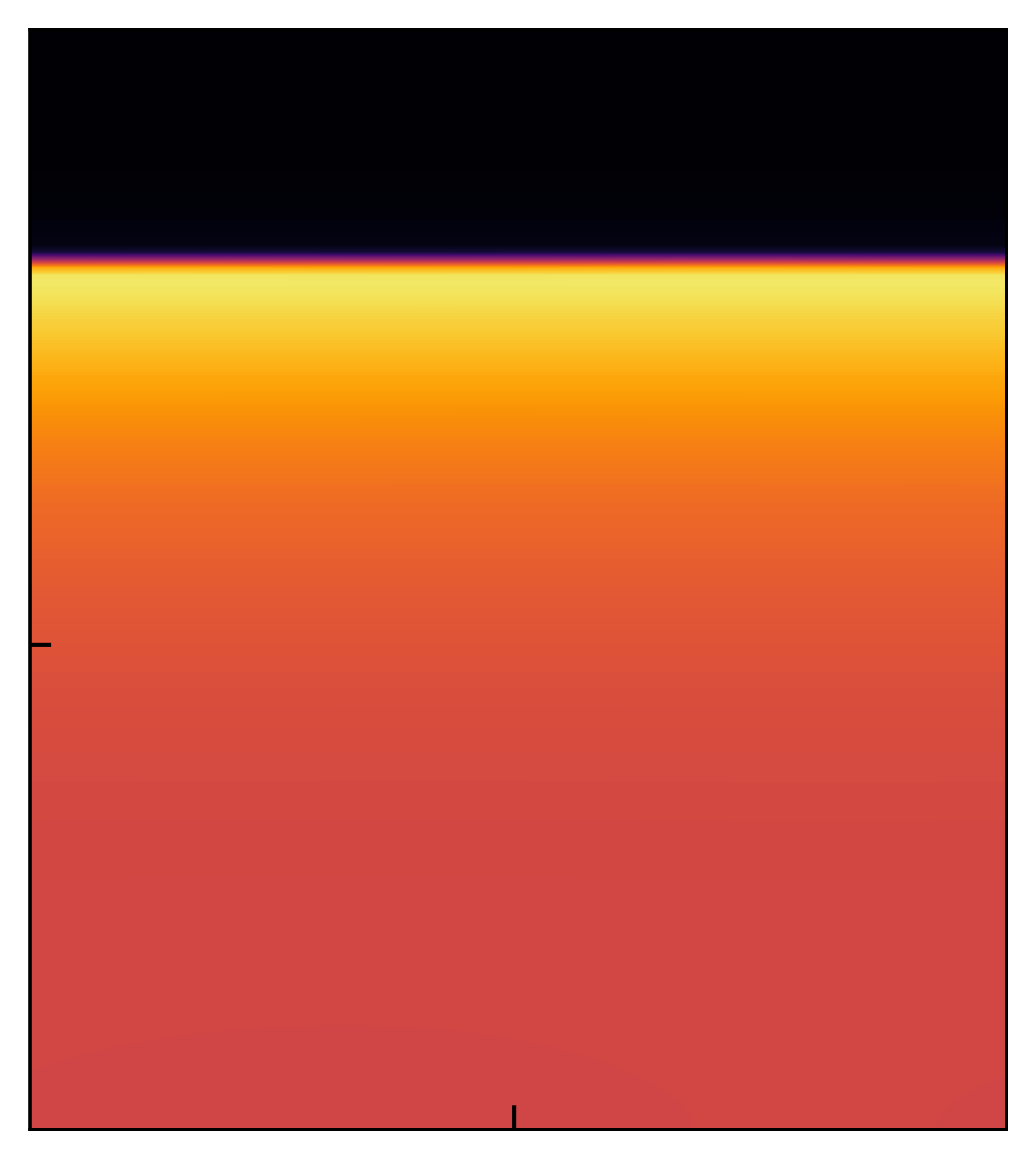}} &
                \raisebox{-0.5cm}{\includegraphics[width=0.15\textwidth]{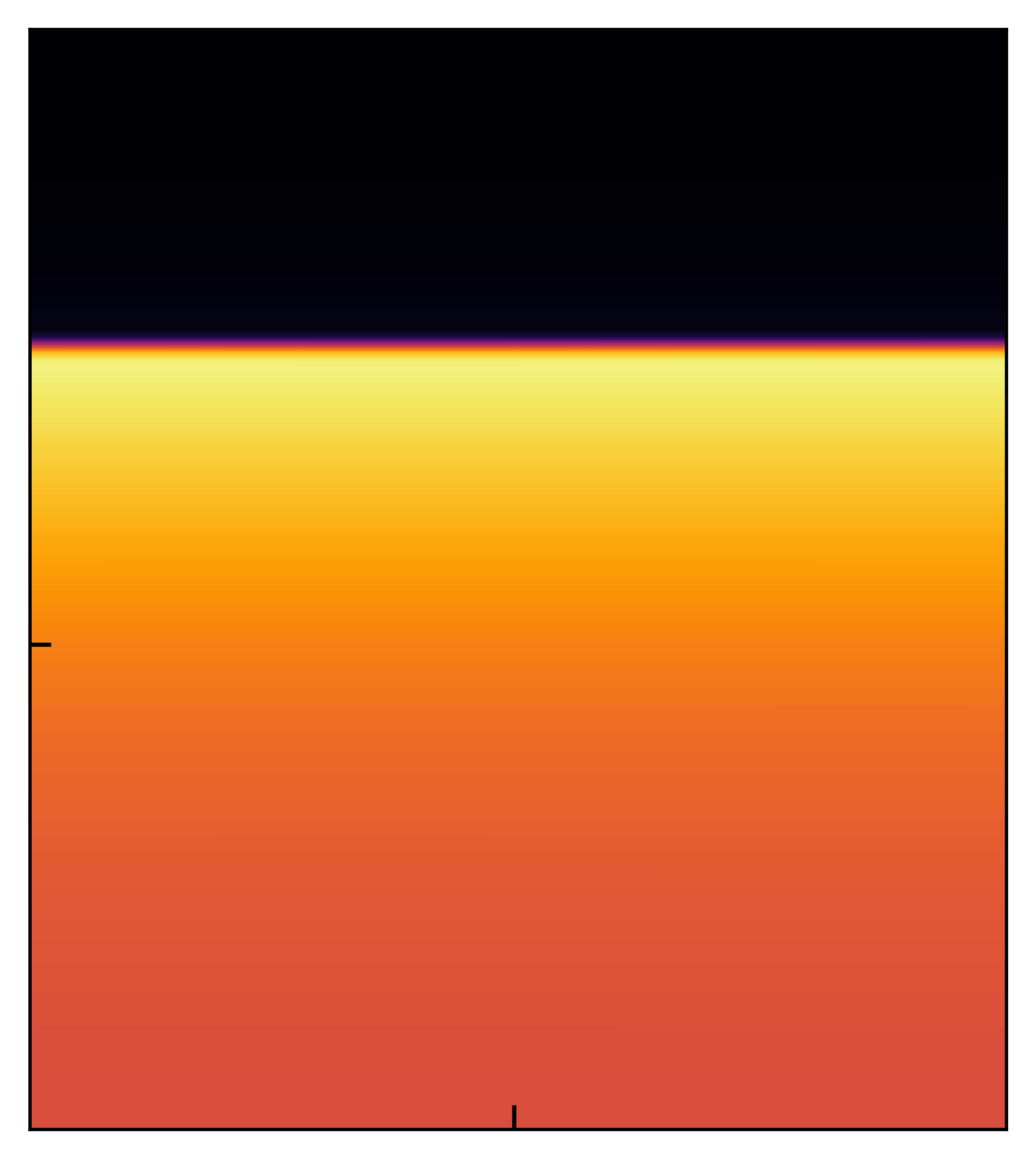}} &
                \raisebox{-0.5cm}{\includegraphics[width=0.15\textwidth]{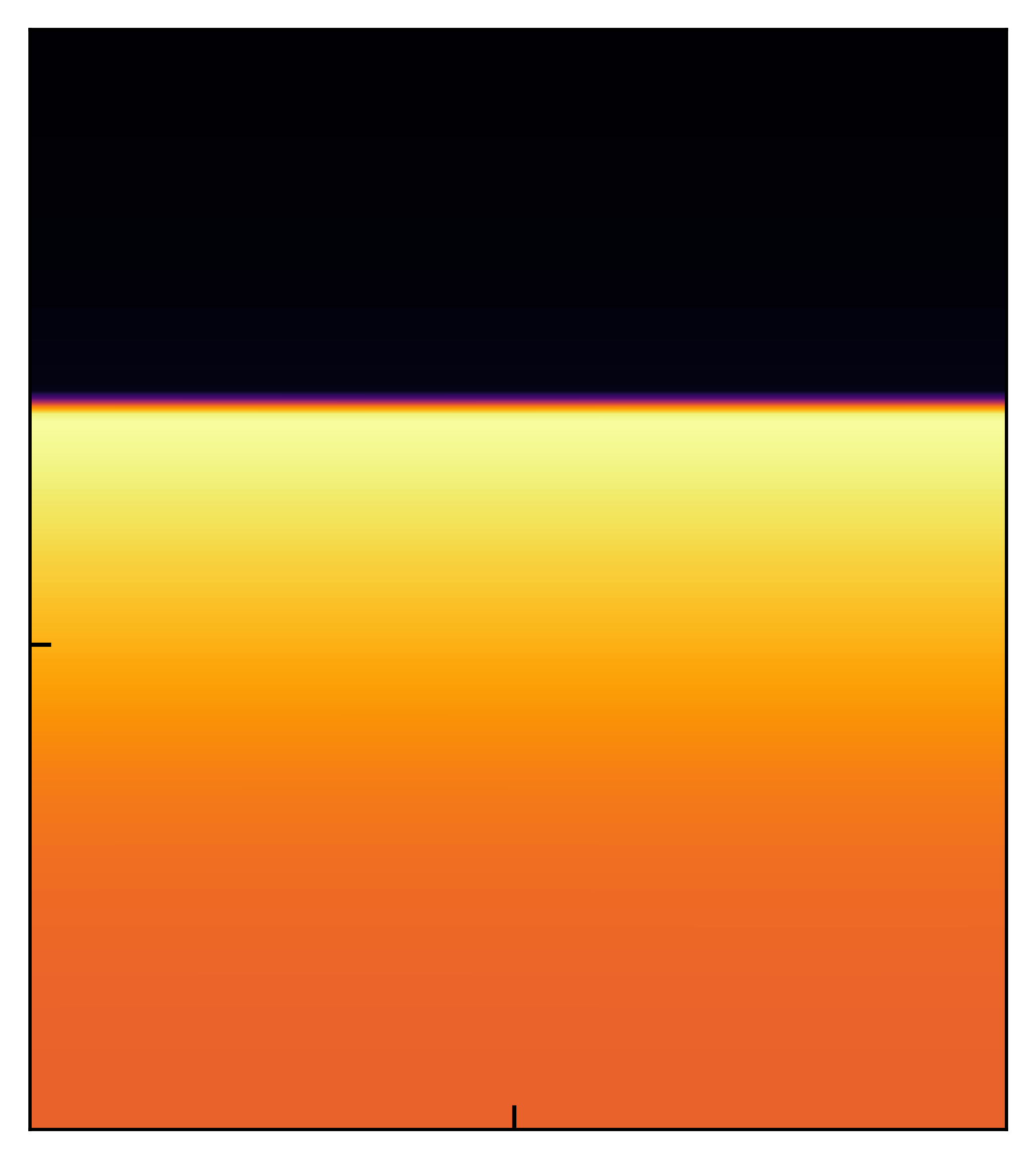}} &
                \raisebox{-0.5cm}{\includegraphics[width=0.15\textwidth]{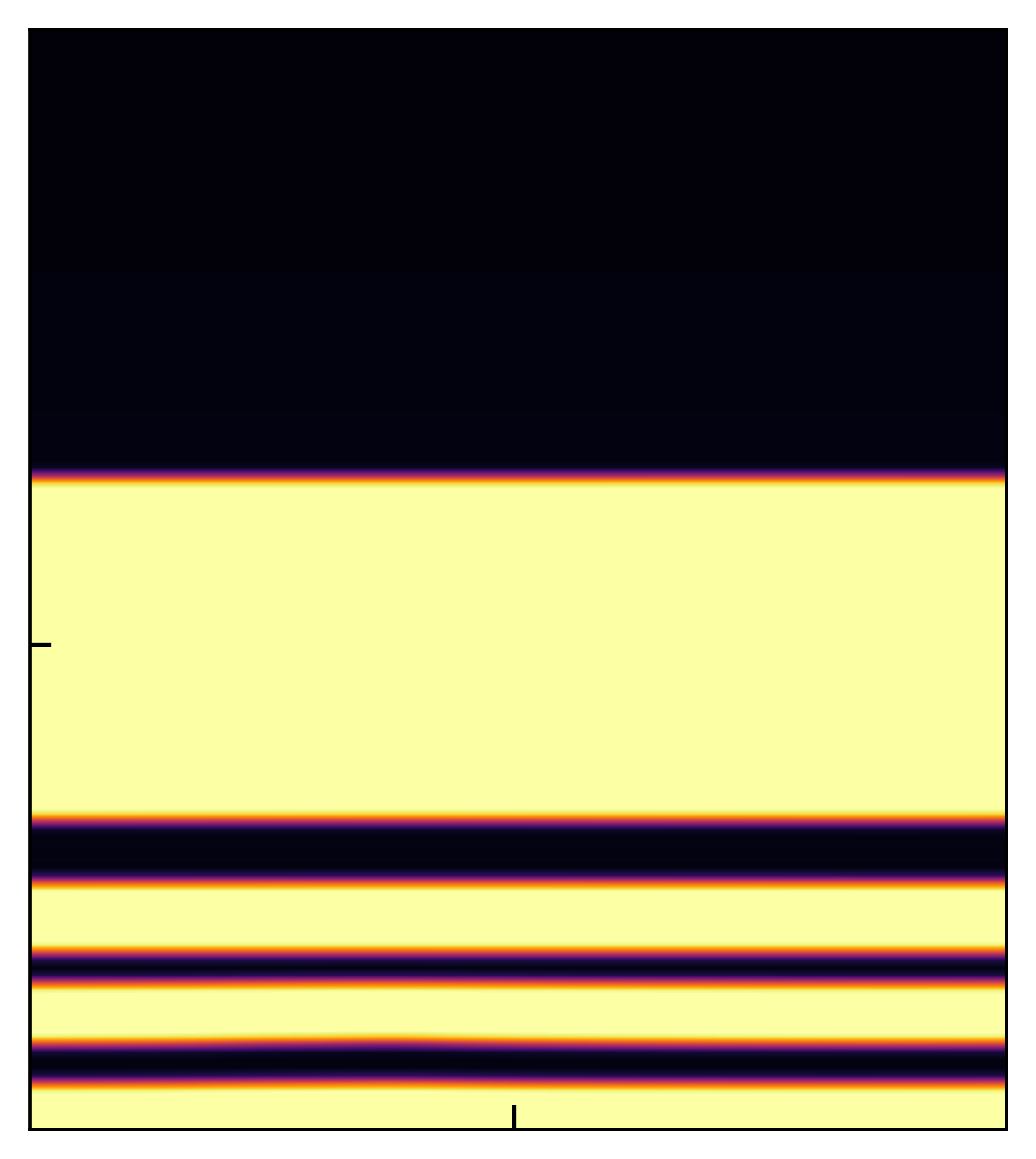}} \\
                \hline
                
            \end{tabular}
            \end{minipage}%
        \hfill
        \begin{minipage}{0.08\textwidth}
            \centering
            \includegraphics[width=\textwidth]{images/colorbar_py_vert.PNG}
        \end{minipage}
        
    \end{subfigure}
    
    \vspace{0.5em}

\begin{subfigure}[b]{1\textwidth}
    \centering
    \begin{minipage}{0.4\textwidth}
        \includegraphics[width=\textwidth]{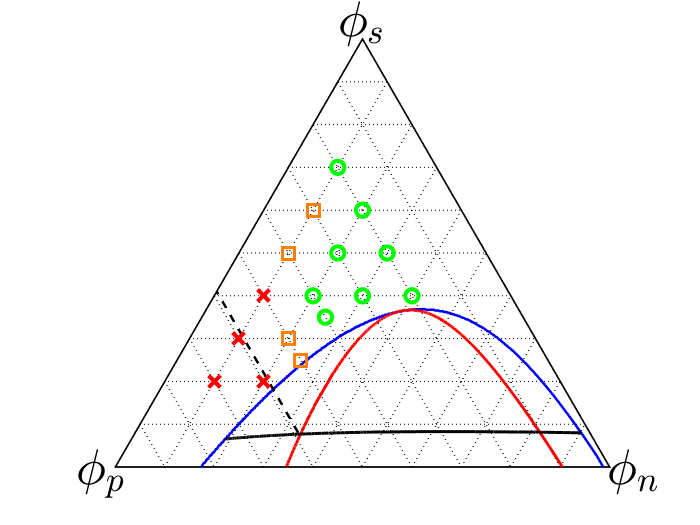}
    \end{minipage}%
    \hfill
    \begin{minipage}{0.6\textwidth}
        \centering
        \renewcommand{\arraystretch}{1.2}
        \textbf{Film initial composition that leads to:}\\[0.3em]
        \begin{tabular}{@{}ll@{}}
            \tikz[baseline=-0.5ex]\draw[green, thick] (0,0) circle [radius=1.5pt]; & Phase separation \\
            \tikz[baseline=-0.5ex]\draw[red, thick] (-1.2pt,-1.2pt) -- (1.2pt,1.2pt) (-1.2pt,1.2pt) -- (1.2pt,-1.2pt); & No phase separation \\
            \tikz[baseline=-0.5ex]\draw[orange, thick] (-1.5pt,-1.5pt) rectangle (1.5pt,1.5pt); & Transitional regime \\
        \end{tabular}
    \end{minipage}
\end{subfigure}
    \caption{ Results from 2D simulations of the NIPS process with an initially perturbed film are shown in the phase diagram, where each point signifies the film's initial composition. The initial composition ($\phi_p^0$,$\phi_n^0$) for each case : a)(0.4,0.25); b)(0.6,0.1); c)(0.4,0.1). The evolution of the polymer volume fraction is depicted at four different simulation times $t$ multiplied by $10^5$. \change{The dotted line is the  $\phi_p = \text{const}$ line that passes through  the intersection point  between the $\mu_{ns} = \mu_{ns}^{\mathrm{bath}}$ curve   and the spinodal line.}}
    \label{fig:2Dnoises}
\end{figure}

 To this purpose, we consider the evolution of an initially perturbed film to test the robustness of the banded regime. The initial volume fraction of polymer, $\phi_p^0$, is perturbed by a small amount of uniformly distributed random noise with an amplitude $\alpha = 10\%$ of $\phi_p$. The results are illustrated in Fig.~\ref{fig:2Dnoises}. Three distinct regimes are observed.\\
 
 First, far from the critical point, the same behavior is observed as in the case without initial perturbation: the film remains in the stable domain, and its composition equilibrates with that of the bath (Fig.~\ref{fig:2Dnoises}b). Second, for compositions close to the critical point, the film is driven out of the stable domain by diffusion (as in the unperturbed case), but the resulting pattern is no longer banded. Instead, a polymer-rich matrix forms, within which polymer-poor inclusions appear. These inclusions are initially organized along lines, and their positions form an alternating array: along the direction normal to the film, one observes an alternation of polymer-rich and polymer-poor regions when following the successive lines of inclusions, as illustrated by the profile along the line shown in Fig.~\ref{fig:2Dnoises}a à $t=1.5$. The subsequent coarsening process leads to a more disordered structure, resembling what is typically observed in spinodal decomposition (SDSD). Finally, as in the case without noise, there exists an intermediate regime between these two behaviors in which a banded pattern emerges (Fig.~\ref{fig:2Dnoises}c).\\

{Figure~\ref{fig:2Dnoises} summarizes the results obtained for different initial film compositions, represented on the phase diagram. Three distinct regions can be identified, each corresponding to one of the behaviors described previously. The first region, marked by green circles, includes initial compositions for which complete phase separation is observed. The second, indicated by red crosses, corresponds to configurations where no phase separation occurs. Finally, the region, marked by orange squares, corresponds to the  intermediate  behavior between these two cases.}\\

{The domain for which there is  phase separation  is small and mostly limited to the vicinity of the critical point. This is related to the fact that the diffusion path is not a straight line in the Gibbs triangle but corresponds to the shortest line when using the metric given  by the multiplication of the Hessian of the free energy  by the mobility matrix \cite{Tree2019}.}\\

{In previous studies, this effect has received limited theoretical attention, even though the mobility of macromolecules such as polymers differs significantly from that of small solvent molecules. In the following section, we analyze in greater detail the impact of the mobility matrix on diffusion trajectories and structure formation.}\\

\subsection{Slow Polymer}
 In this section, we consider the effect of varying mobilities on the pattern formation process. This is motivated by the fact that in previous studies, the equal mobility hypothesis is always used. Here we consider that chemical species can have different mobilities\footnote{Mobilities are not parameters that can be tuned, but there is no a priori reason to assume they are equal}. We limit ourselves to the simple case where the polymer mobility is taken smaller than the other chemical species. This choice is physically justified, since the polymer molecules are generally much less mobile than the solvent and nonsolvent molecules. First, we briefly discuss a limiting case that will clearly show the possible effect of mobility on the microstructure, because of the difference with the results presented in the previous section.

 If the polymer is infinitely slow,  the film will equilibrate with the bath at a composition that lies at the intersection of the lines $\phi_p=cst$ and $\mu_{ns}=\mu_{ns}^{bath}$. For a wide range of initial film composition this point lies in the spinodal domain.  Thereafter, at the infinitely long timescale of the polymer diffusion, phase separation will occur in the whole film, with a SDSD-like pattern due to the presence of the interface with the bath.  This contrasts with the case of the polymer mobility being similar to the one of solvent, for which there is a significant part of the initial film composition that does not lead to phase separation or to banded patterns. We now turn to the numerical results. \\

\begin{figure}[H]
    \centering
    \begin{subfigure}[b]{0.45\textwidth}
        \centering
        \includegraphics[width=1\linewidth]{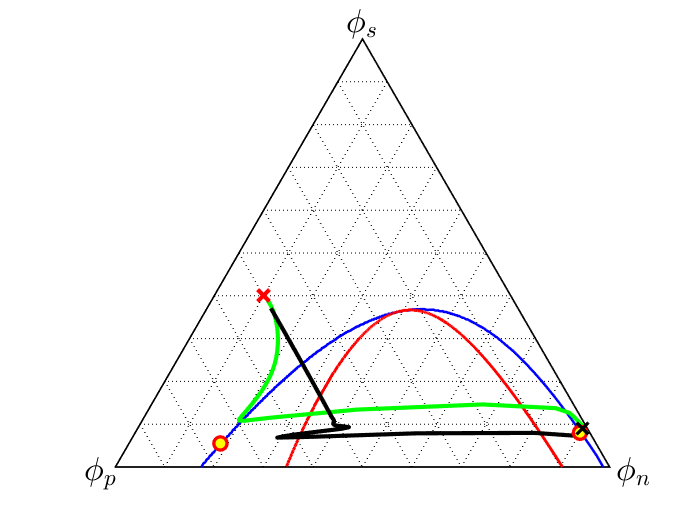}
        \caption{}
        \label{fig:GIBS_1D}    
    \end{subfigure}%
    \begin{subfigure}[b]{0.5\textwidth}
        \centering
        \includegraphics[width=1\linewidth]{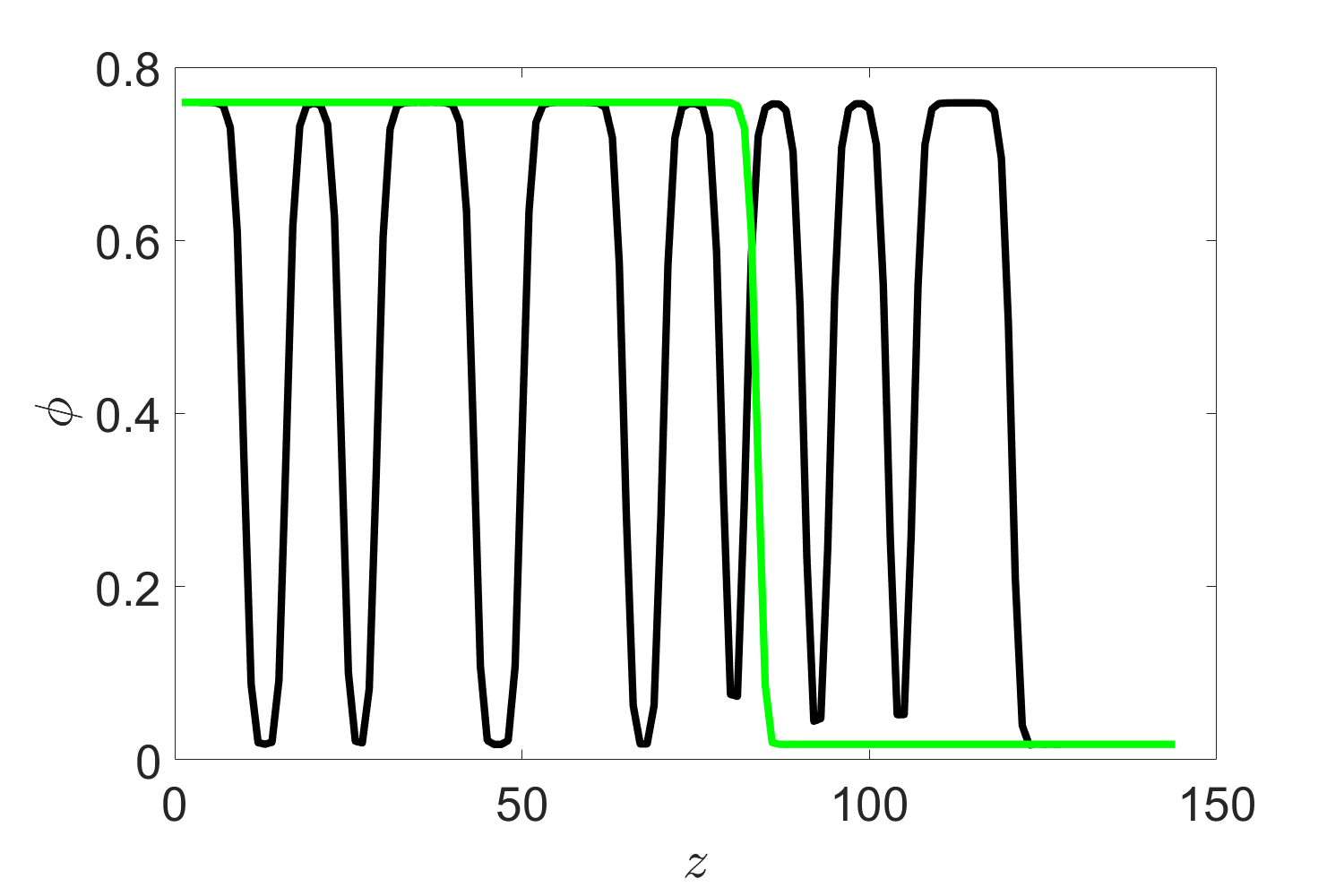}
        \caption{}
        \label{fig:Simulation_1D}    
    \end{subfigure}%
    \caption{\centering \change{Phase separation in 1D: effect of polymer mobility. Both cases start from the same initial film composition, $(\phi_p^0, \phi_n^0) = (0.5, 0.1)$, with polymer mobility set to $M_p = M$ (green) and $M_p = \frac{M}{100}$ (black).  
	(a) Film composition within the ternary phase diagram at an intermediate stage. One can see that the black line presents significant oscillations that correspond to uphill diffusion in the spinodal domain.
	(b) Polymer volume fraction profiles $\phi_p(z)$ at a late time.  
	The yellow point on the phase diagram indicates the composition of the bath.}}

    \label{fig:1D}
\end{figure}

 First we consider a film with a composition from the no phase separation region of the previous section and consider its evolution for two different polymer mobilities, $M_p = M$ for which no phase separation occurs  and $M_p = \frac{M}{100}$. \change{ In order to allow a clear visualization of the composition trajectory within the phase diagram during the film’s evolution we have used the results from 1D simulations.  In Figure~\ref{fig:1D} the polymer composition of the film at a late time is plotted as a function of the spatial coordinate $z$, while the composition of the entire film at a chosen time $t$ is shown in the Gibbs triangle in the same figure. The time $t$ has been selected to clearly illustrate the trajectory of the film composition as it evolves within the phase diagram.} In the case of equal mobilities, the film composition remains outside the unstable zone (except for a thin interfacial region) and stabilizes on the binodal curve in equilibrium with the bath. The film remains monophasic, as shown by the green curve in figure \ref{fig:1D}. In the case of a polymer mobility reduced by a factor of 100, the film composition remains close to the line  $\phi_p=\mathrm{constant}$, and eventually enters the unstable zone. At later times it  separate into two phases, as can be seen on the orange curve on the right.  This indicates clearly that, as expected, the pattern that appears is strongly dependent on the polymer mobility. \\

\begin{figure}
    \centering
    \renewcommand{\arraystretch}{1.5}
    \begin{tabular}{|c|c|c|c|c|}
        \hline
        t & \textbf{$t_1$} & \textbf{$t_2$} & \textbf{$t_3$} & \textbf{$t_4$} \\
        \hline
        {\textbf{$M_p=M$}} & 
        \raisebox{-0.5cm}{\includegraphics[width=0.1\textwidth]{images/With_noise/three_regions/separation/t10.png}} &
        \raisebox{-0.5cm}{\includegraphics[width=0.1\textwidth]{images/With_noise/three_regions/separation/t30.png}} &
        \raisebox{-0.5cm}{\includegraphics[width=0.1\textwidth]{images/With_noise/three_regions/separation/t50.png}} &
        \raisebox{-0.5cm}{\includegraphics[width=0.1\textwidth]{images/With_noise/three_regions/separation/t100.png}} \\
        \hline
        
        {\textbf{$M_p=\frac{M}{10}$}} & 
        \raisebox{-0.5cm}{\includegraphics[width=0.1\textwidth]{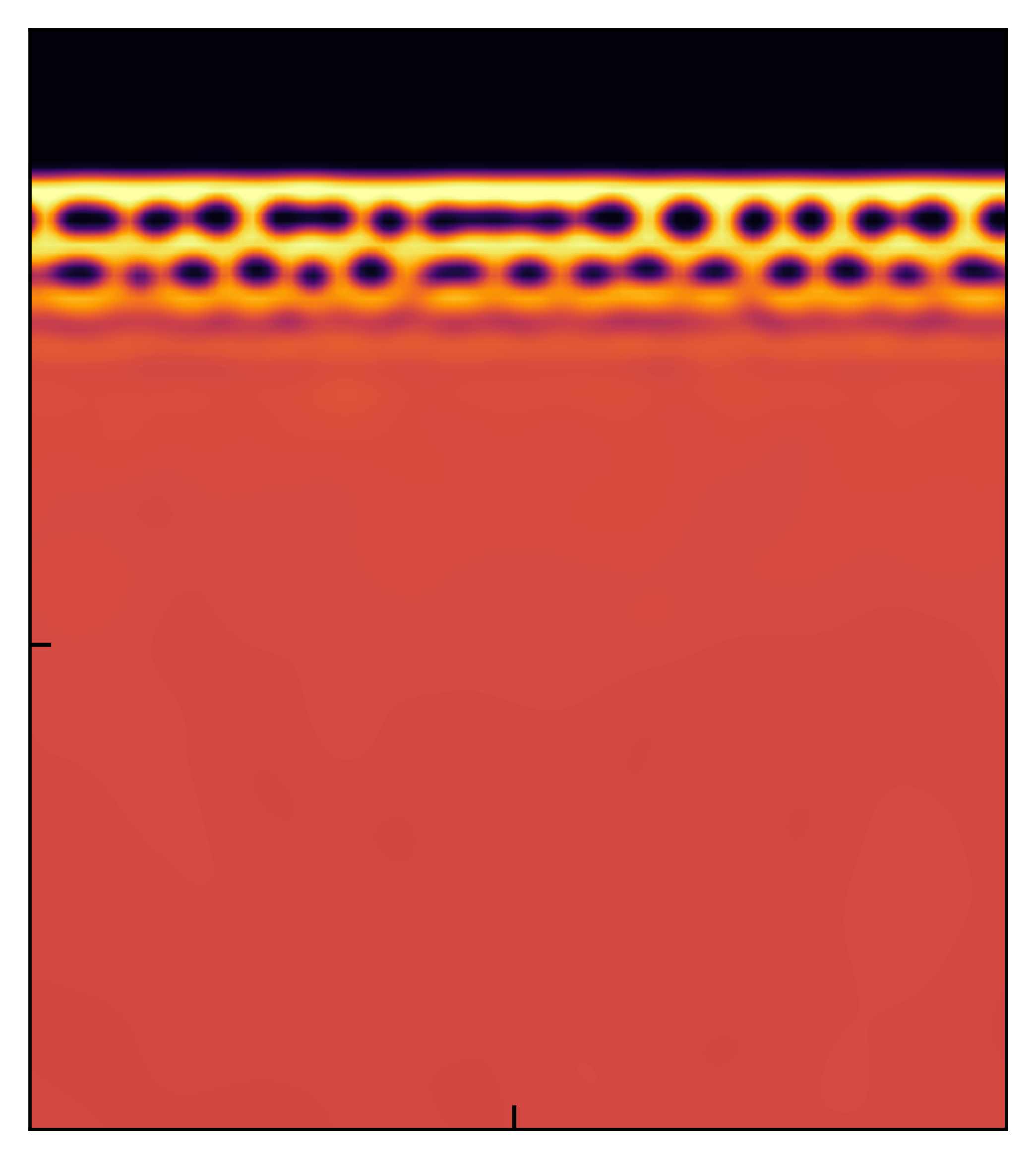}} &
        \raisebox{-0.5cm}{\includegraphics[width=0.1\textwidth]{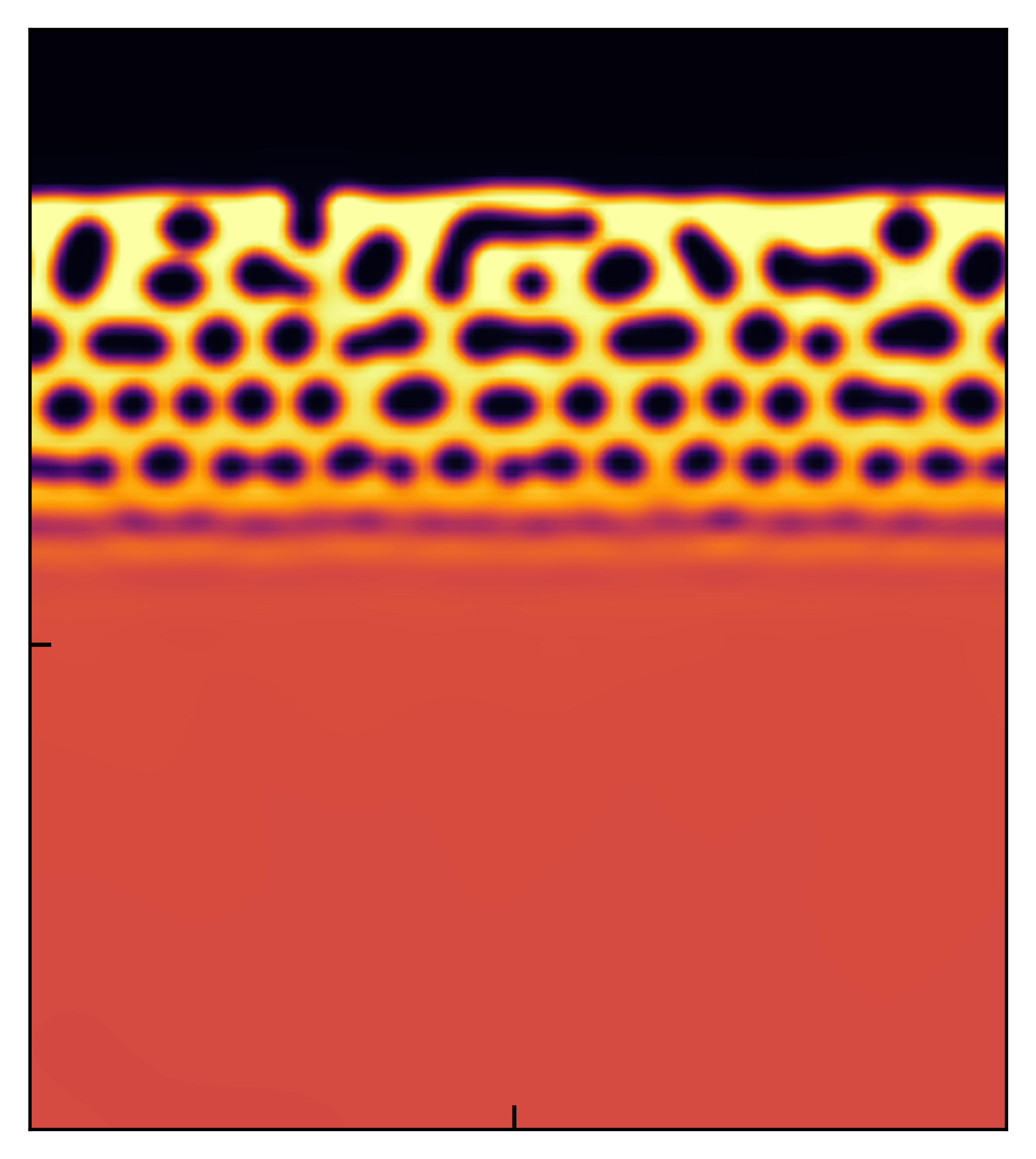}} &
        \raisebox{-0.5cm}{\includegraphics[width=0.1\textwidth]{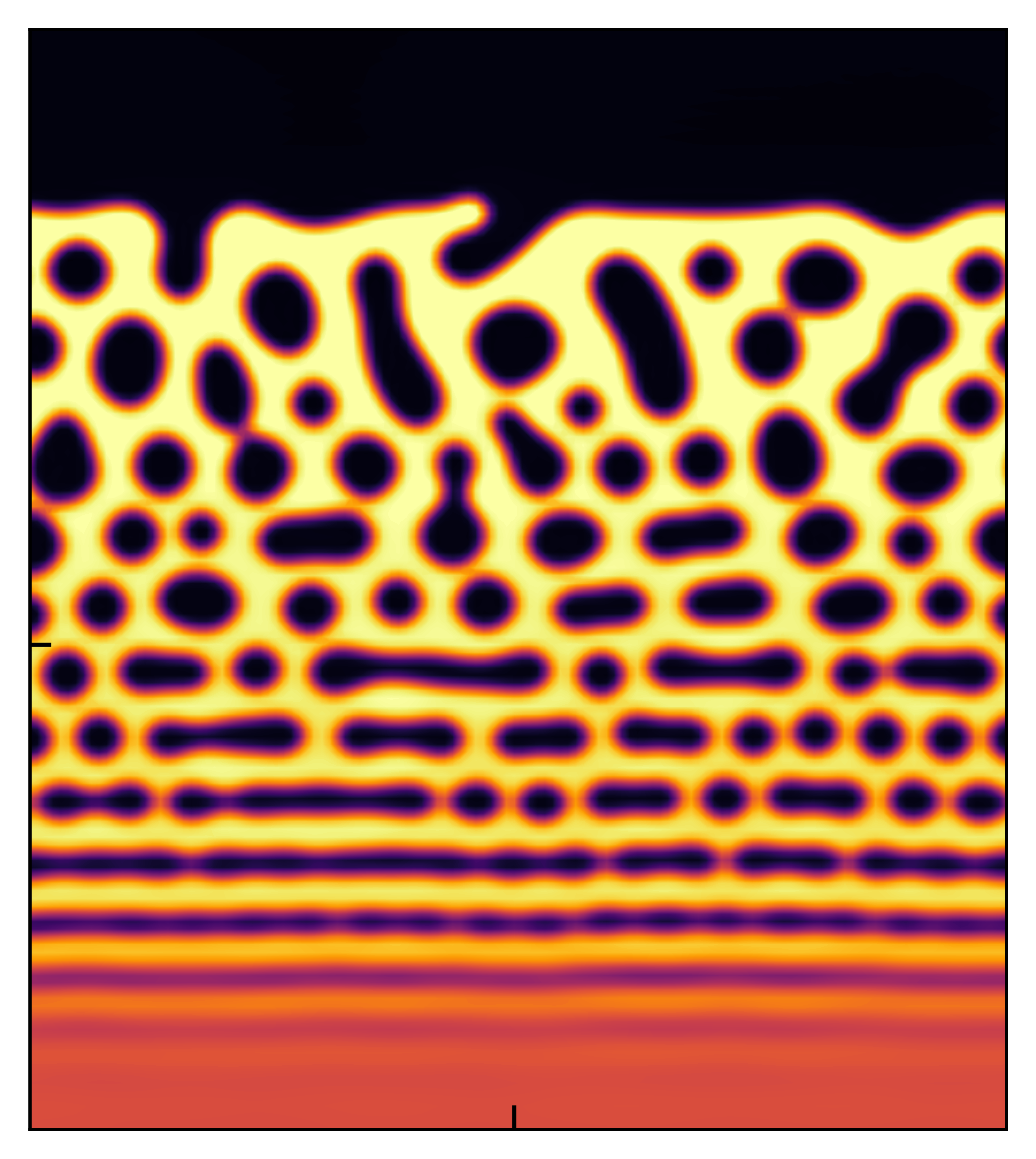}} &
        \raisebox{-0.5cm}{\includegraphics[width=0.1\textwidth]{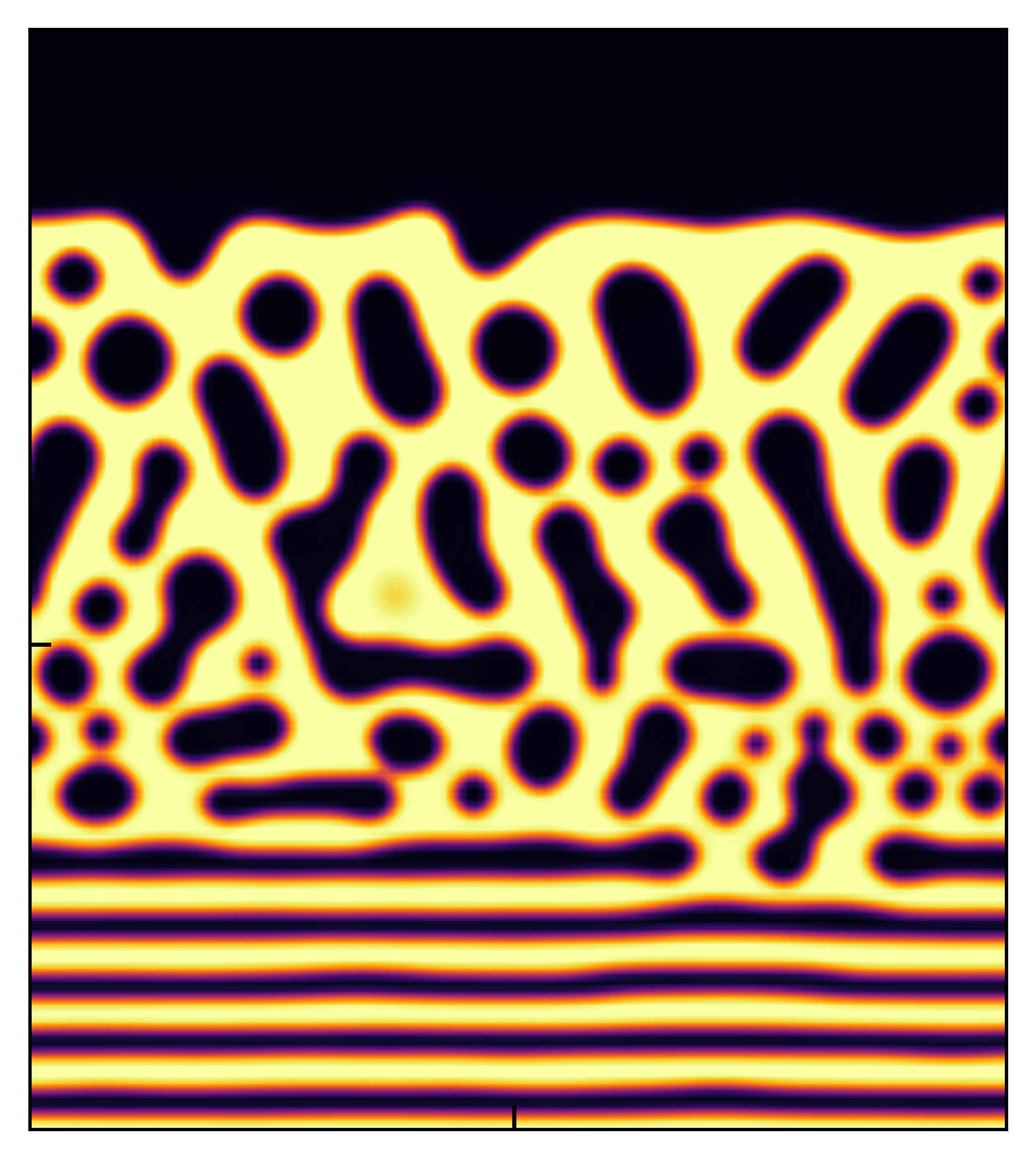}} \\
        \hline
        
        {\textbf{$M_p=\frac{M}{100}$}} & 
        \raisebox{-0.5cm}{\includegraphics[width=0.1\textwidth]{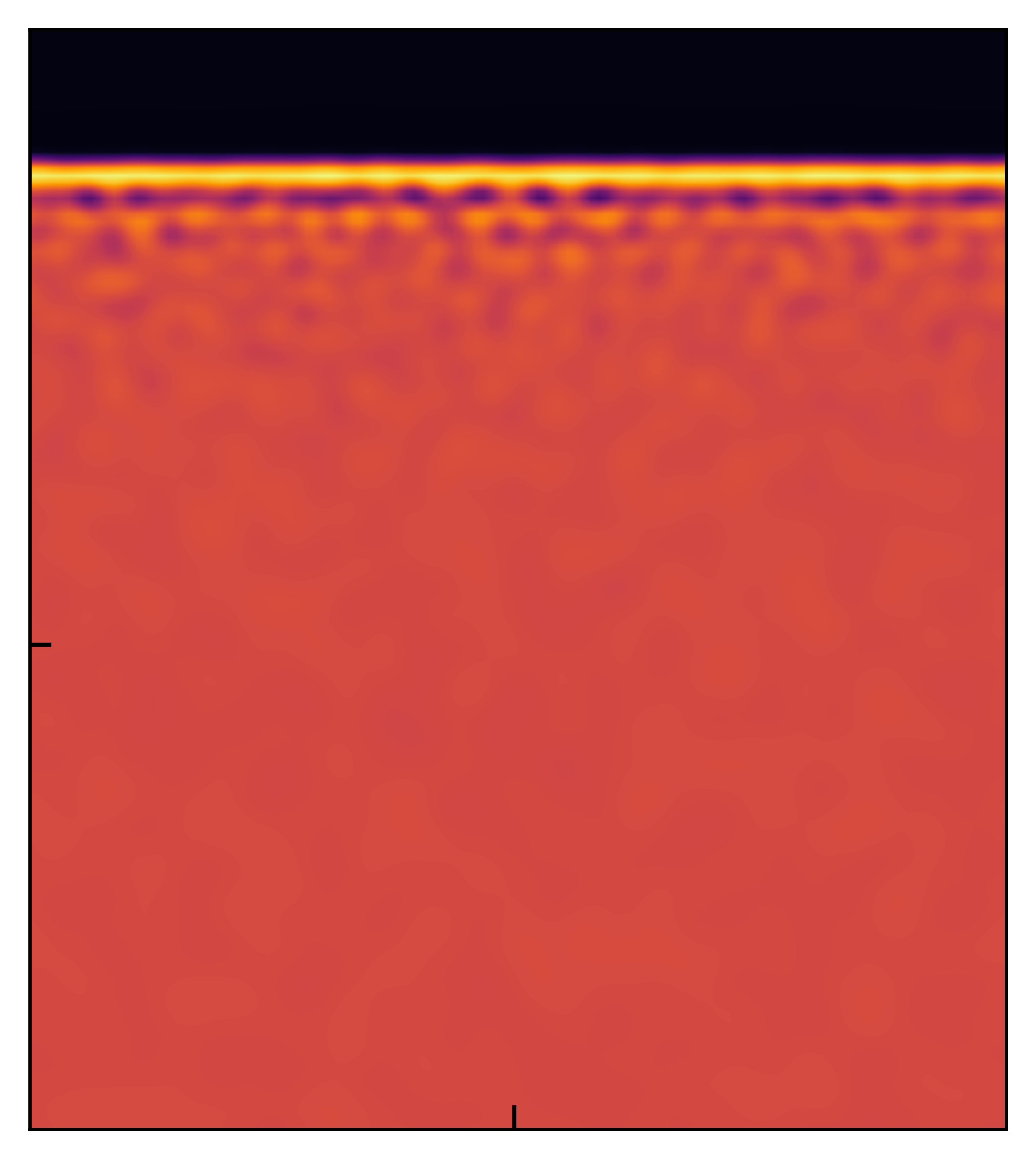}} &
        \raisebox{-0.5cm}{\includegraphics[width=0.1\textwidth]{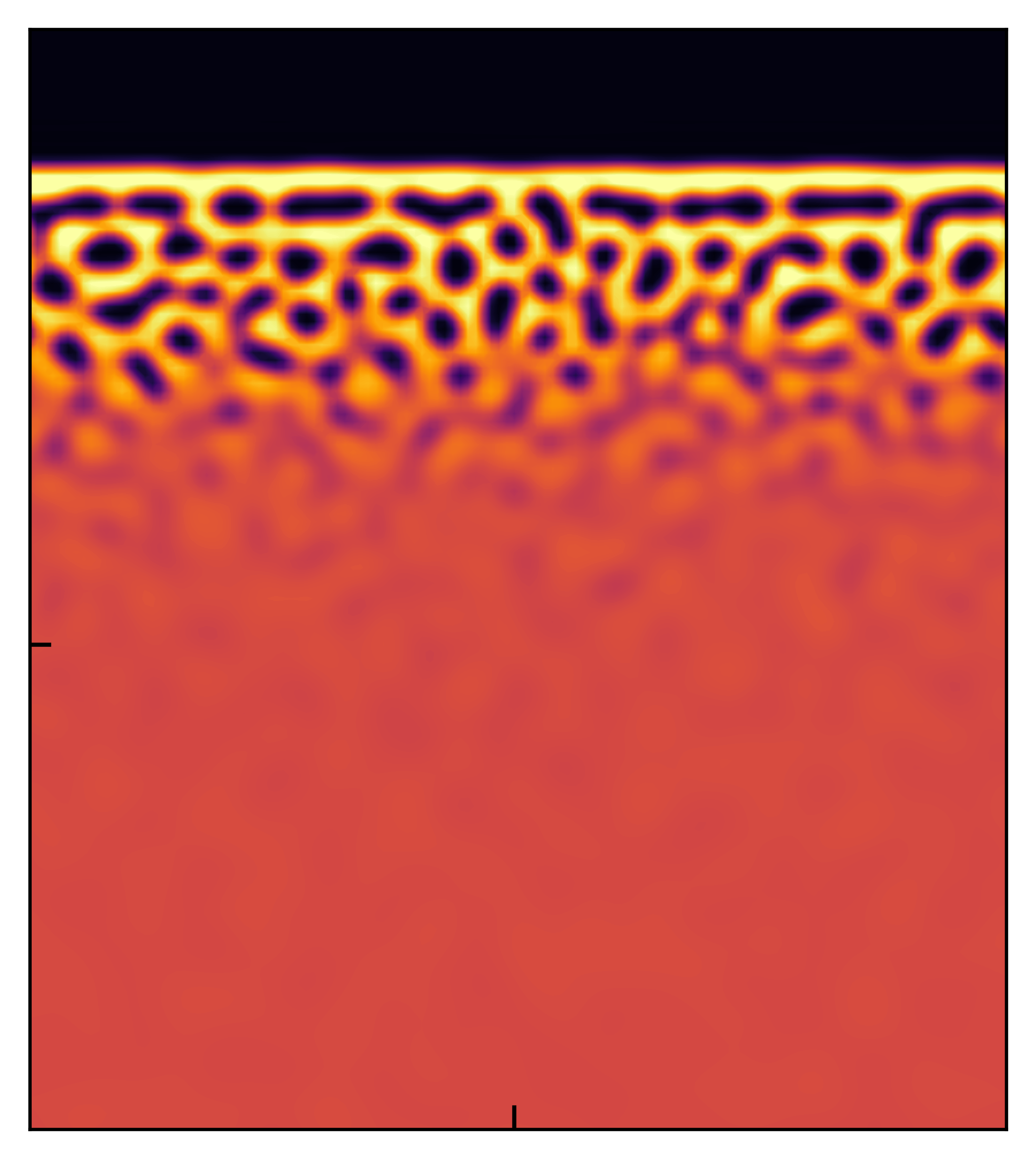}} &
        \raisebox{-0.5cm}{\includegraphics[width=0.1\textwidth]{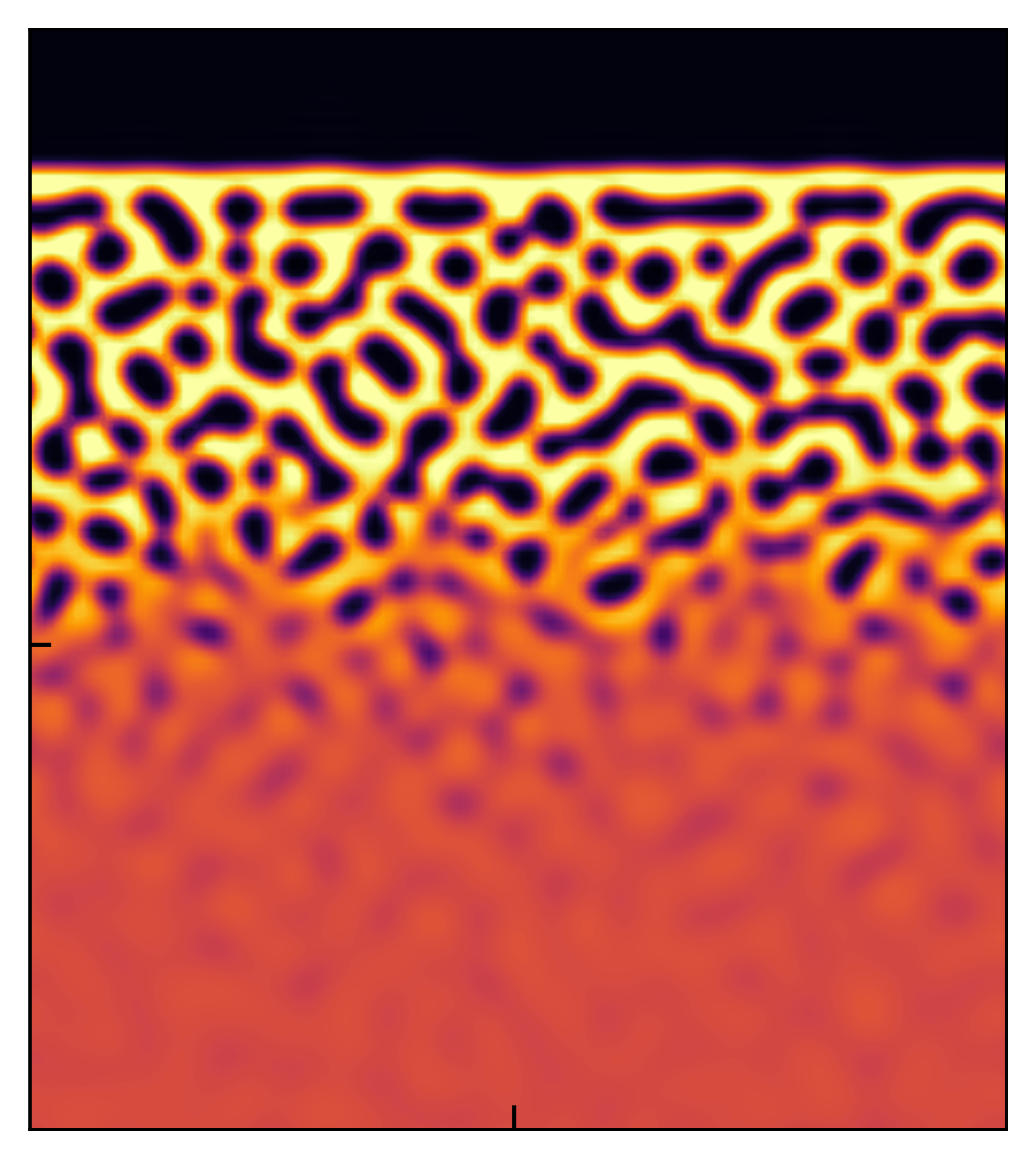}} &
        \raisebox{-0.5cm}{\includegraphics[width=0.1\textwidth]{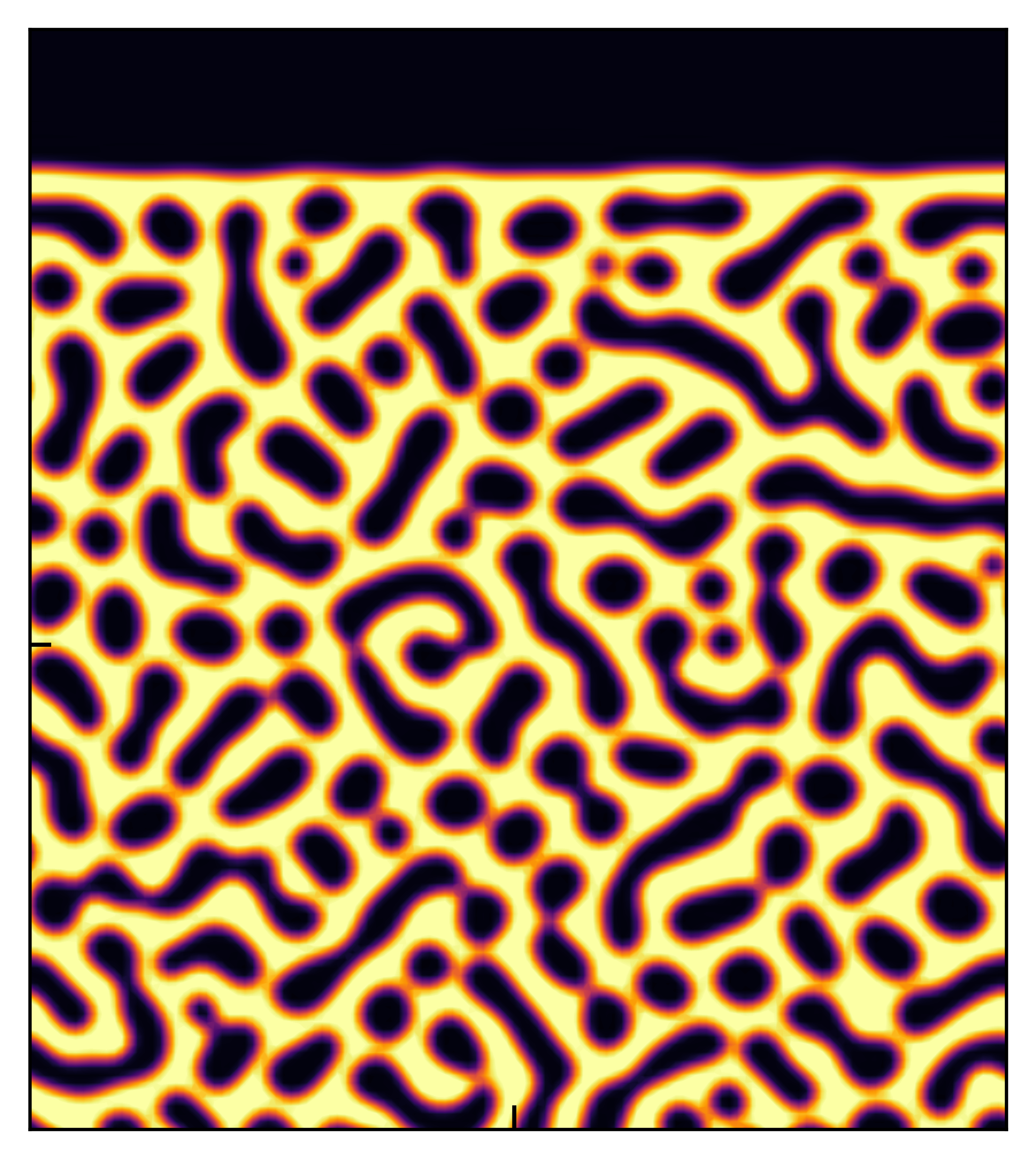}} \\
        \hline
    \end{tabular}
    \vspace{0.1cm}
    
    \begin{minipage}{0.35\textwidth}
        \centering
        \includegraphics[width=\textwidth]{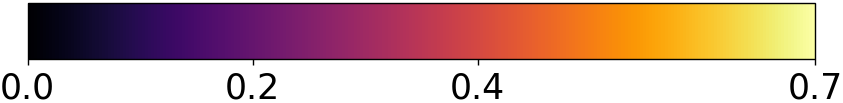}
    \end{minipage}
    \caption{ Effect of reducing polymer mobility. The evolution of polymer volume fraction for three simulations with the same initial composition of the film (0.4;0.25) and three different mobilities  a) $M_p=M$; b)$M_p=\frac{M}{10}$; c)$M_p=\frac{M}{100}$. The simulation times for each case, multiplied by $10^5$, are as follows: (a) $[0.3,0.9,1.5,3]$; (b) $[0.3,0.9,2.1,3]$; (c) $[0.6,1.8,3,6]$. It is worth mentioning that the values of $t_i$ for different values of $M_p$ differ by a factor of at most 2 while the mobilities differ by a factor of 100.}
    \label{fig:mobility_pattern}
\end{figure}

The reduction of polymer mobility also alters the dynamics observed in the phase separation. Figure \ref{fig:mobility_pattern} presents simulation results for the same initial film composition but with three different polymer mobilities : $M_p = M$, $M_p = \frac{M}{10}$ and $M_p = \frac{M}{100}$. When \( M_p \) is reduced, the polymer becomes less mobile, which helps preserve the film thickness near its initial value. Under these conditions, phase separation leads to the formation of maze-like patterns.\\

In the case \( M_p = \frac{M}{10} \) at time \( t_1 \), we observe a polymer-rich matrix in which polymer-poor inclusions form. These inclusions are initially organized along lines, similar to the case where \( M_p = M \), forming alternating layers. The subsequent coarsening process leads to a disordered pattern (see \( t_2 \) and \( t_3 \)), which is closer to what is expected from SDSD rather than a maze-like pattern. At the bottom of the film, this behavior evolves into a clearer SDSD structure. This is probably due to  the decreasing amplitude of the noise over time that cannot dominate the effect of the  directional  transport that induces phase separation (from the bath to the bottom of the film).\\

When $M_p$ is strongly reduced ($M_p = M/100$), the polymer diffusion is slow enough to preserve the film thickness close to its initial value. The system evolves through the formation of an interface between the film and the bath, where the first polymer-rich band forms. Subsequently, a complex maze-like morphology develops throughout the film, and no banded structures are observed. The fact that the values of $t$ for which similar pattern are obtained varies by a factor of 2 while the mobilities vary by a factor of 100 is surprising. It can be attributed to the fact that when diffusion drives the film outside of the stable domain, the further it drives it inside the spinodal domain the higher  the thermodynamic force is (the increase of the driving force is governed by the solvent/non solvent exchange).  As a result the low mobility of the polymer  is \change{partially} compensated by the higher driving force and as a result characteristic time scales \change{scale sublinearly with the mobilities( a   short rationale for this in the framework of a  toy model is given in appendix \ref{sublinear})}. It must be noted that this is partly in contradiction with the \textit{naïve} picture presented at the beginning of this section for extremely slow polymer diffusion. \\

\begin{figure}
    \centering
    \begin{subfigure}[b]{0.33\textwidth}
        \centering
        \includegraphics[width=1\linewidth]{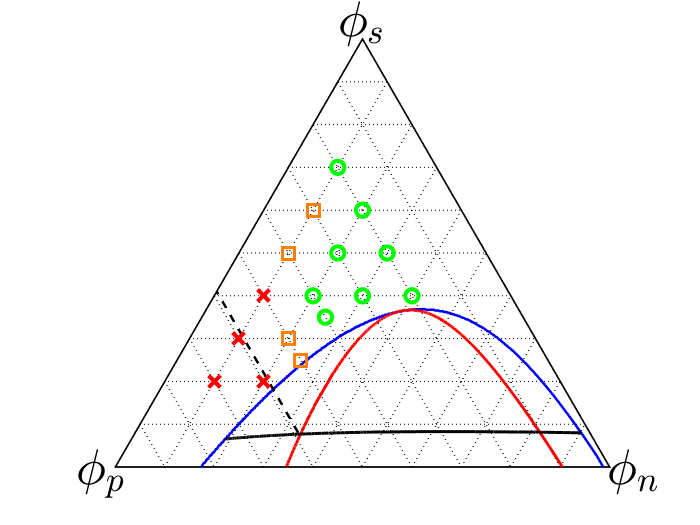}
        \caption{$M_p=M$}
        \label{fig:gibbs_behavior1}    
    \end{subfigure}%
    \begin{subfigure}[b]{0.33\textwidth}
        \centering
        \includegraphics[width=1\linewidth]{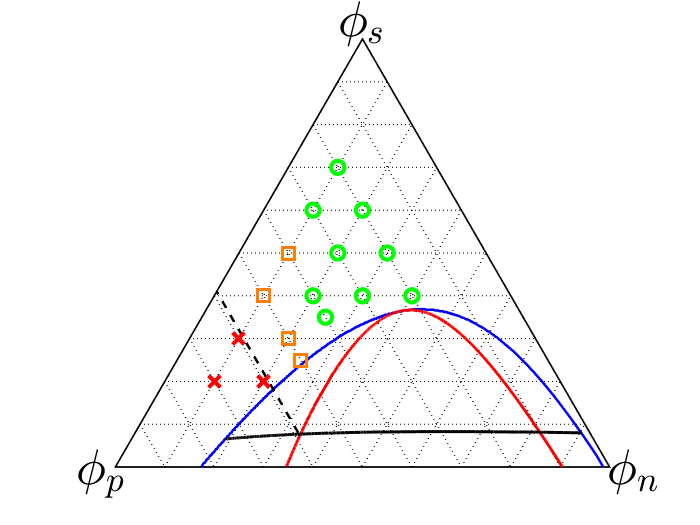}
        \caption{$M_p=\frac{M}{2}$}
        \label{fig:gibbs_behavior2}    
    \end{subfigure}%
    \begin{subfigure}[b]{0.33\textwidth}
        \centering
        \includegraphics[width=1\linewidth]{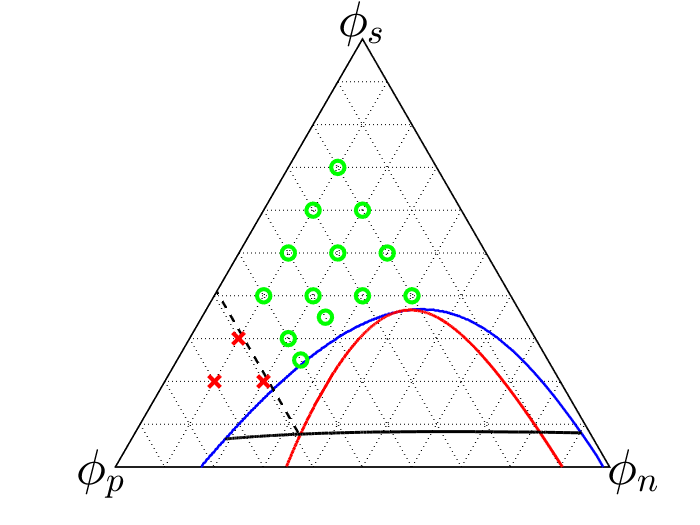}
        \caption{$M_p=\frac{M}{100}$}
        \label{fig:gibbs_behavior3}    
    \end{subfigure}
    \caption{ The effects of polymer mobility on the three regions: Green circles represent phase separation, orange squares indicate delayed phase separation, and red crosses correspond to no phase separation. The diagrams summarize the results for three different polymer mobility values: (a) $M_p = M$, (b) $M_p = \frac{M}{2}$, and (c) $M_p = \frac{M}{100}$.}

    \label{fig:three_mobilities}
\end{figure}

Obviously for different values of polymer mobilities, different behavior are observed and decreasing the polymer mobility induces an increase in the size of the initial composition domain for which phase separation is observed. This is illustrated in Figure \ref{fig:three_mobilities}. The domains of initial film composition for which   phase separation is induced is  represented for decreasing values of the mobility of the polymer. The domain of film composition increases significantly, especially on the line $\phi_n=0.1$. It must also be noted that below a polymer mobility of $M/100$, there should be little change. Indeed the points that do not lead to phase separation are situated in a region for which the line $\phi_p=\mathrm{cst}$ does intersect the line $\mu{ns}=\mu_{bath}$before the boundary of the spinodal domain. This implies that the invasion of the solvent in the film will lead to a equlibration of the fast chemical species  before the film is unstable.\\

\begin{figure}
    \centering
    \includegraphics[width=0.5\textwidth]{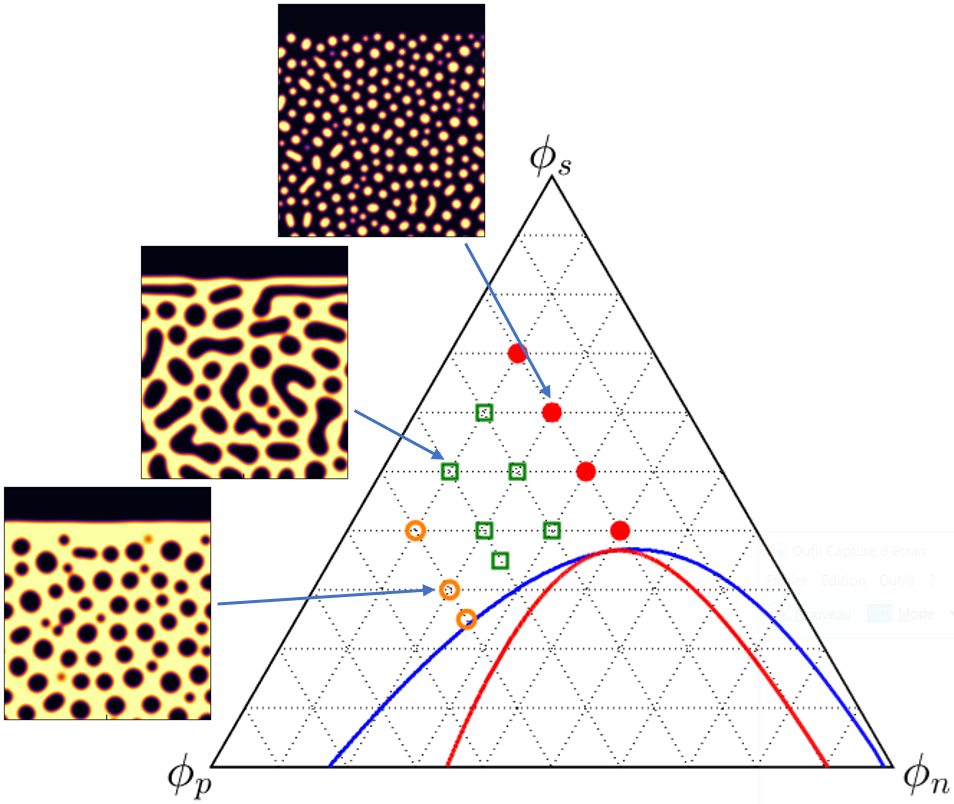}
    \caption{Polymer volume fraction effects on phase separation}
    \label{fig:Three_forms}
\end{figure}

 This set of simulations  shows that the phase separation in the film  is  related to the relative mobilities of the three chemical species.  However  the effects of mobility on the pattern observed in  2D simulations are limited. Indeed, as can be seen in  Figure \ref{fig:Three_forms} the morphologies are mostly related to the initial volume fraction of polymer. \\
 
 At low polymer concentration, we observe polymer-rich droplets in a non-solvent matrix: a discontinuous 2D structure. By increasing the volume fraction of the polymer, the resulting structure is  a discontinuous network of voids that are similar to maze-like structures observed in 2D phase separation with a 0.5 volume fraction. When more polymer is added: non-solvent droplets form in a polymer solution. In all cases, no bicontinuous structure where both polymer-rich and polymer-poor regions are continuous  is observed\footnote{In 2D such pattern cannot be observed because if one phase is a percolating cluster, it is drawing an infinite  line through space that the other phase cannot cross. As a result the other phase is not a percolating cluster}. In contrast, in 3D systems, a truly bicontinuous microstructure can be observed. Therefore, in the next section, we will discuss more precisely the pattern formation, focusing on 3D simulation results.   \\

\subsection{Simulations in 3D}
 In order to better understand the effect of the relative mobilities of polymer and solvent on the microstructure in 3D  we have performed simulations using two values of the polymer mobilities: $M_p=M/10$ and $M_p=M/100$ for which most initial film composition lead to an actual phase separation and we have used a wide range of initial film composition so that the volume fraction of the polymer rich phase is varied. At first we have focused on the visual aspect of the pattern and on its bicontinuity. This feature  is the minimal requirement to have a flow conducting porous structure  that is also  mechanically stable.  The simulations were performed under the same conditions as the 2D simulations discussed earlier.  The continuity of the pattern cannot be easily seen through simple imaging. However it can be easily checked using image processing software as Paraview. Here, we have used the following method: the isosurface $\phi_p=0.5$ was computed, then using a \textit{connectivity} filter each continuous domain was attributed an index\footnote{Using Paraview.simple.Connectivity, the domains are numerated in the increasing order of their lowest $z$. \change{The filter is applied with the "Cool to Warm" colormap; no legend is shown, as the colors only indicate the number of bicontinuous regions.}}. In the following we present results for $M_p=M/100$ and varying values of the initial polymer volume fraction in the film $\phi_p^0$. The first point we want to stress is the difference between 2D and 3D patterns.\\

\begin{figure}
    \centering
    \renewcommand{\arraystretch}{1.5}
    \begin{tabular}{|c|c|c|c|c|c|}
        \hline
        & \textbf{$\phi_p^0=0.15$} & \textbf{$\phi_p^0=0.2$} & \textbf{$\phi_p^0=0.3$} & \textbf{$\phi_p^0=0.4$} & \textbf{$\phi_p^0=0.5$}\\
        \hline
        \raisebox{1cm}{\textbf{2D}} & 
        \includegraphics[width=0.12\textwidth]{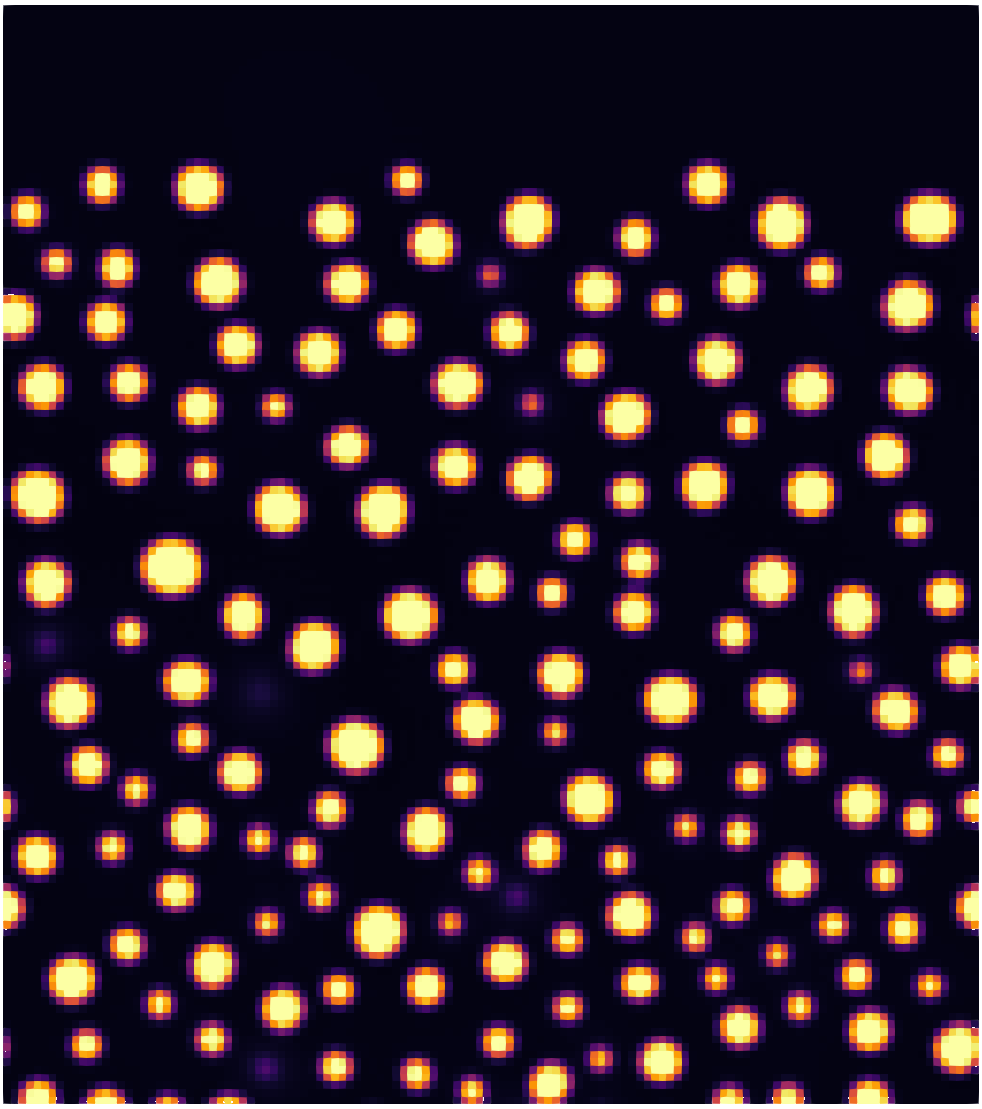} &
        \includegraphics[width=0.12\textwidth]{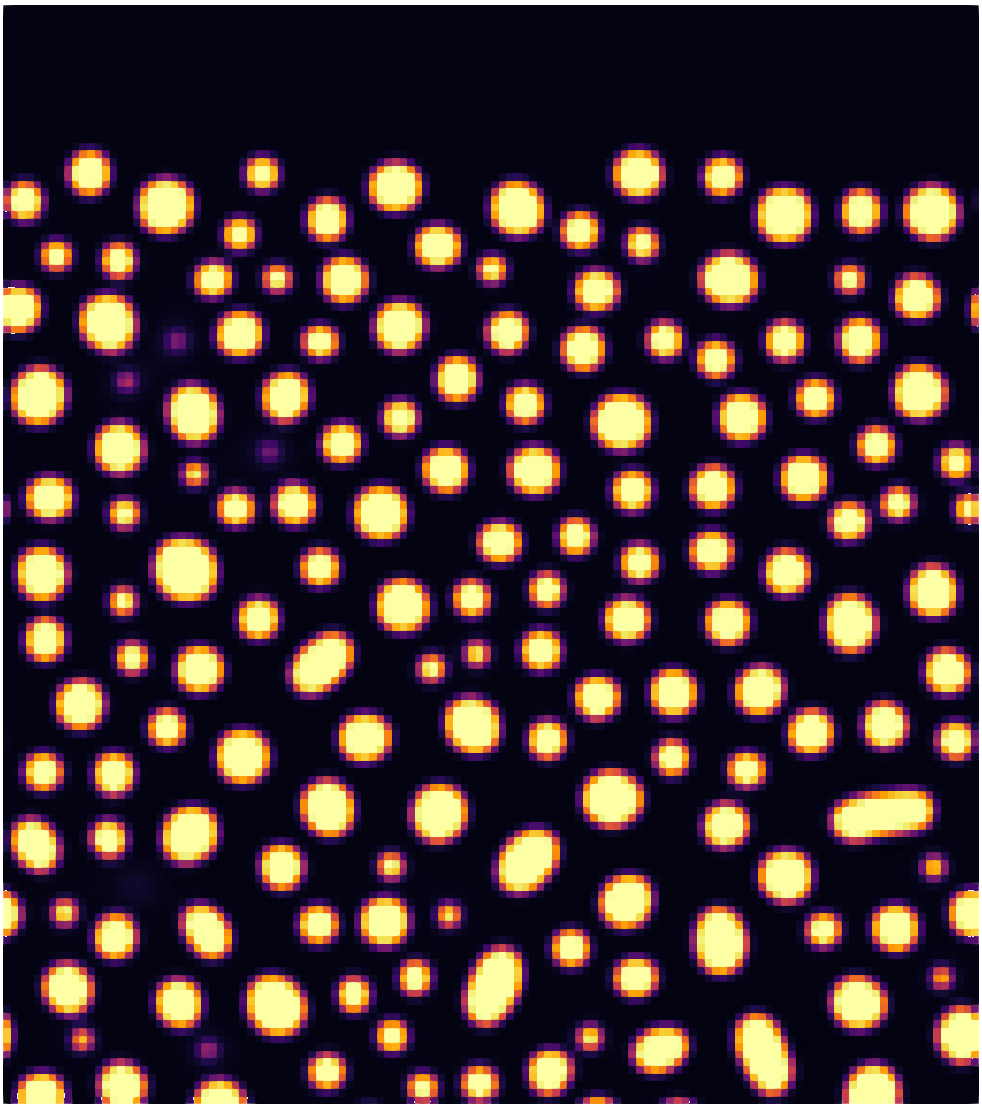} &
        \includegraphics[width=0.12\textwidth]{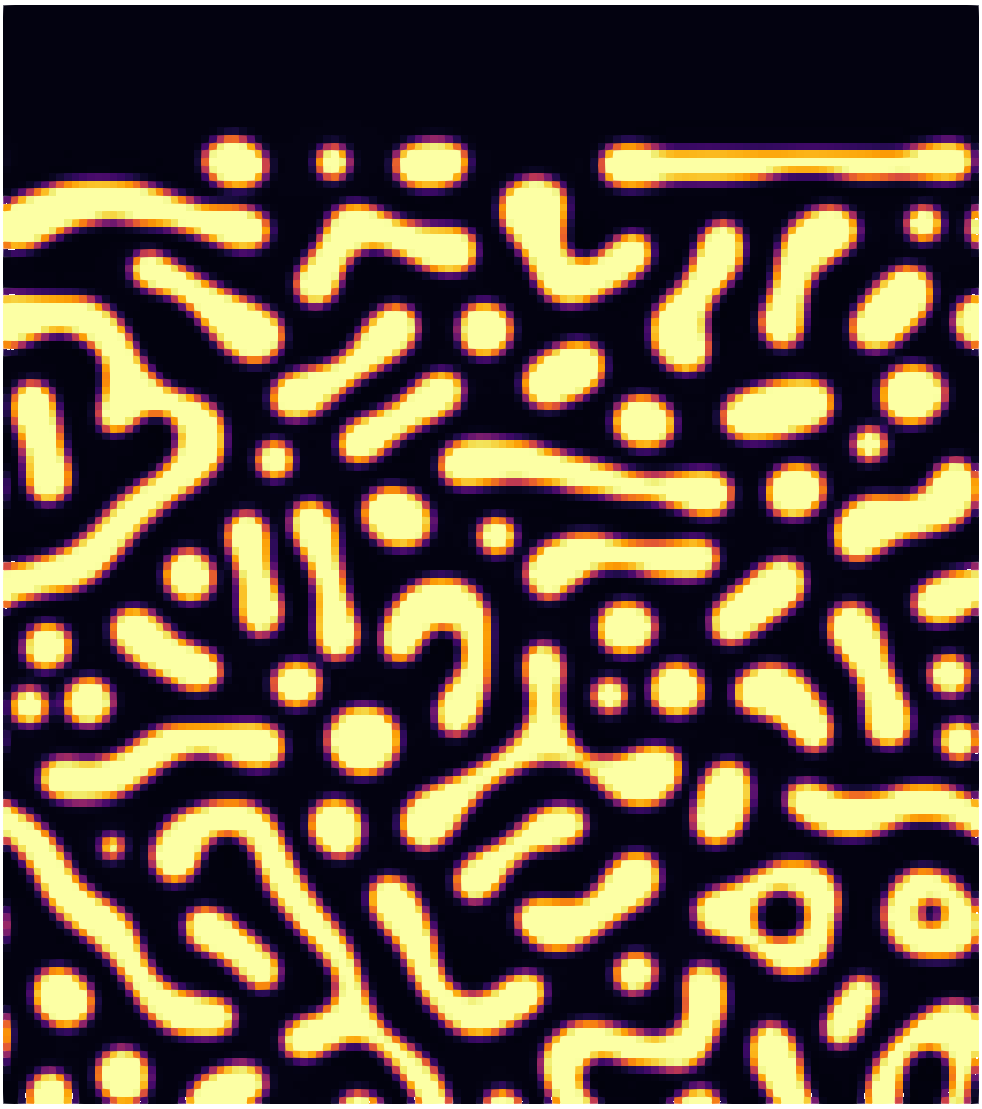} &
        \includegraphics[width=0.12\textwidth]{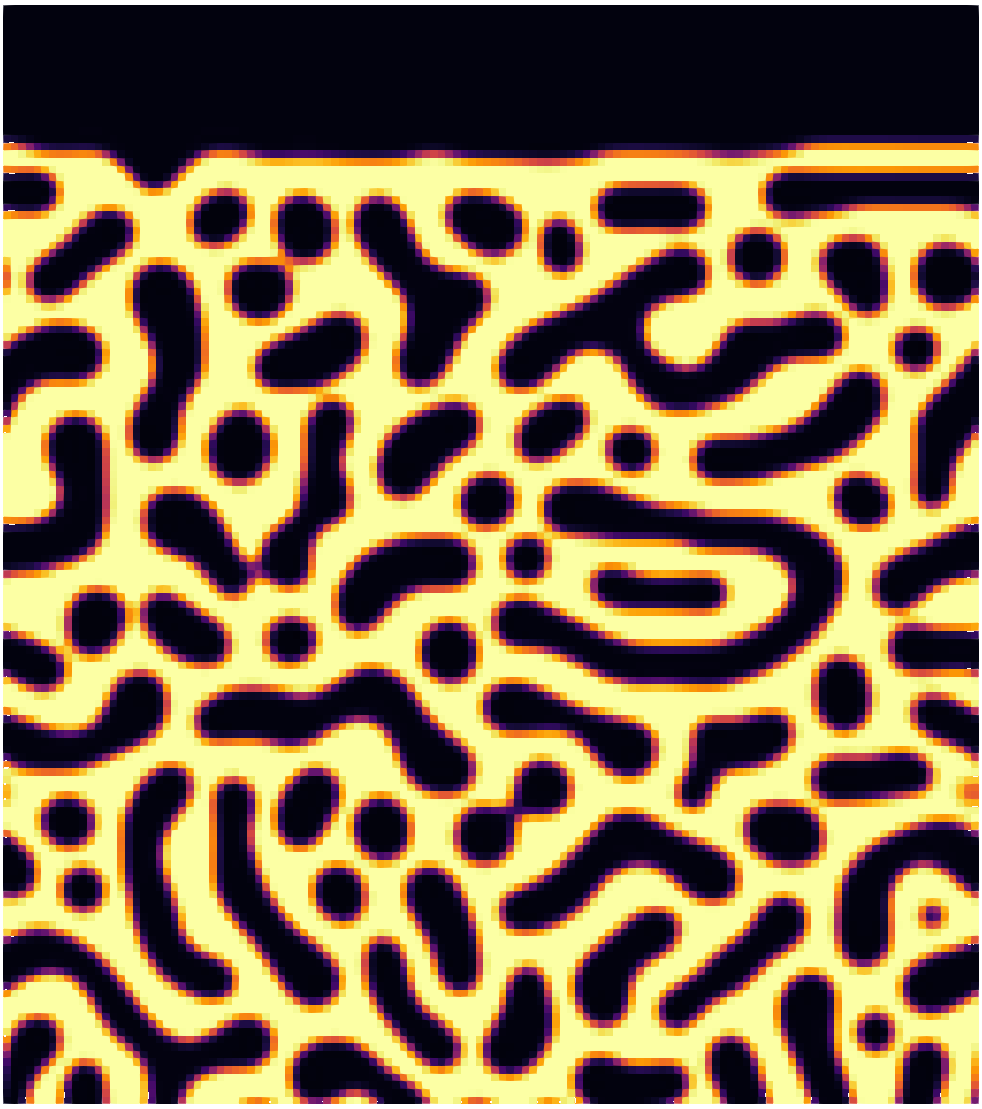} &
        \includegraphics[width=0.12\textwidth]{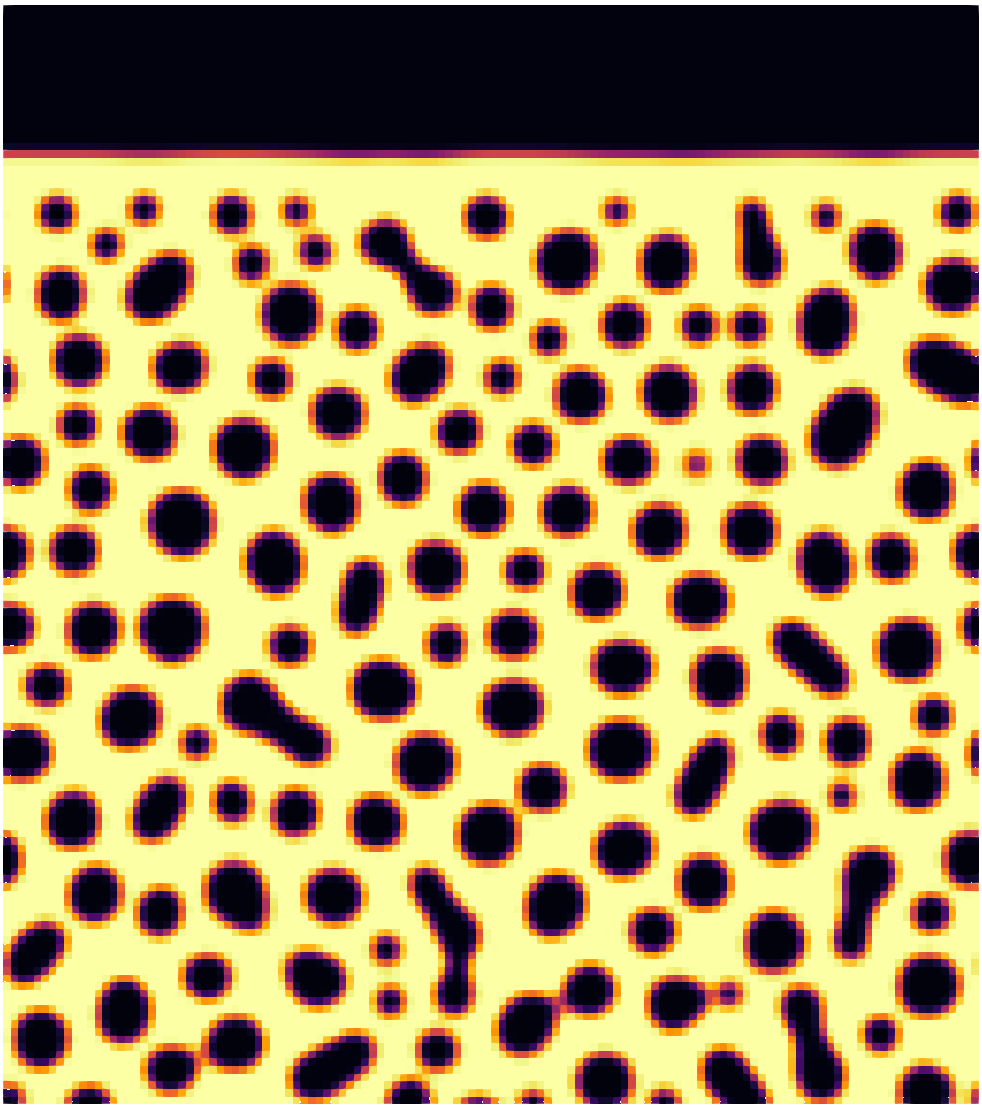}\\
        \hline
        
        \raisebox{1cm}{\textbf{3D}} & 
        \includegraphics[width=0.15\textwidth,trim={4cm 4cm 4cm 5cm},clip]{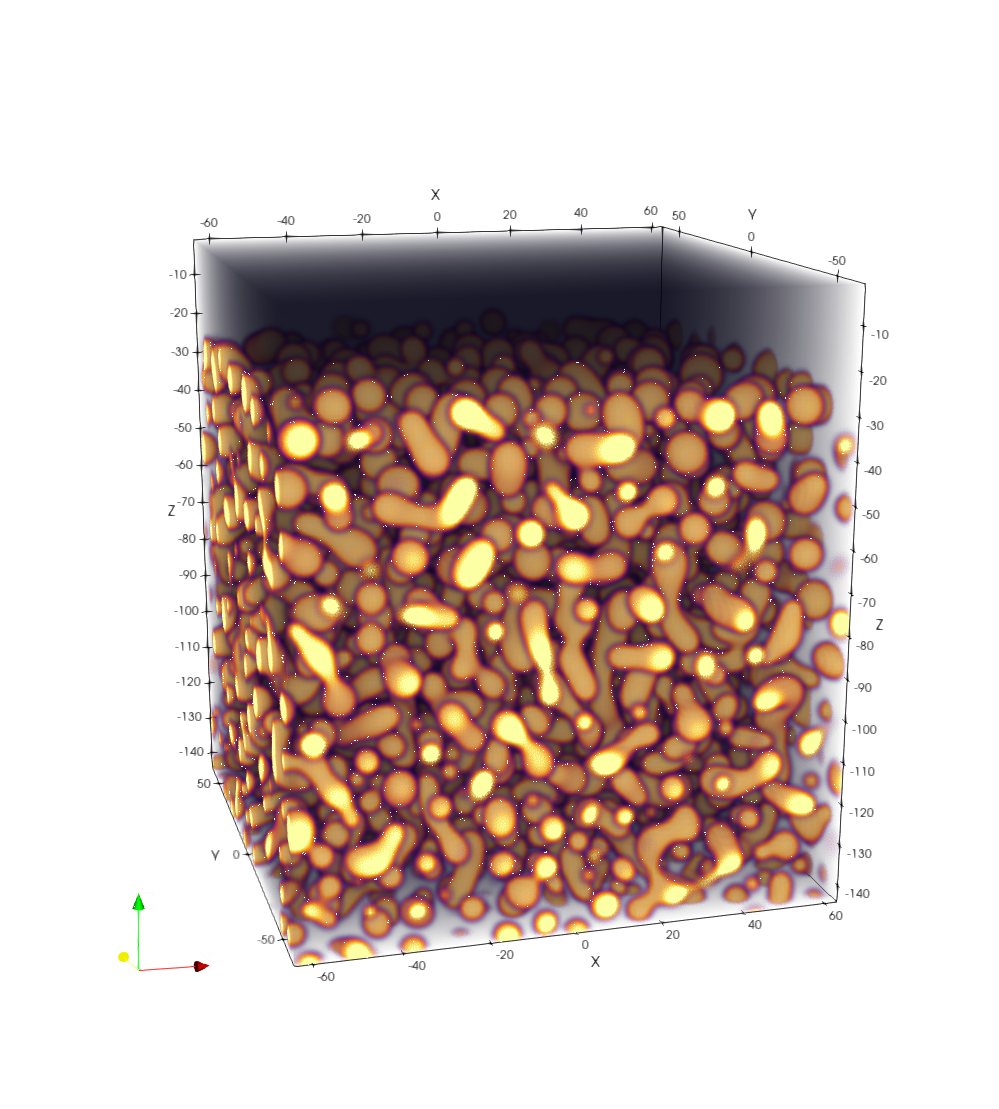} &
        \includegraphics[width=0.15\textwidth,trim={4cm 4cm 4cm 5cm},clip]{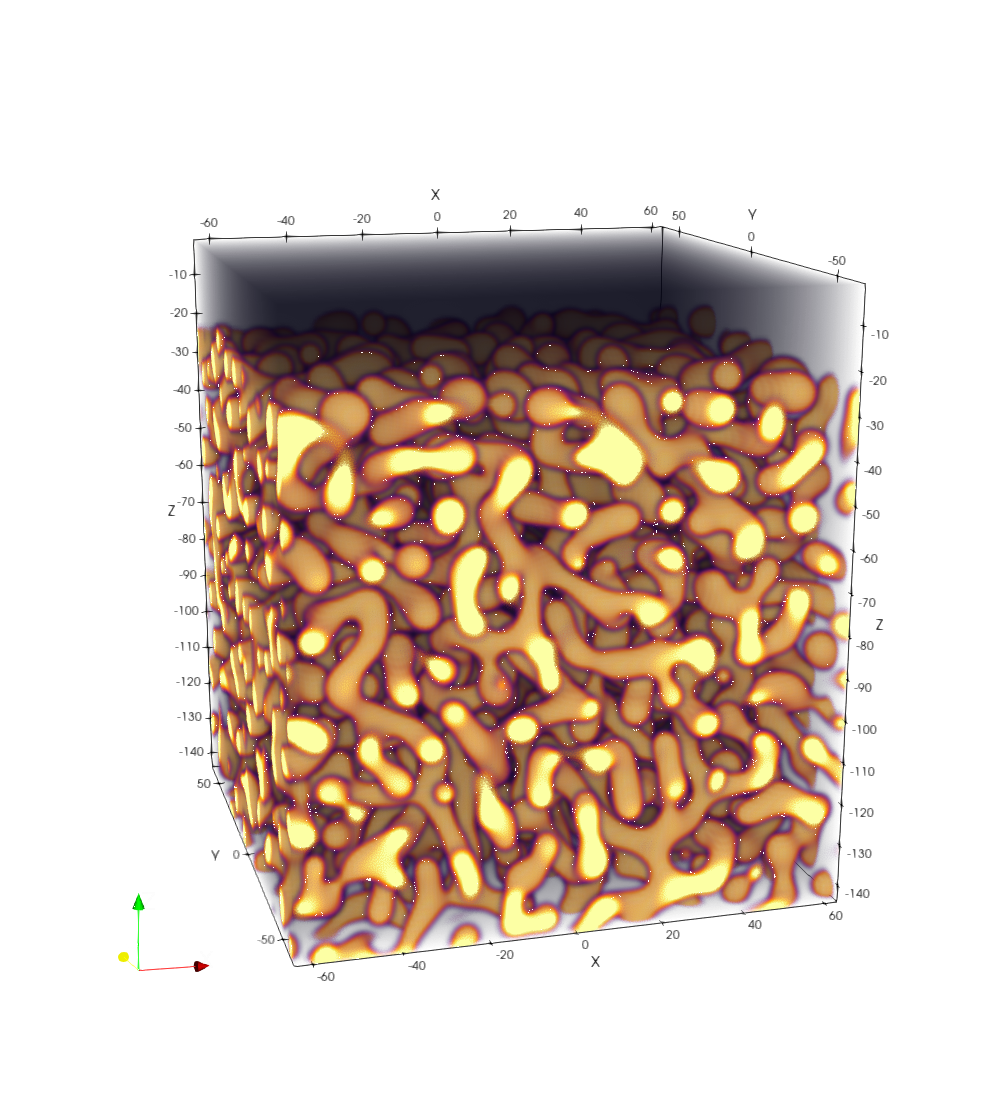} &
        \includegraphics[width=0.15\textwidth,trim={4cm 4cm 4cm 5cm},clip]{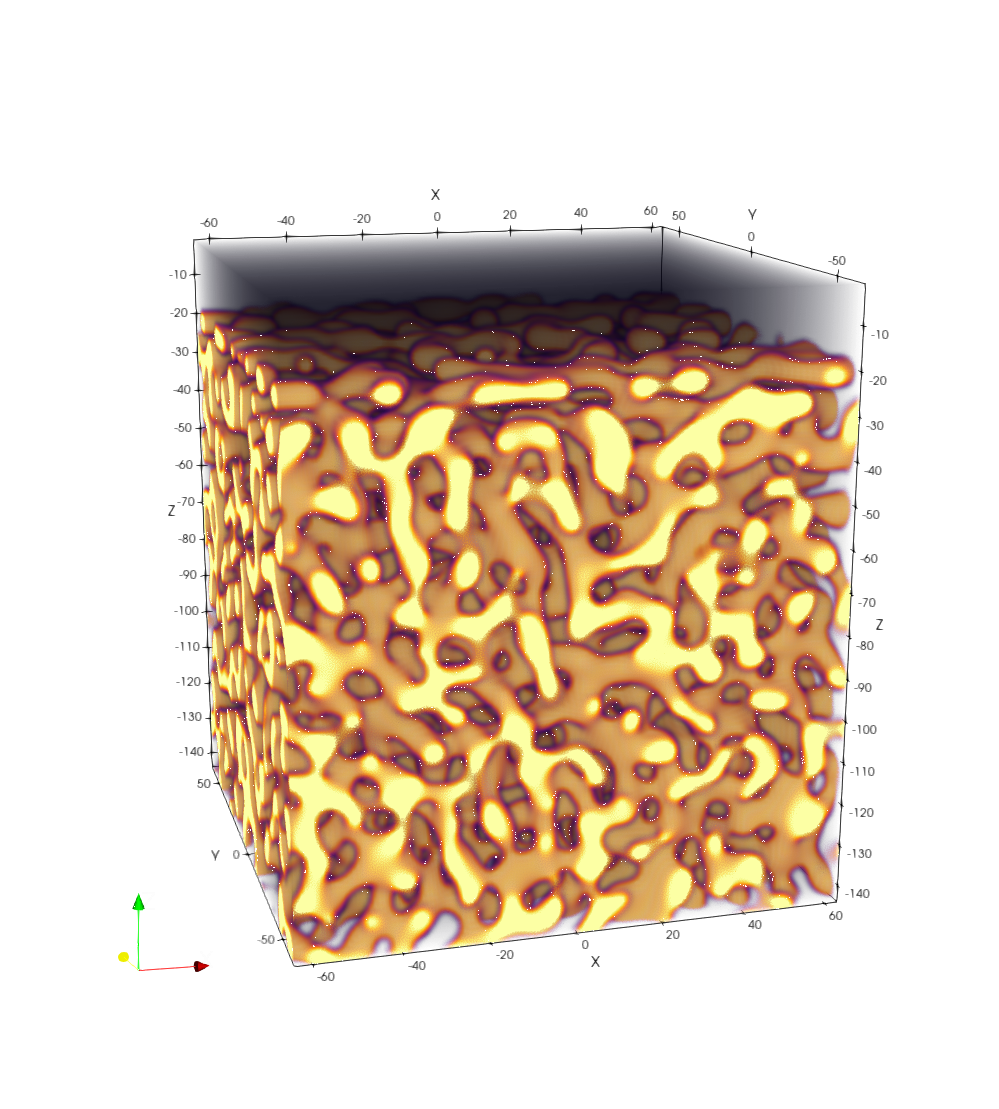} &
        \includegraphics[width=0.15\textwidth,trim={4cm 4cm 4cm 5cm},clip]{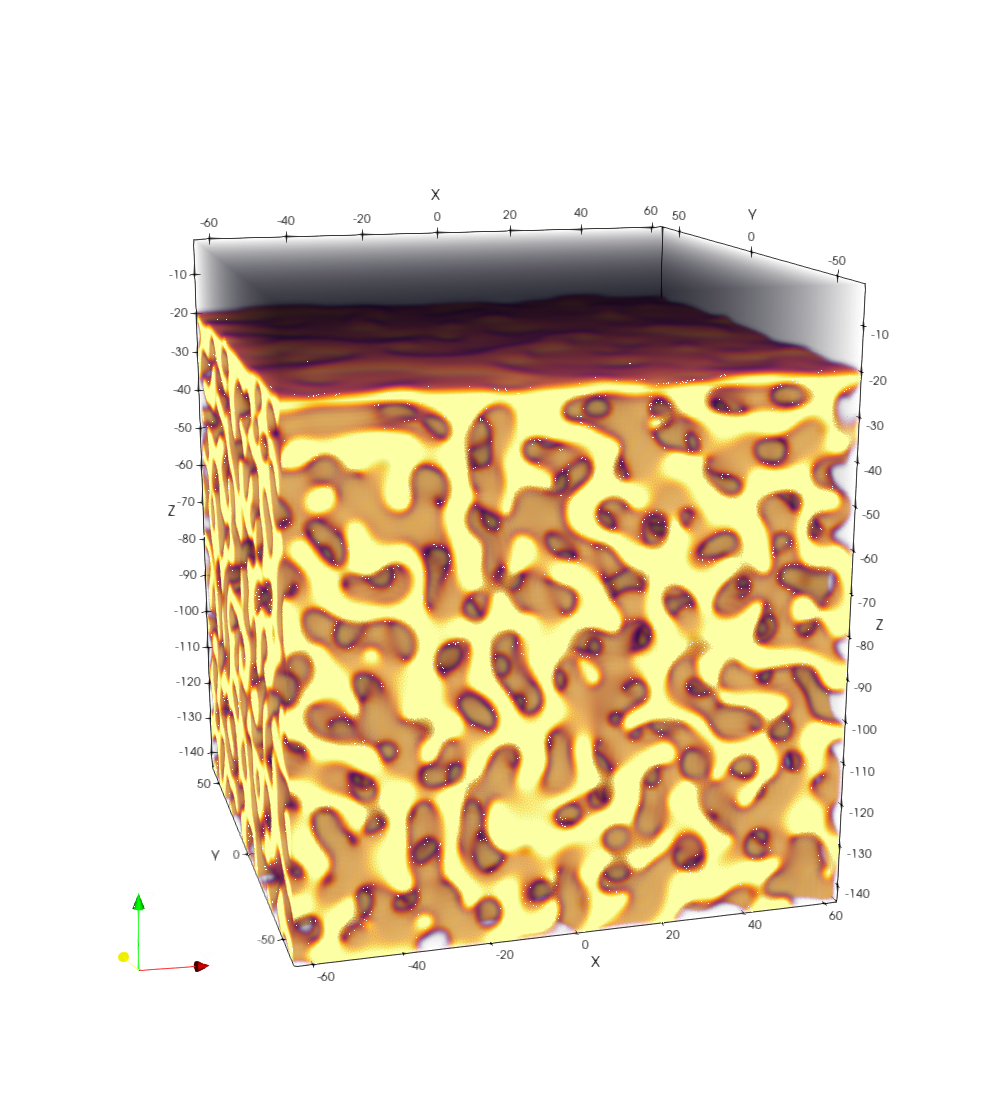} &
        \includegraphics[width=0.15\textwidth,trim={4cm 4cm 4cm 5cm},clip]{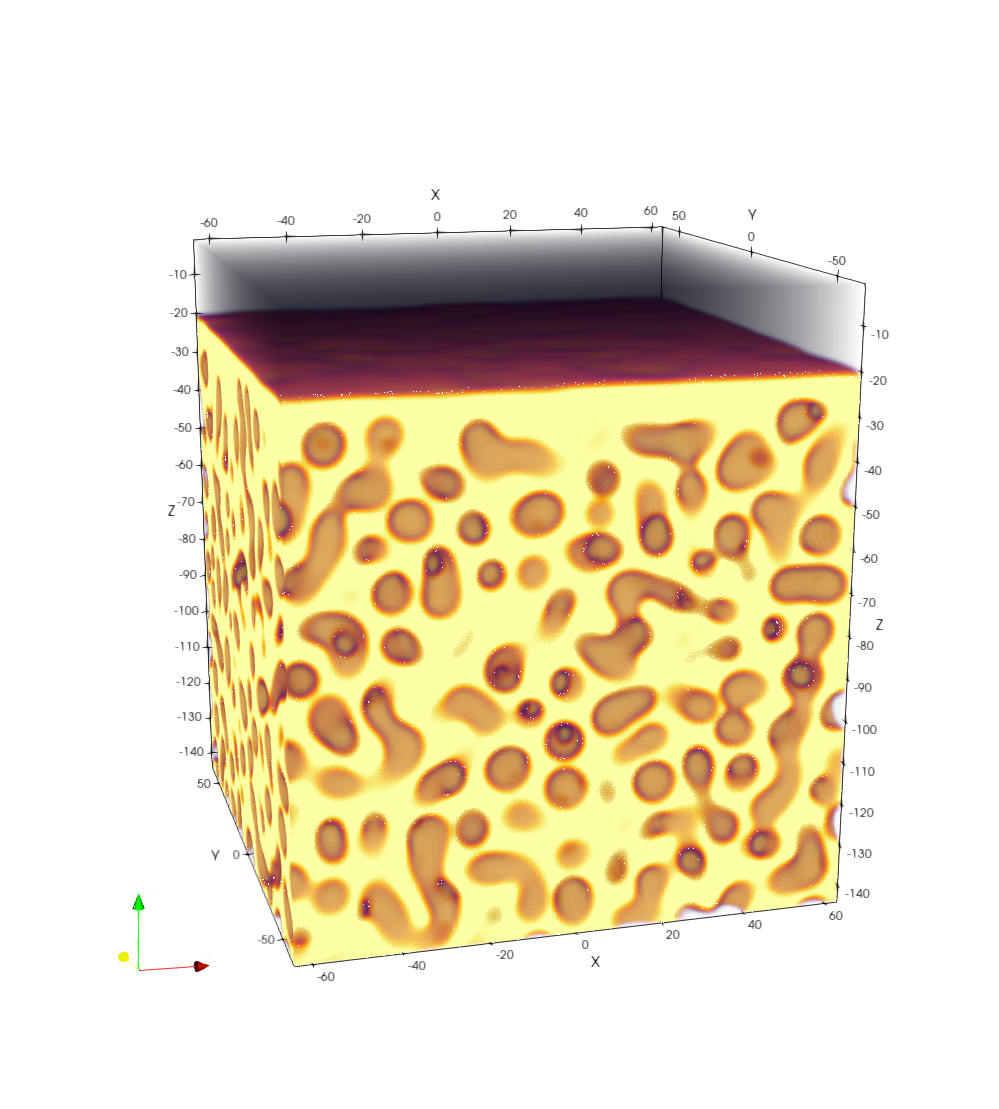} \\
        \hline

        \raisebox{1cm}{\textbf{}} & 
        \includegraphics[width=0.15\textwidth,trim={2.2cm 1.5cm 2.2cm 0.5cm},clip]{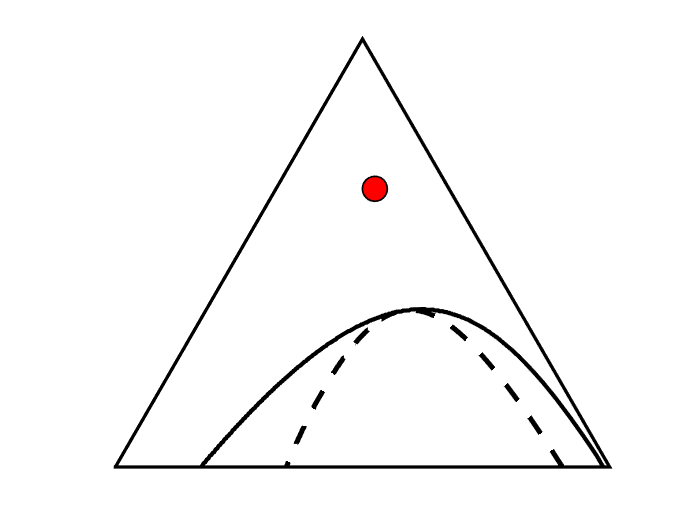} &
        \includegraphics[width=0.15\textwidth,trim={2.2cm 1.5cm 2.2cm 0.5cm},clip]{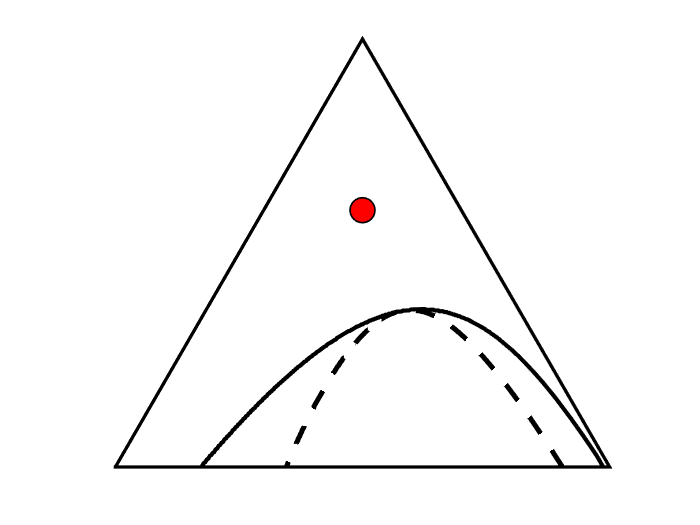} &
        \includegraphics[width=0.15\textwidth,trim={2.2cm 1.5cm 2.2cm 0.5cm},clip]{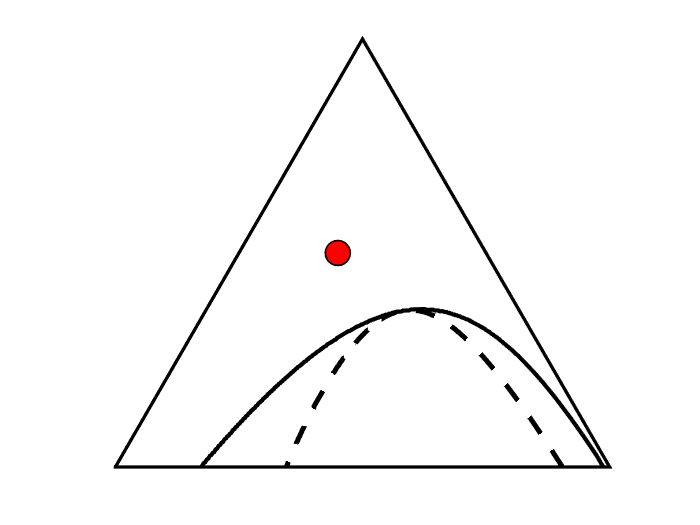} &
        \includegraphics[width=0.15\textwidth,trim={2.2cm 1.5cm 2.2cm 0.5cm},clip]{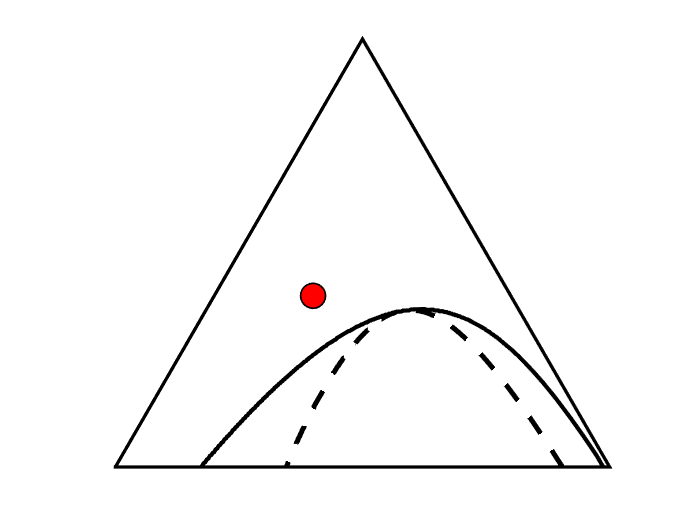} &
        \includegraphics[width=0.15\textwidth,trim={2.2cm 1.5cm 2.2cm 0.5cm},clip]{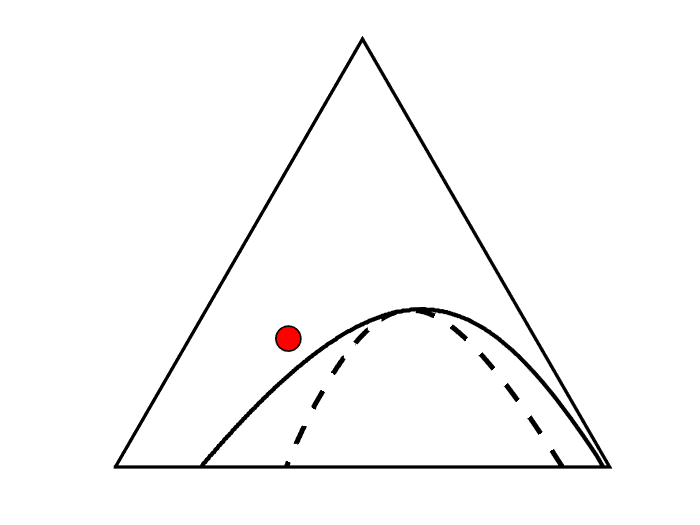} \\
        \hline
    \end{tabular}
    
    \vspace{0.01cm}
    \includegraphics[width=0.3\textwidth]{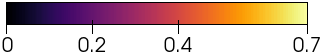}
    \caption{ Effect of $\phi_p$ on the microstructure: Comparison of 2D and 3D results for different values of the initial composition of the film $\phi_p^0$. The 2D plots are colormaps (the bath is at the top and is not shown completely) of the polymer volume fraction $\phi_p$ using the color scale shown below.  The 3D plots are plots of the isosurface $\phi_p=0.5$.}

    \label{fig:2D3D}
\end{figure}

 This is illustrated in Figure \ref{fig:2D3D}. For  5 different values of $\phi_p^0$, the initial polymer concentration in the film, the microstructure is shown with 2D colorplot of the 2D simulation and 3D images of the isosurface $\phi_p$=0.5. The plots correspond to the moment where there is a visible phase separation at the bottom. It can be seen as the final stage of the initial phase separation. In 2D, for high (resp. low)  initial polymer concentration in the film, the final microstructure consists of circular droplets of the polymer poor (resp. rich) phase in a polymer rich (resp. poor) phase. For intermediate values of the initial polymer concentration ($\phi_p^0= 0.3$) the 2D pattern consists also of isolated droplets of the polymer rich phase in a polymer poor matrix.  However, the droplet are no longer circular and have a more elongated shape. Finally for $\phi_p^0$ the pattern may be seen as maze-like and reminiscent of a bicontinuous microstructure. Nevertheless it still consists of isolated domains. \\

 This is not the case in 3D. Indeed,  while for either low or high values of  $\phi_p^0$, a pattern consisting of isolated domains in a matrix can be identified clearly. On the contrary  for intermediate values, whether both polymer rich and polymer poor regions are continuous is  unclear. However the use of the tools described above allows to get a better understanding of the final pattern and of the evolution of the continuity of the phases.\\

\begin{figure}[H]
    \centering
    \renewcommand{\arraystretch}{1.5}
    \begin{minipage}{0.85\textwidth}
    \centering
    \begin{tabular}{|c|c|c|c|c|}
        \hline
        t & \textbf{$4.8$} & \textbf{$7.2$} & \textbf{$9$} & \textbf{$12$} \\
        \hline
        \raisebox{1cm}{\textbf{2D}} & 
        \includegraphics[width=0.13\textwidth]{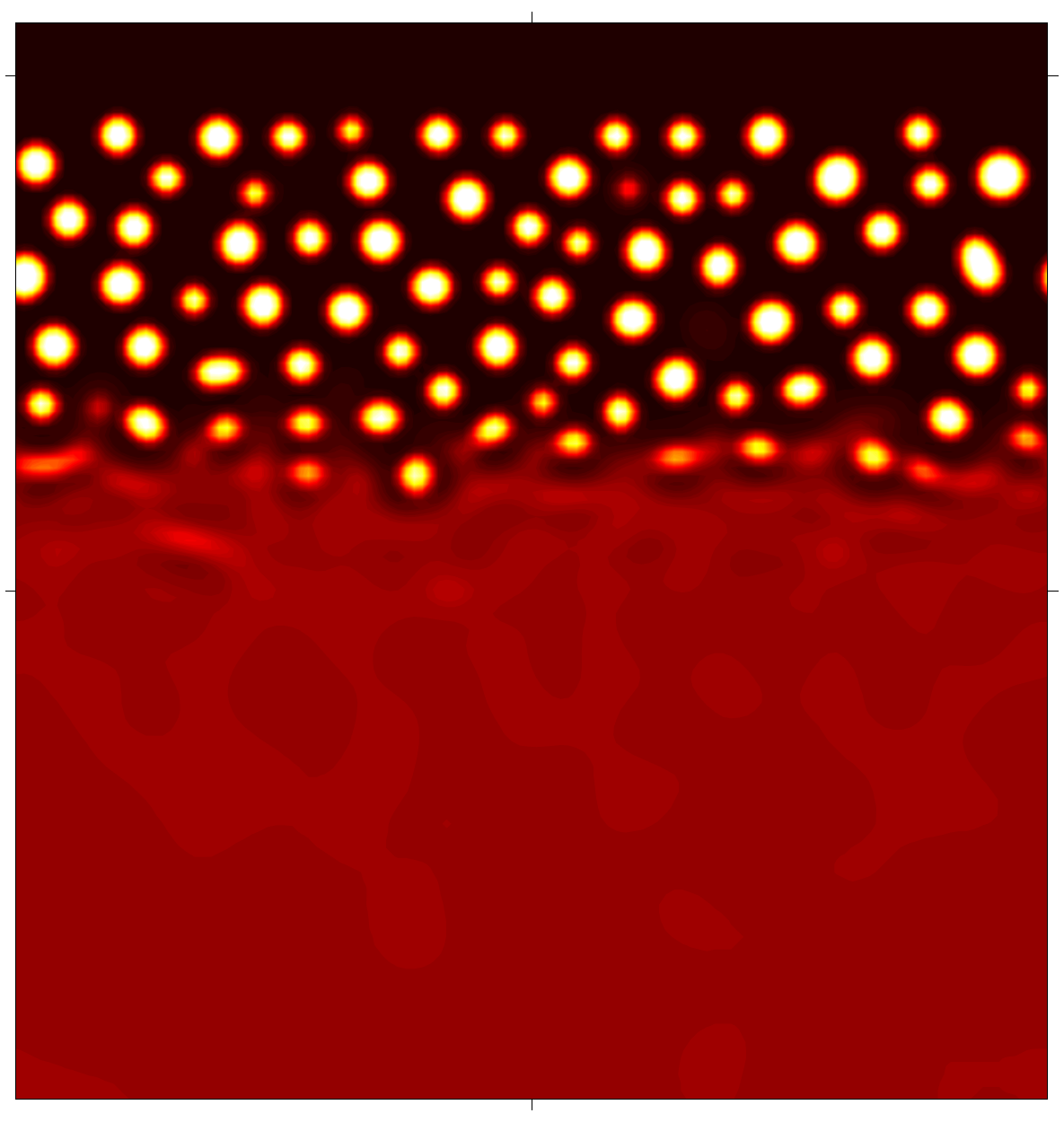} &
        \includegraphics[width=0.13\textwidth]{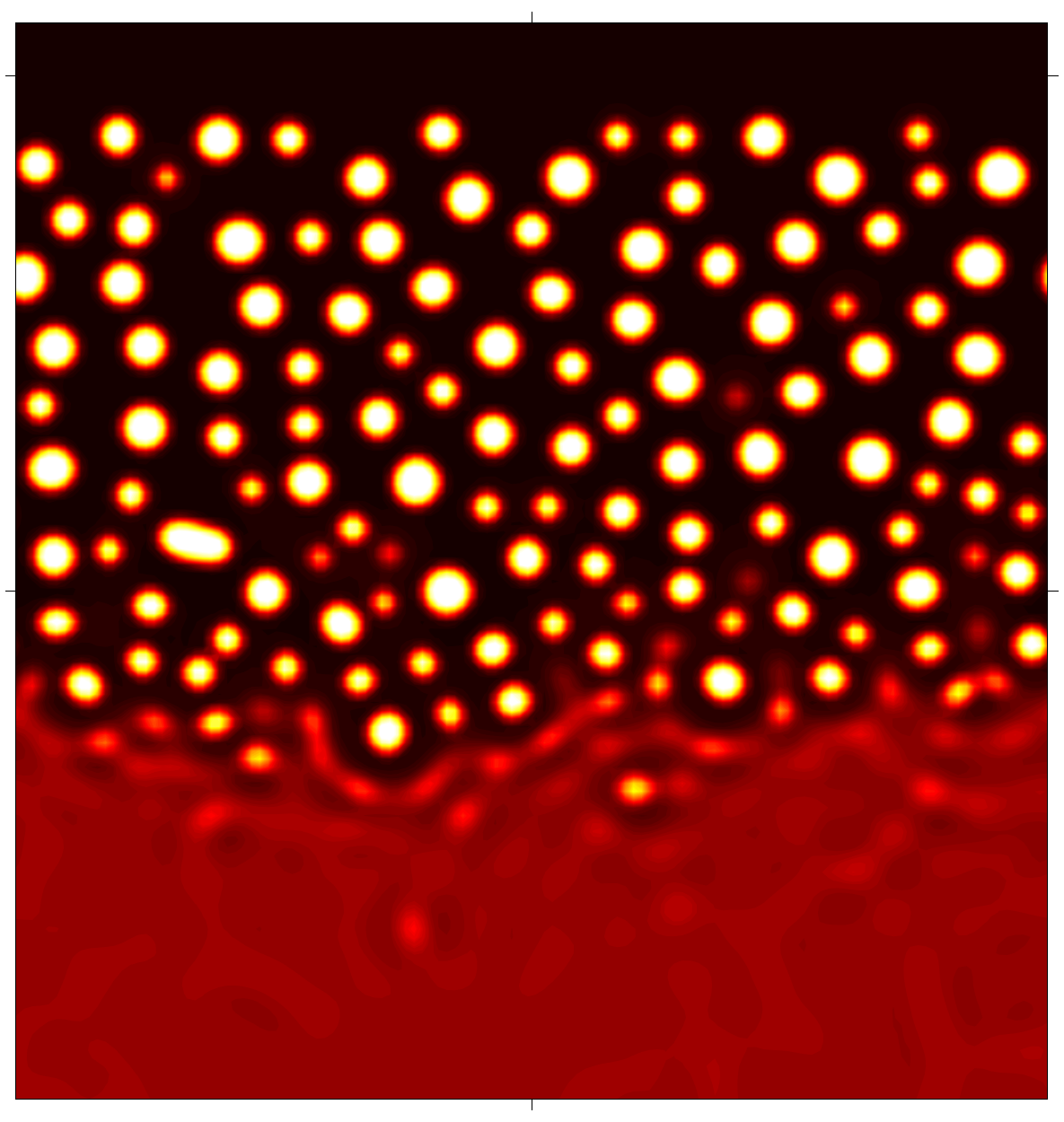} &
        \includegraphics[width=0.13\textwidth]{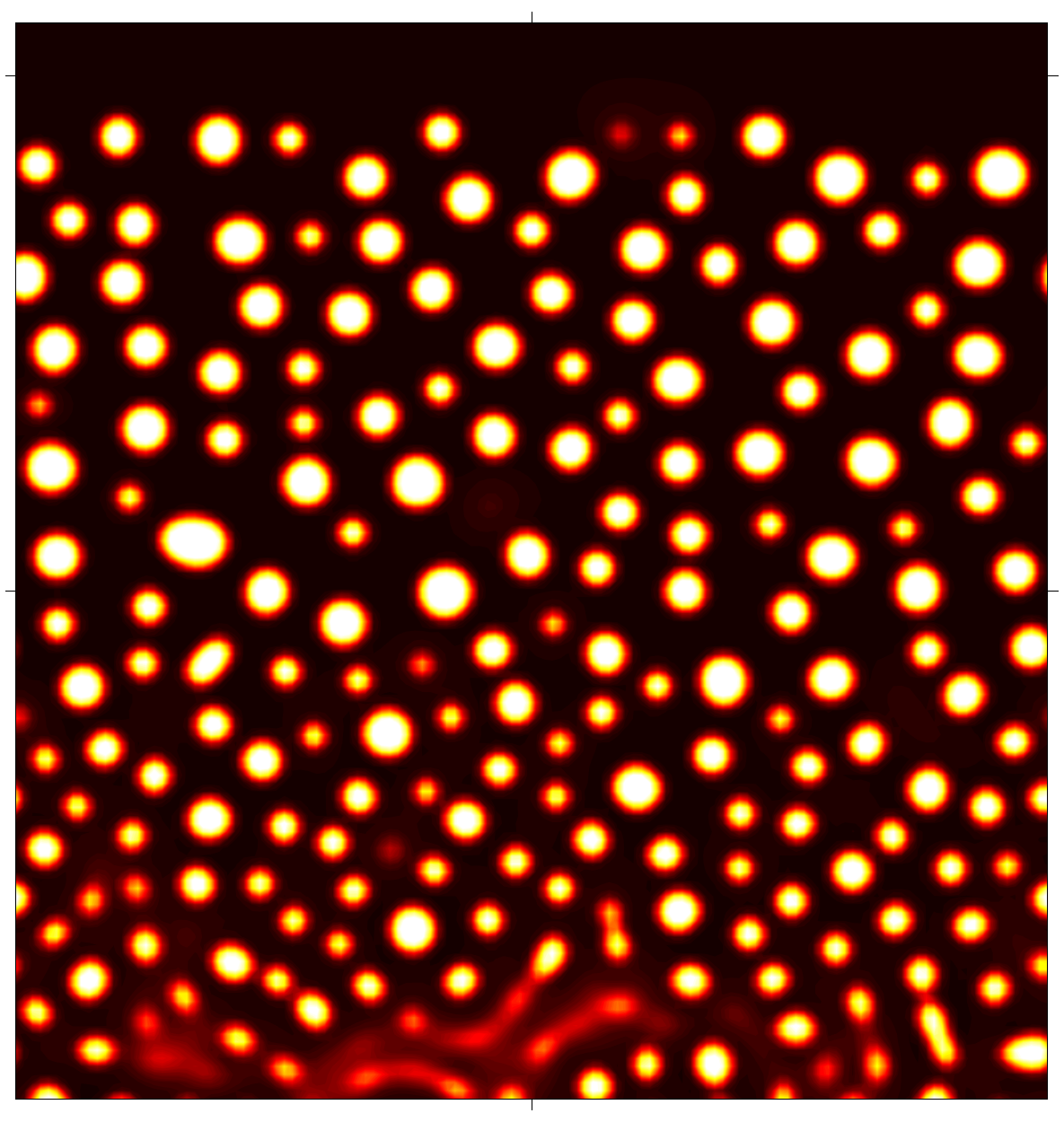} &
        \includegraphics[width=0.13\textwidth]{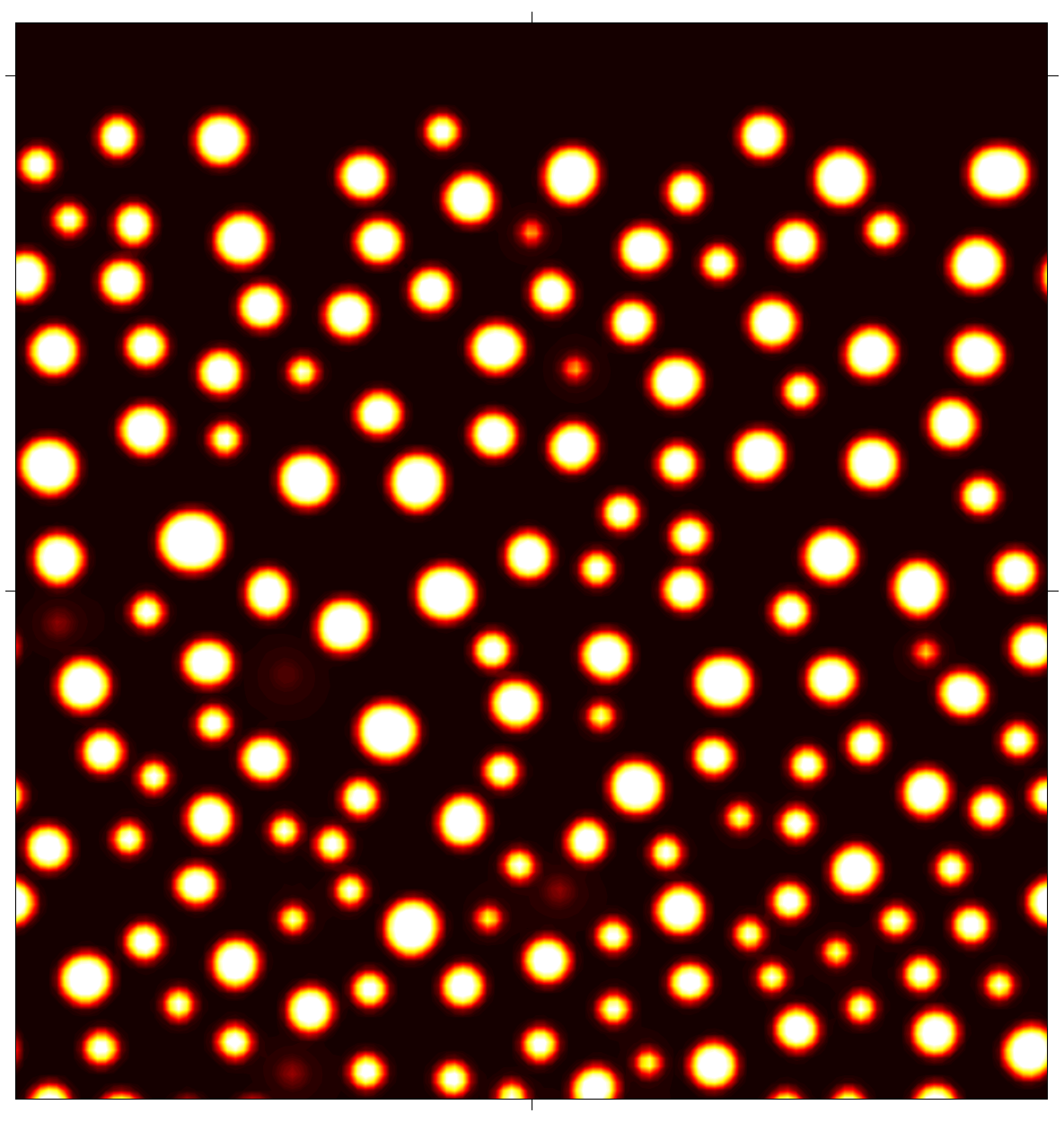} \\
        \hline

        \raisebox{1cm}{\textbf{3D slice}} & 
        \includegraphics[width=0.13\textwidth]{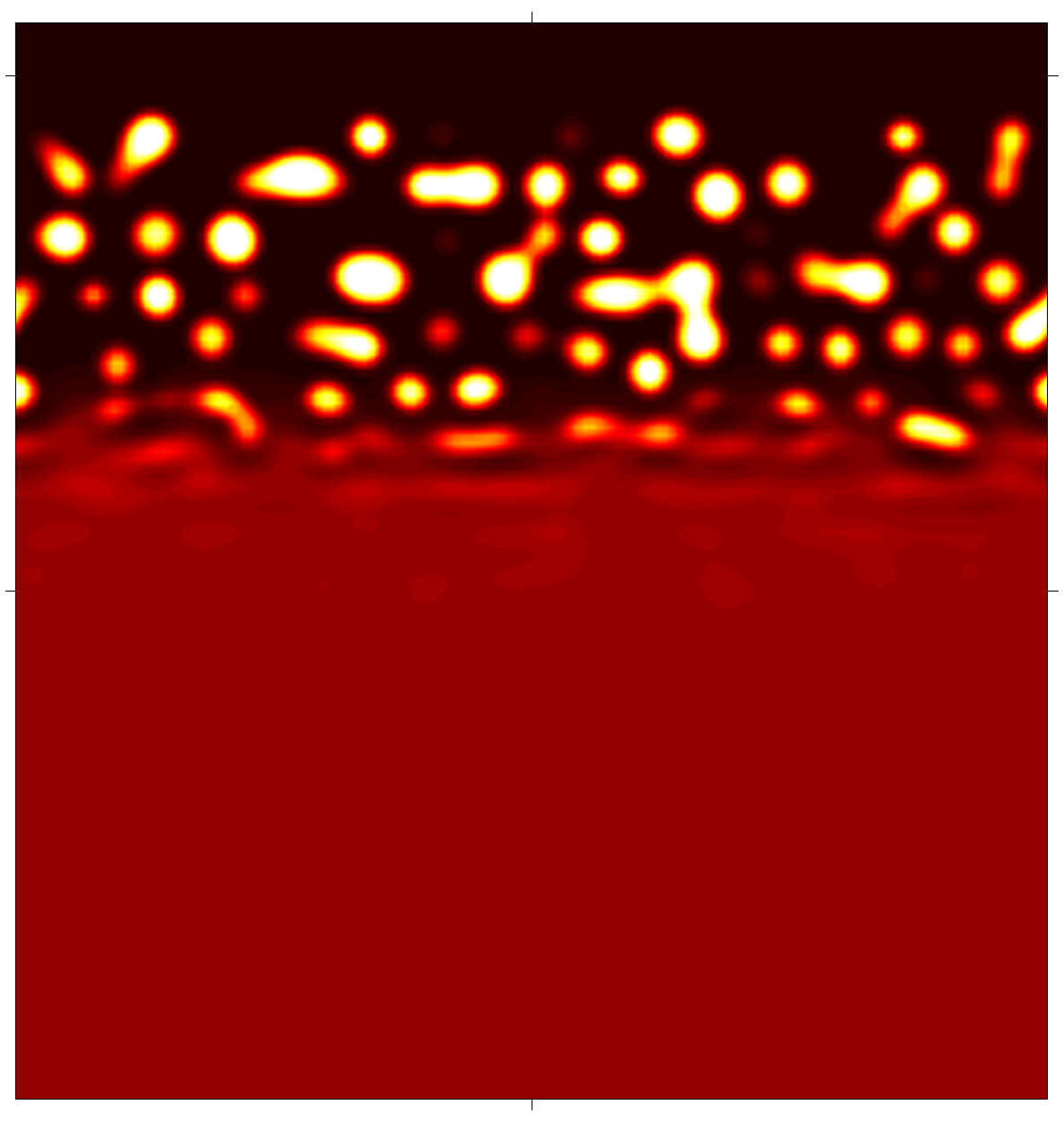} &
        \includegraphics[width=0.13\textwidth]{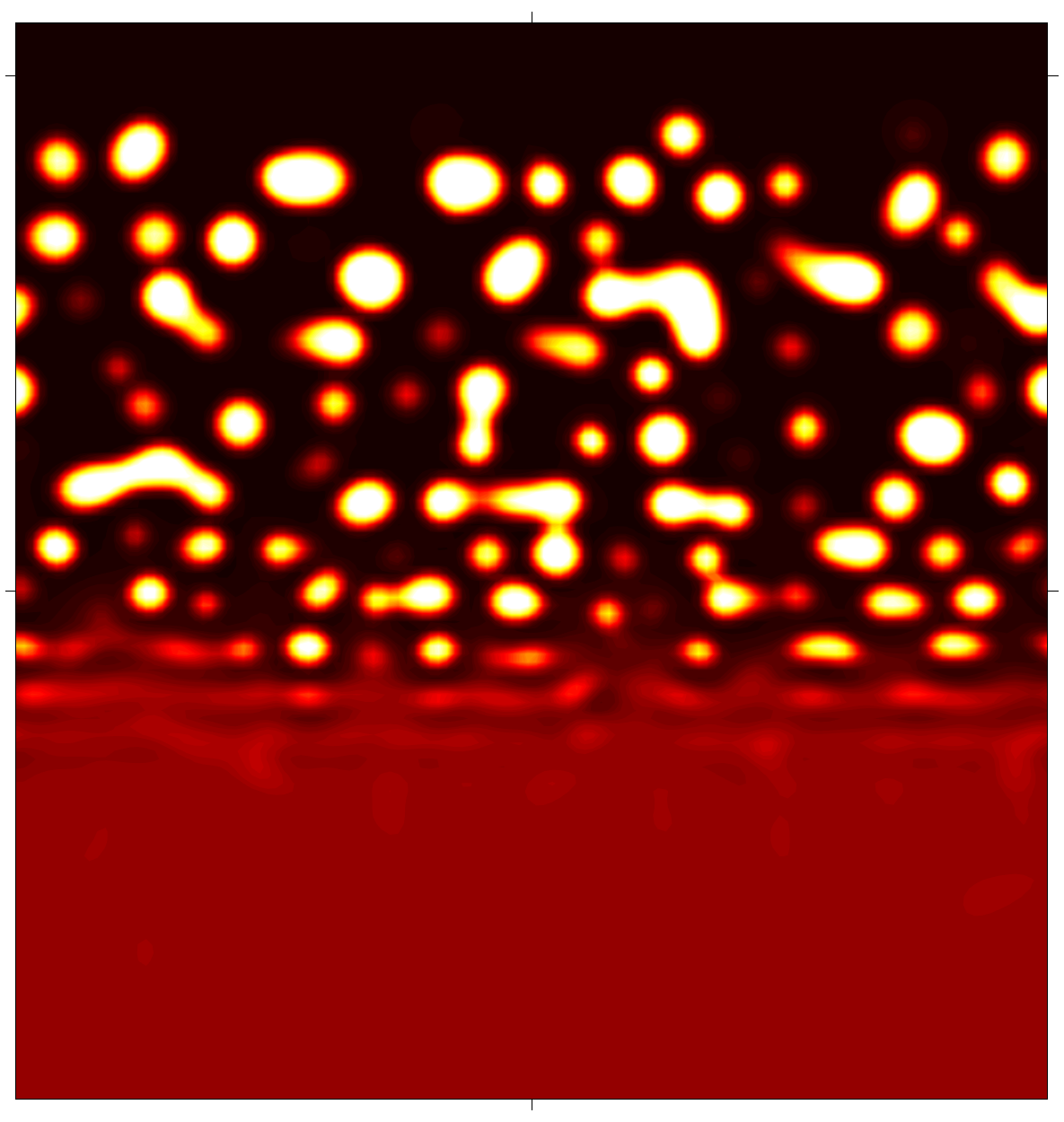} &
        \includegraphics[width=0.13\textwidth]{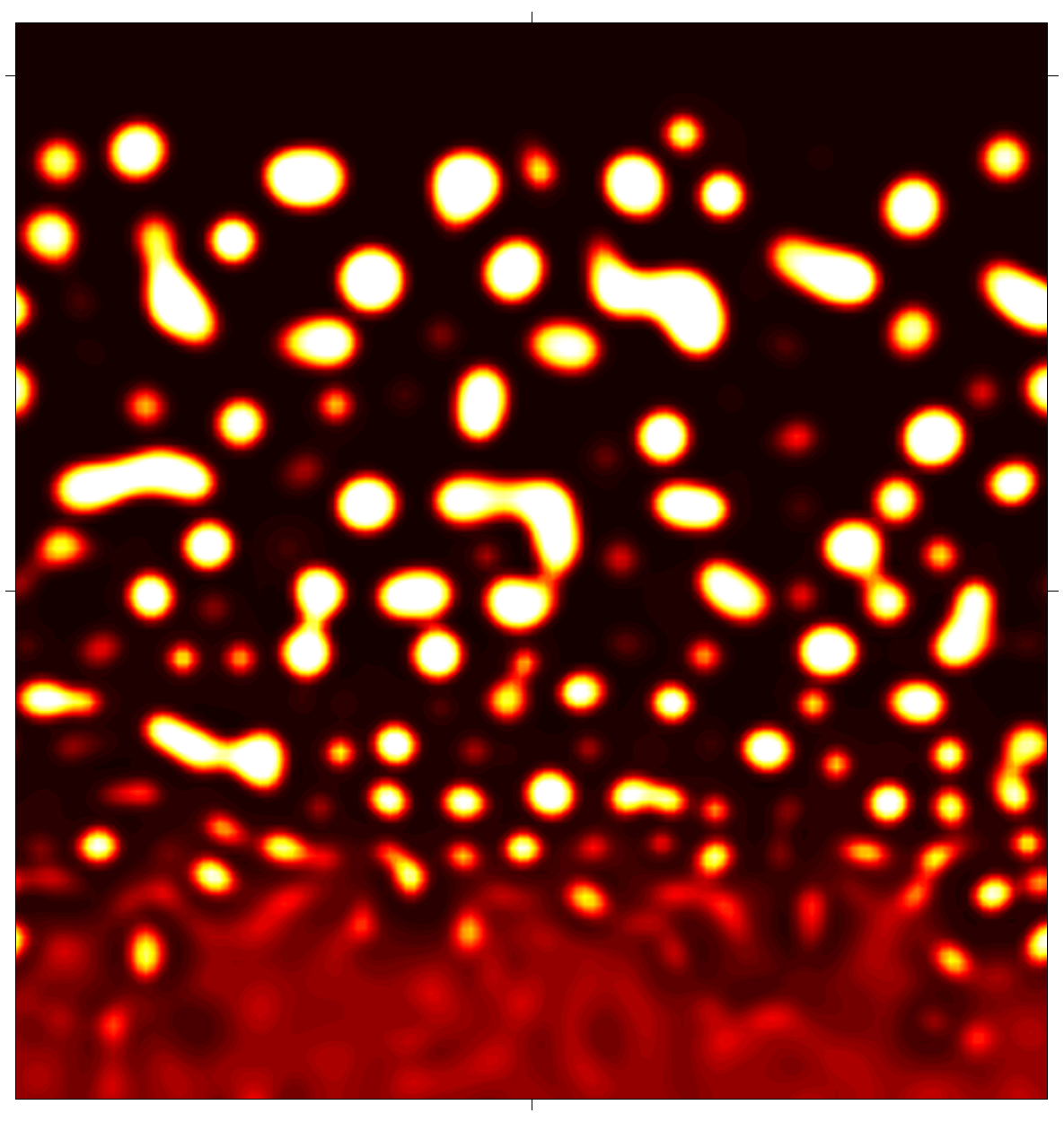} &
        \includegraphics[width=0.13\textwidth]{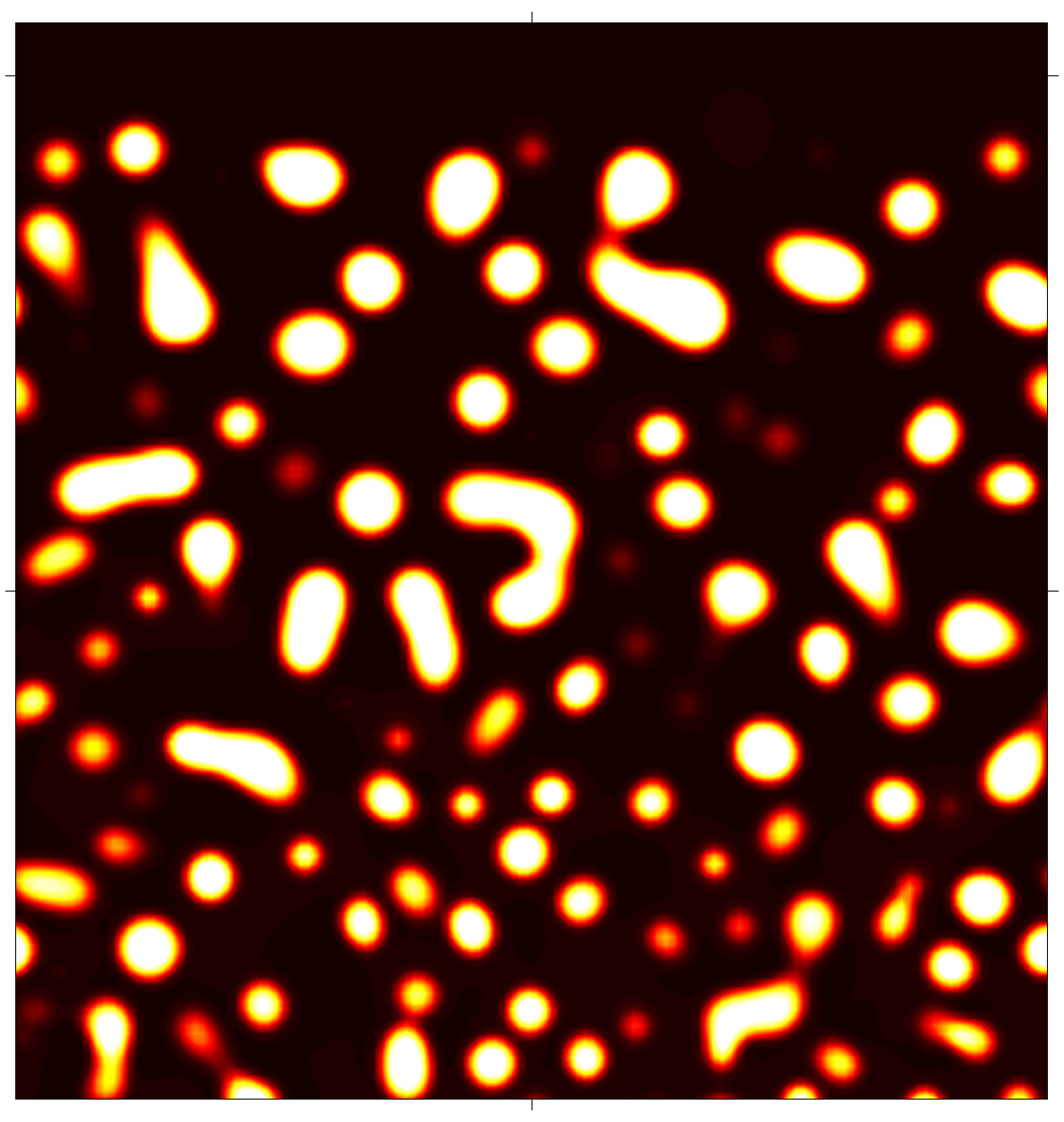} \\
        \hline

        \raisebox{1cm}{\textbf{Connectivity}} & 
        \includegraphics[width=0.16\textwidth,trim={4cm 4cm 4cm 5cm},clip]{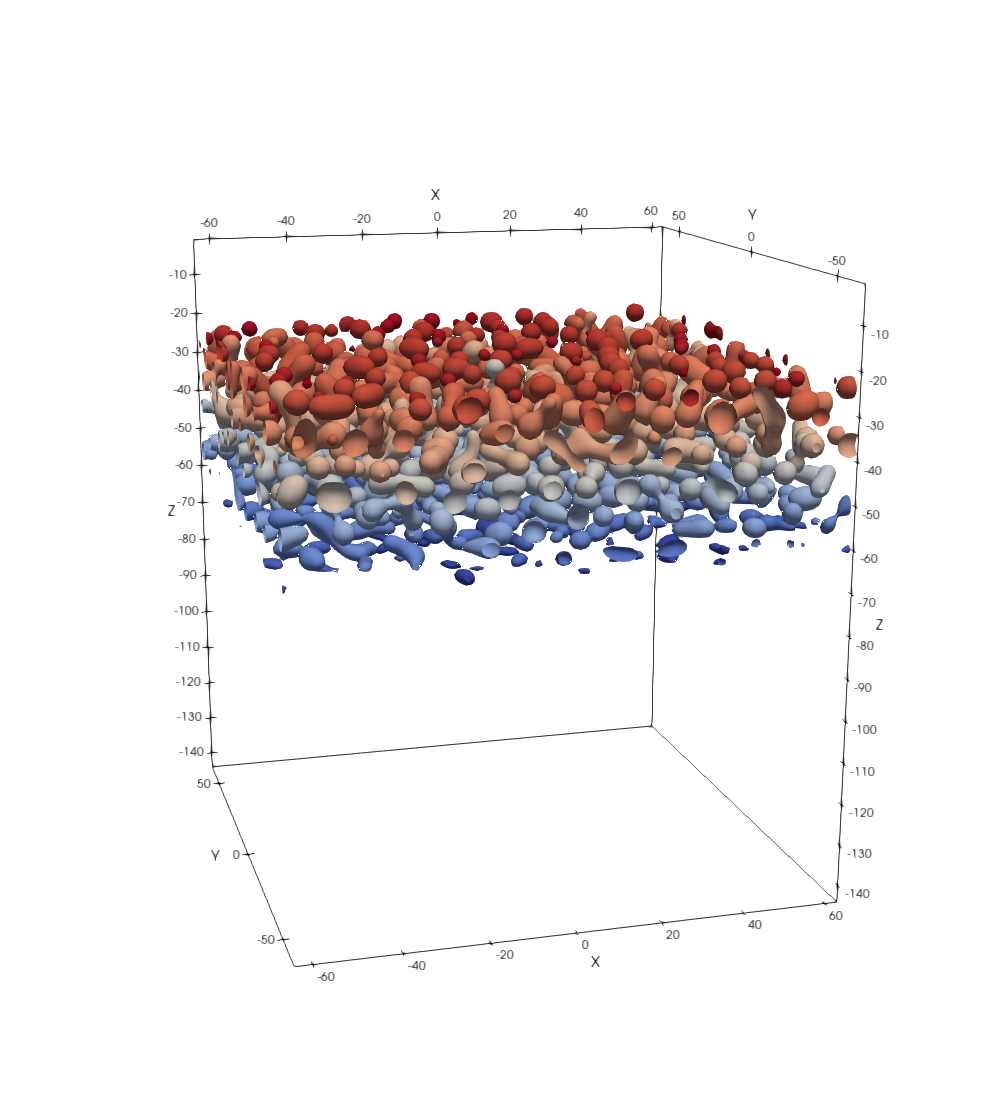} &
        \includegraphics[width=0.16\textwidth,trim={4cm 4cm 4cm 5cm},clip]{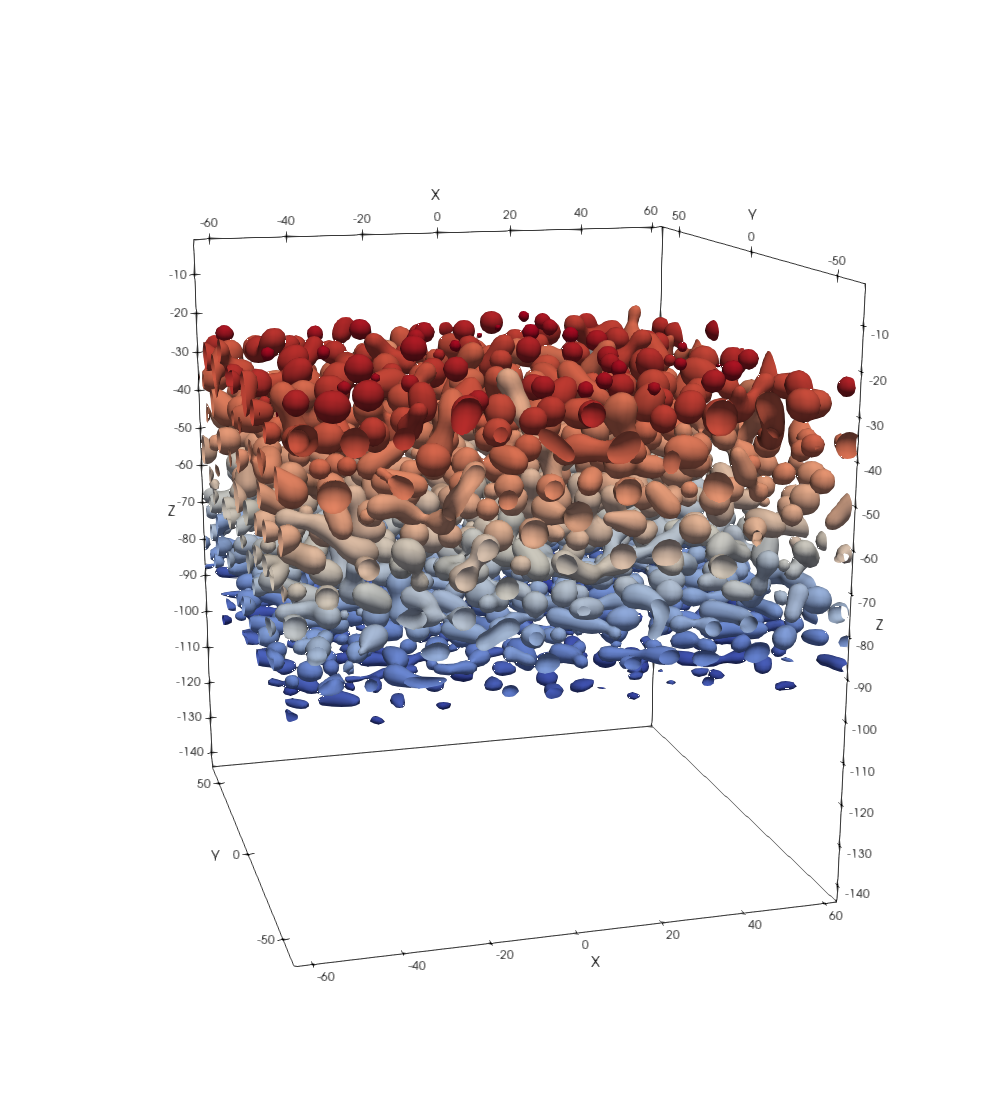} &
        \includegraphics[width=0.16\textwidth,trim={4cm 4cm 4cm 5cm},clip]{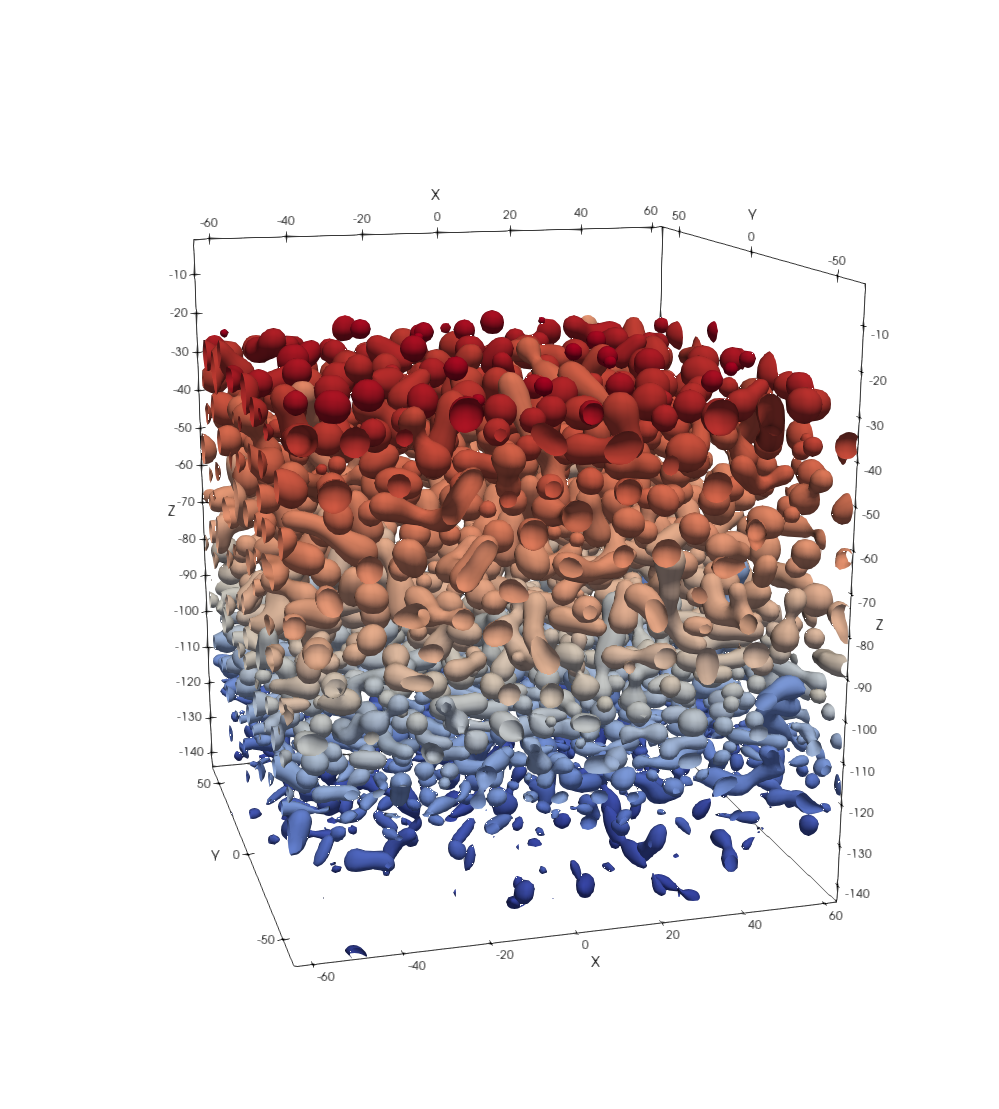} &
        \includegraphics[width=0.16\textwidth,trim={4cm 4cm 4cm 5cm},clip]{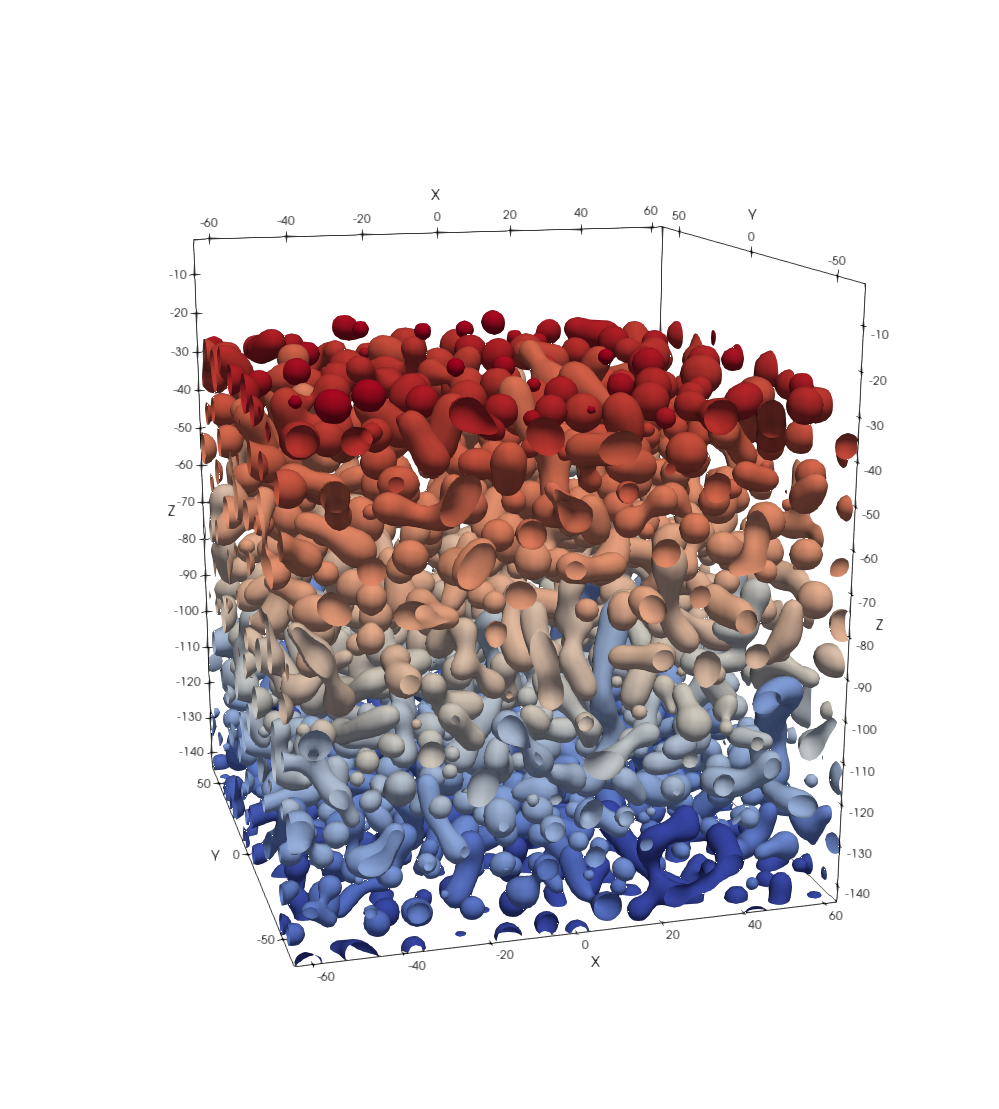} \\
        \hline
    \end{tabular}

    \end{minipage}%
    \hfill
    \raisebox{0.65cm}{
    \begin{minipage}{0.07\textwidth}
        \centering
        \includegraphics[width=\textwidth]{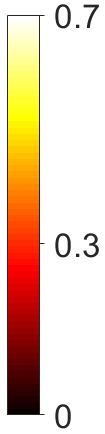}
    \end{minipage}
    }
    
	\caption{Simulation results for $\phi_p^0 = 0.15$, illustrating the temporal evolution of the microstructure at four time values ($t \times 10^4$). The first row shows 2D simulations, the second row presents a planar slice extracted from the 3D simulations (whose position is shown above), and the third row displays the connectivity analysis obtained using a ParaView filter: \change{ the color corresponds to the index of the region and has no physical meaning}. The bath is located at the top of each image.} 
    \label{fig:3D_phi_0.15}

\end{figure}

First we consider the situation where both 2D and 3D system give rise to a discontinuous polymer structure. In fig.  \ref{fig:3D_phi_0.15} we represent  the evolution of the pattern in a film with an initial low polymer concentration:$\phi_p^0=0.15$ in 2D (top row) and in 3D (2 bottom rows). The first two rows corresponds to 2D color maps of $\phi_p$ of a 2D simulation and of a plane of a 3D simulation that  goes from the top to the bottom of the simulation box. The bottom row is a perspective view of the isosurface $\phi_p=0.5$ where each isolated interface is given a  color.  The 2D color map show that there are clear differences between the patterns. First  in images from the 2D simulations, the initial structure (at the very early stage of phase separation) before the polymer concentration reaches its equilibrium value is reminiscent of a maze  and can be seen deep below the phase separated domain. It rapidly evolves toward a pattern of almost circular droplet in a polymer poor matrix. In the case of 3D simulations,  the initial pattern that is appearing just below the phase separated region is  a layered structure normal to the vertical axis and it evolves towards complex shapes that can be elongated along any direction. This  can be seen for instance at $t=12$. However the pattern remains discontinuous. This is evidenced in the bottom row where the index of the continuous object is not the same at top and bottom.  Hence, the process in 2D and 3D differ significantly despite the nature of the final patterns are the same.\\
\begin{figure}
    \centering
    \renewcommand{\arraystretch}{1.5}
    \begin{minipage}{0.85\textwidth}
    \centering
    \begin{tabular}{|c|c|c|c|c|}
        \hline
        t & \textbf{$3$} & \textbf{$4.2$} & \textbf{$6$} & \textbf{$12$}  \\
        \hline
        \raisebox{1cm}{\textbf{2D}} & 
        \includegraphics[width=0.13\textwidth]{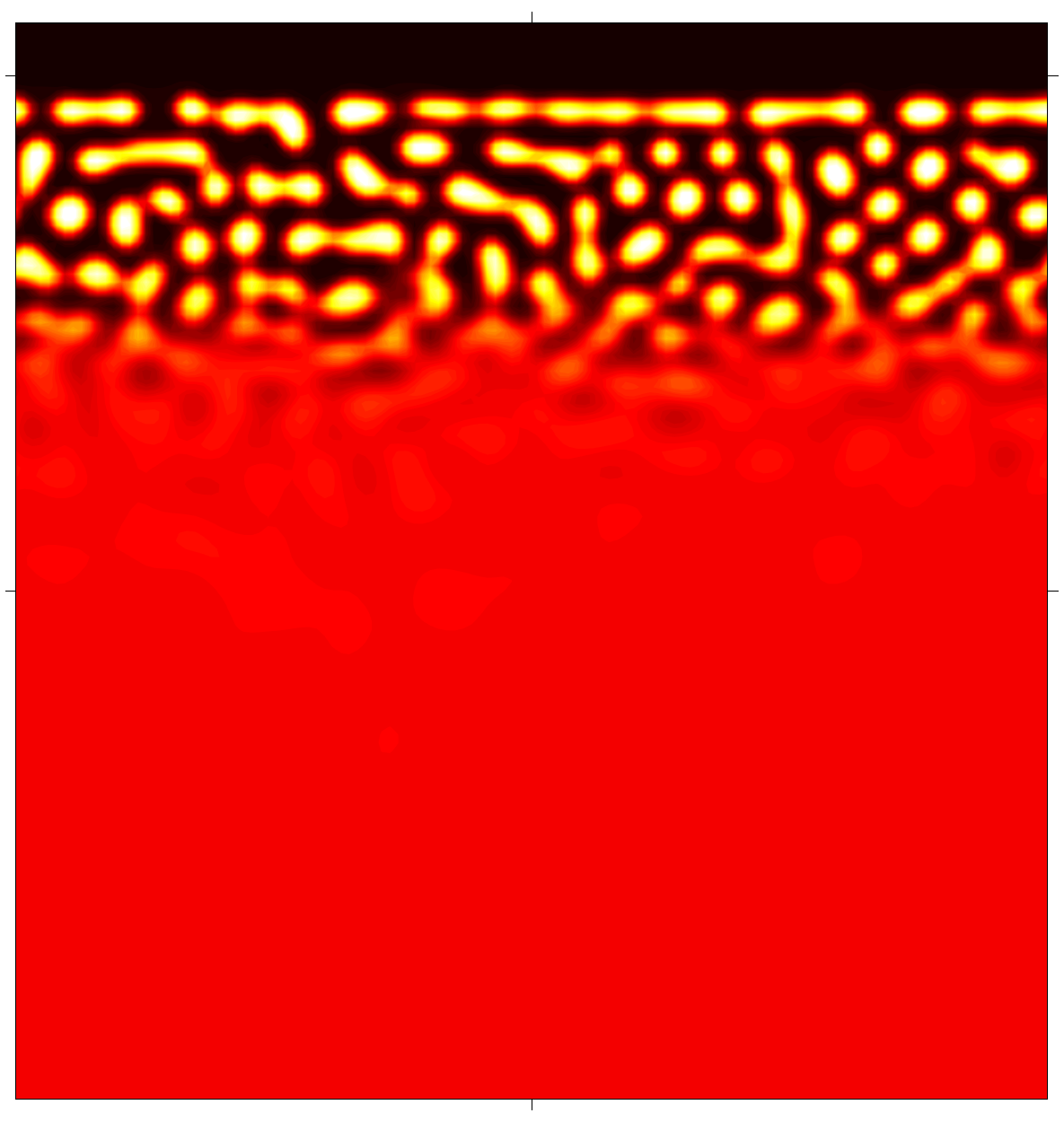} &
        \includegraphics[width=0.13\textwidth]{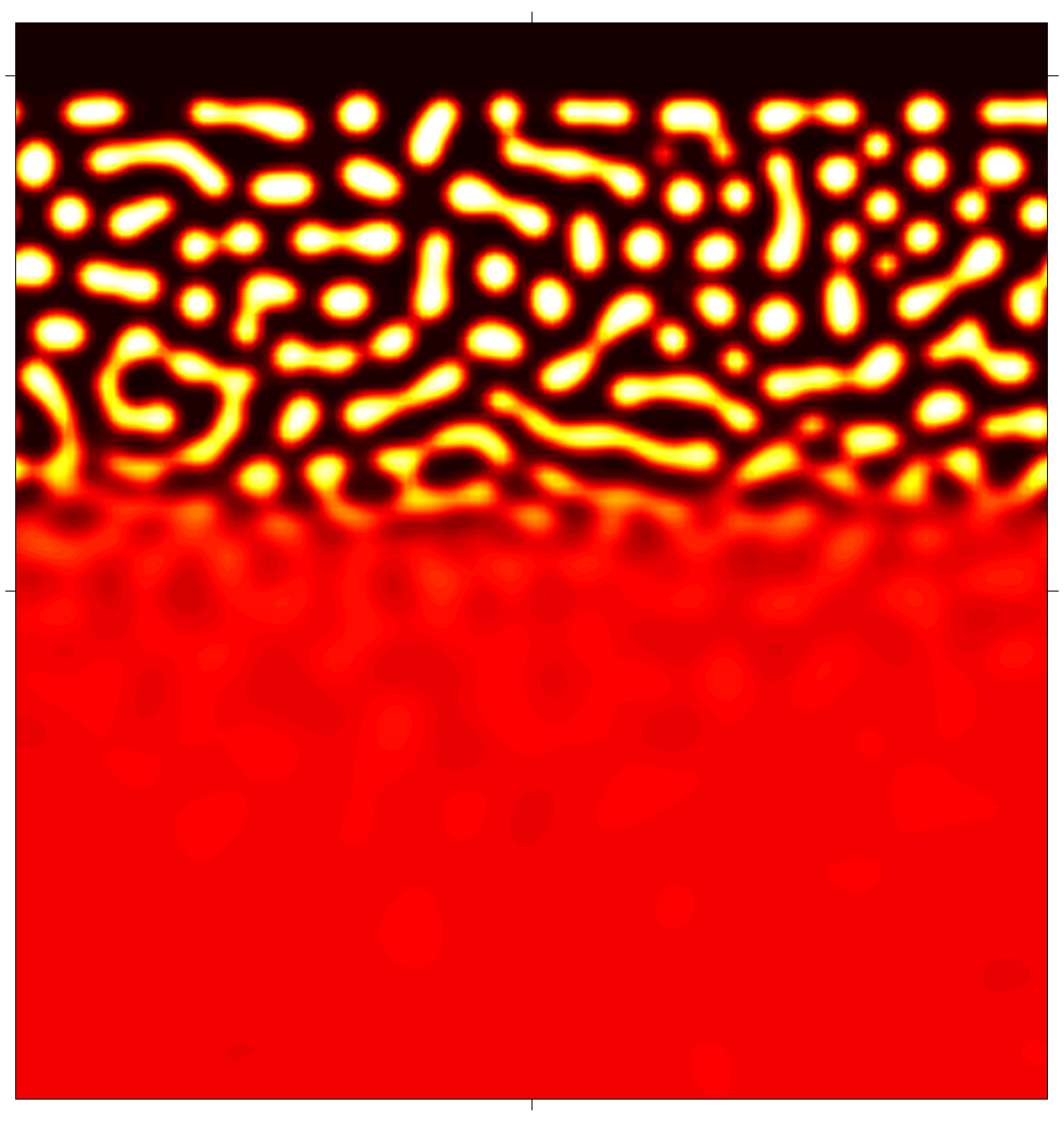} &
        \includegraphics[width=0.13\textwidth]{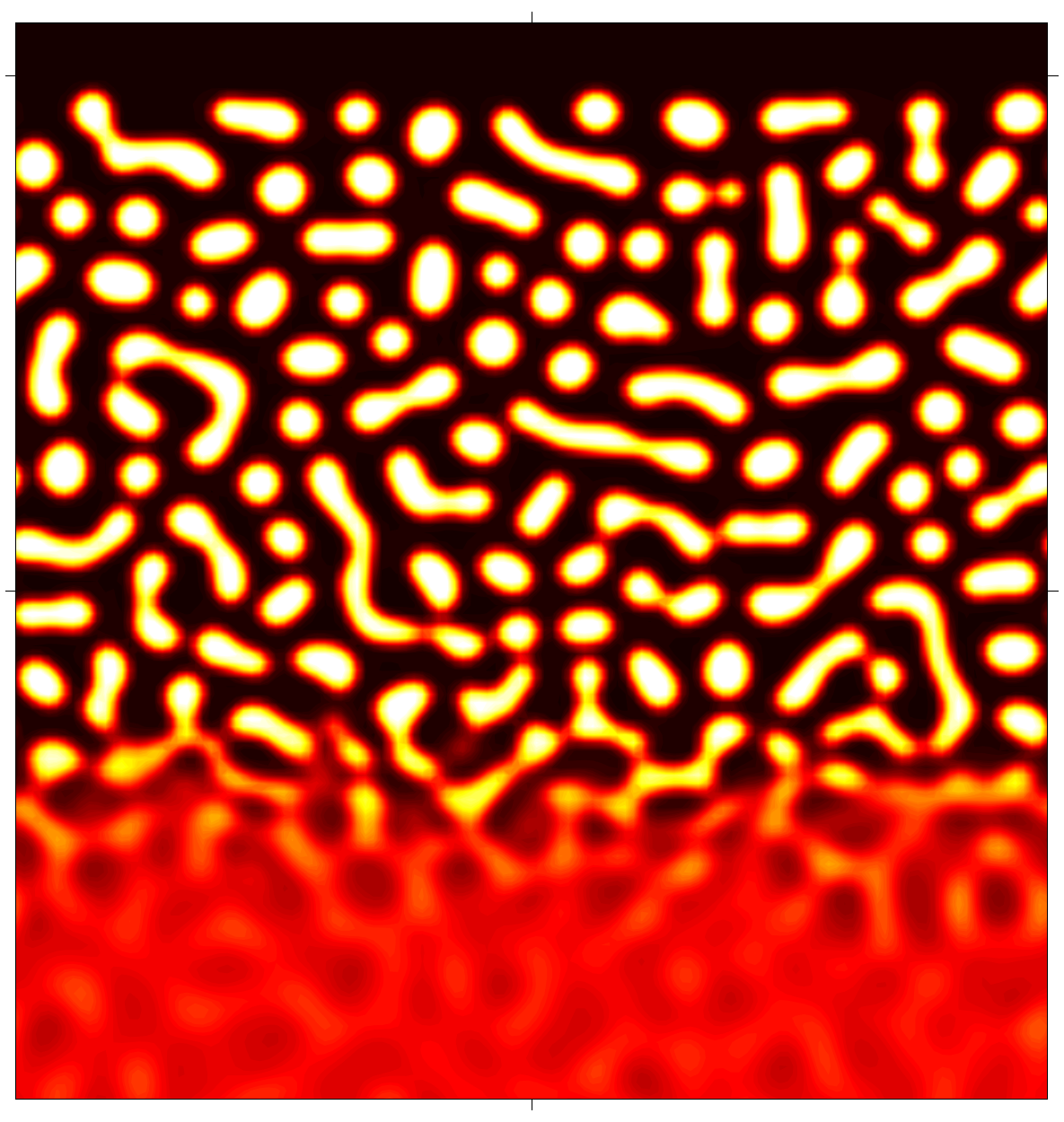} &
        \includegraphics[width=0.13\textwidth]{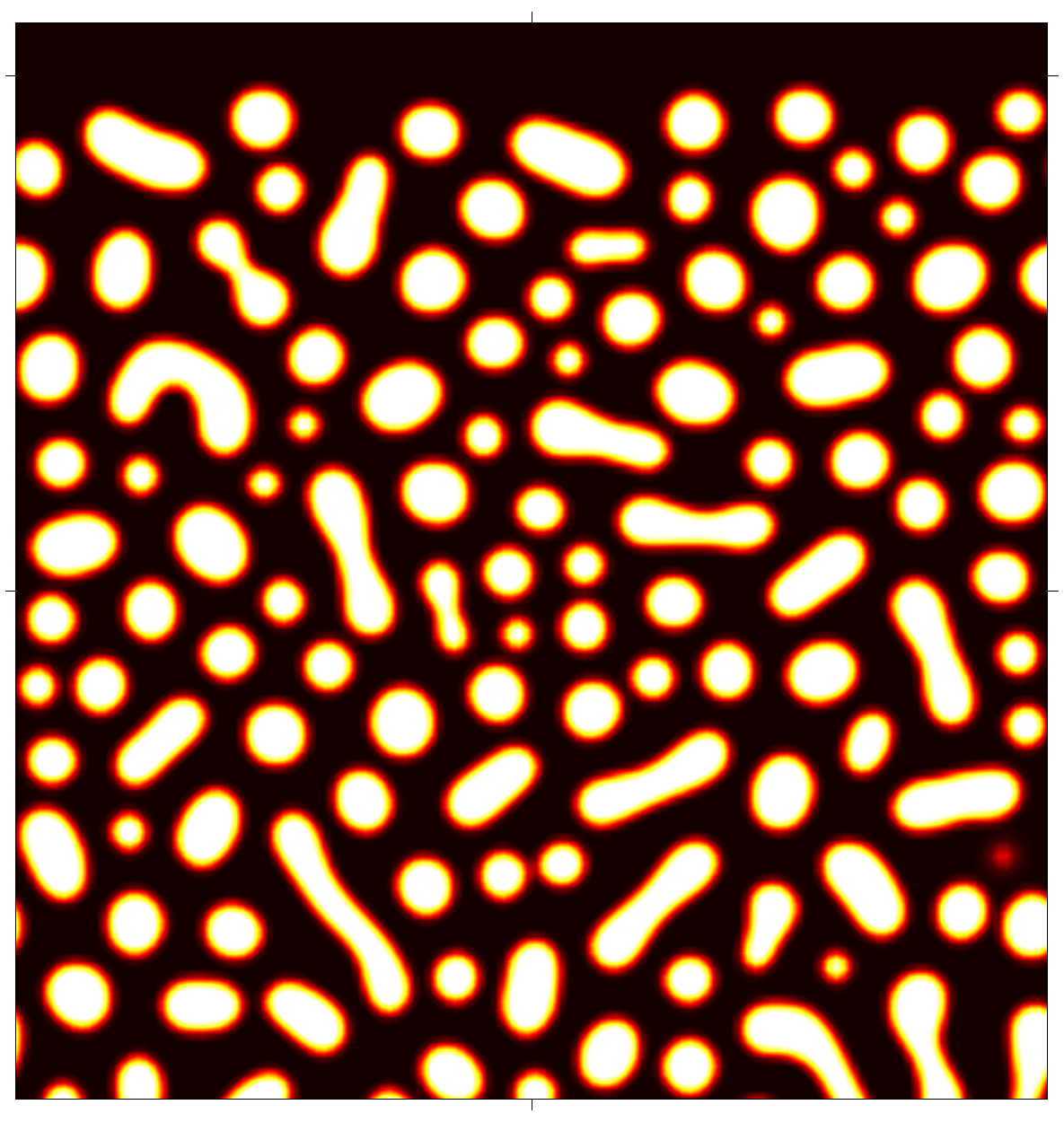} \\
        \hline
        
        \raisebox{1cm}{\textbf{3D slice}} & 
        \includegraphics[width=0.13\textwidth]{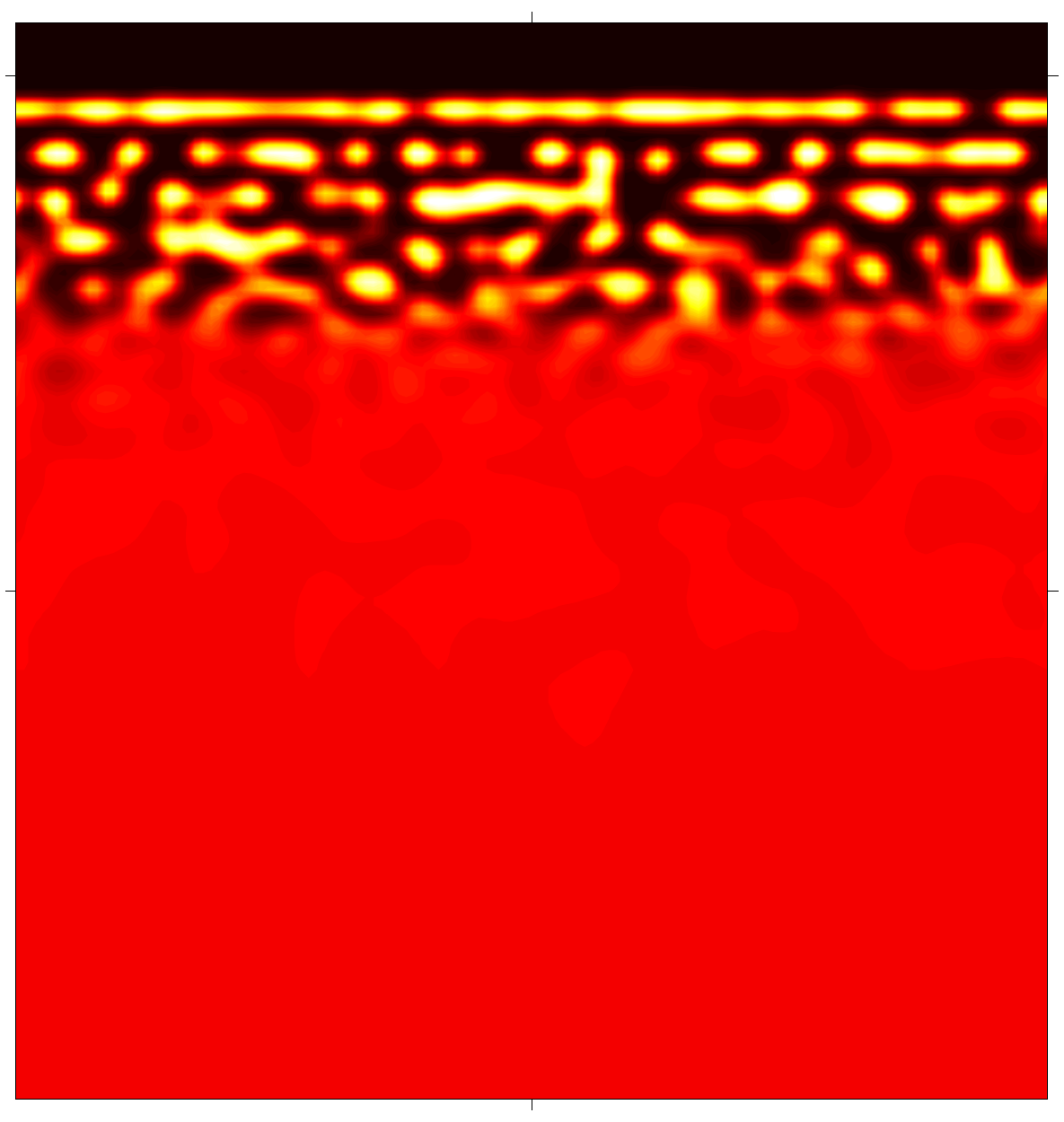} &
        \includegraphics[width=0.13\textwidth]{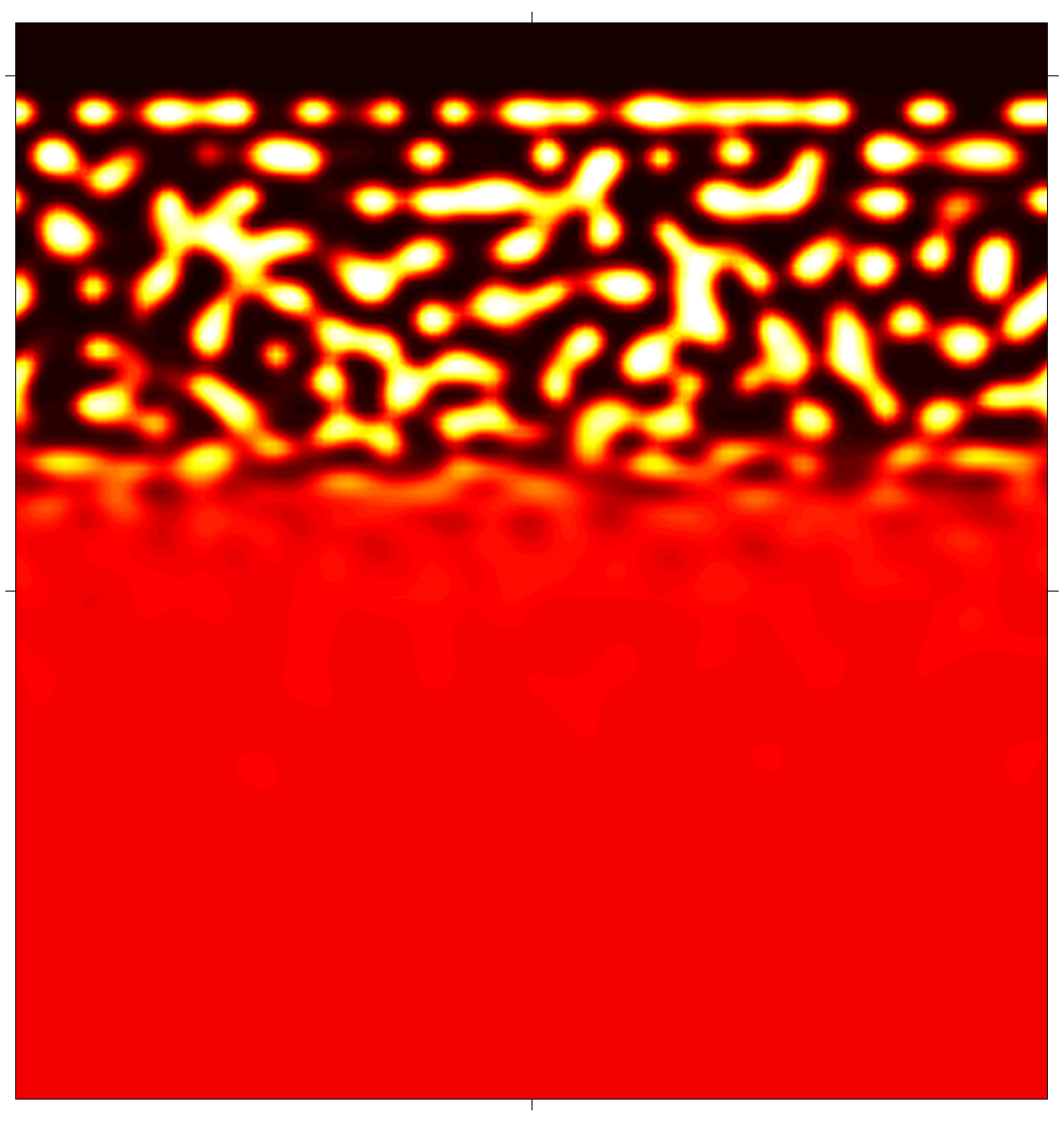} &
        \includegraphics[width=0.13\textwidth]{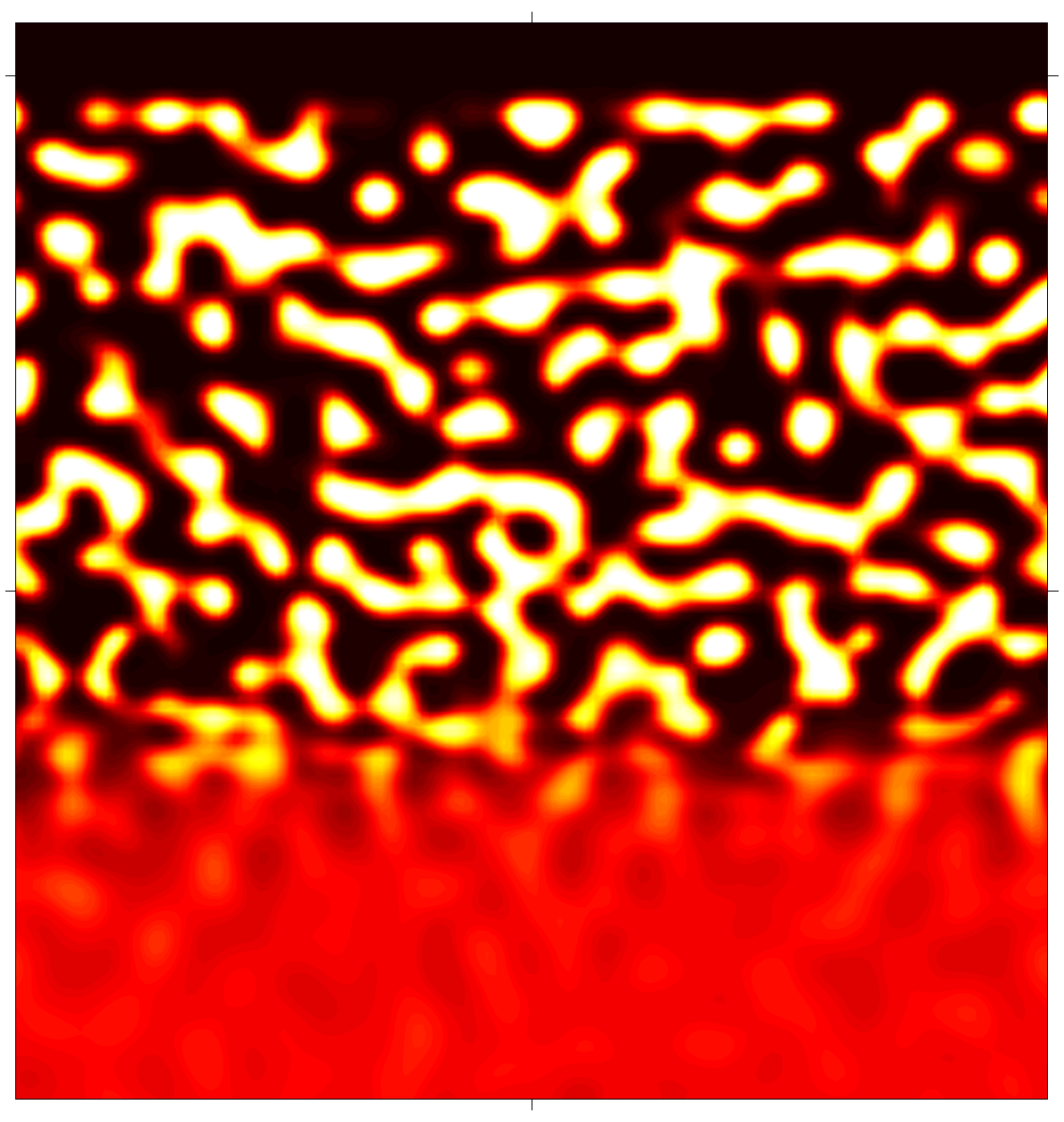} &
        \includegraphics[width=0.13\textwidth]{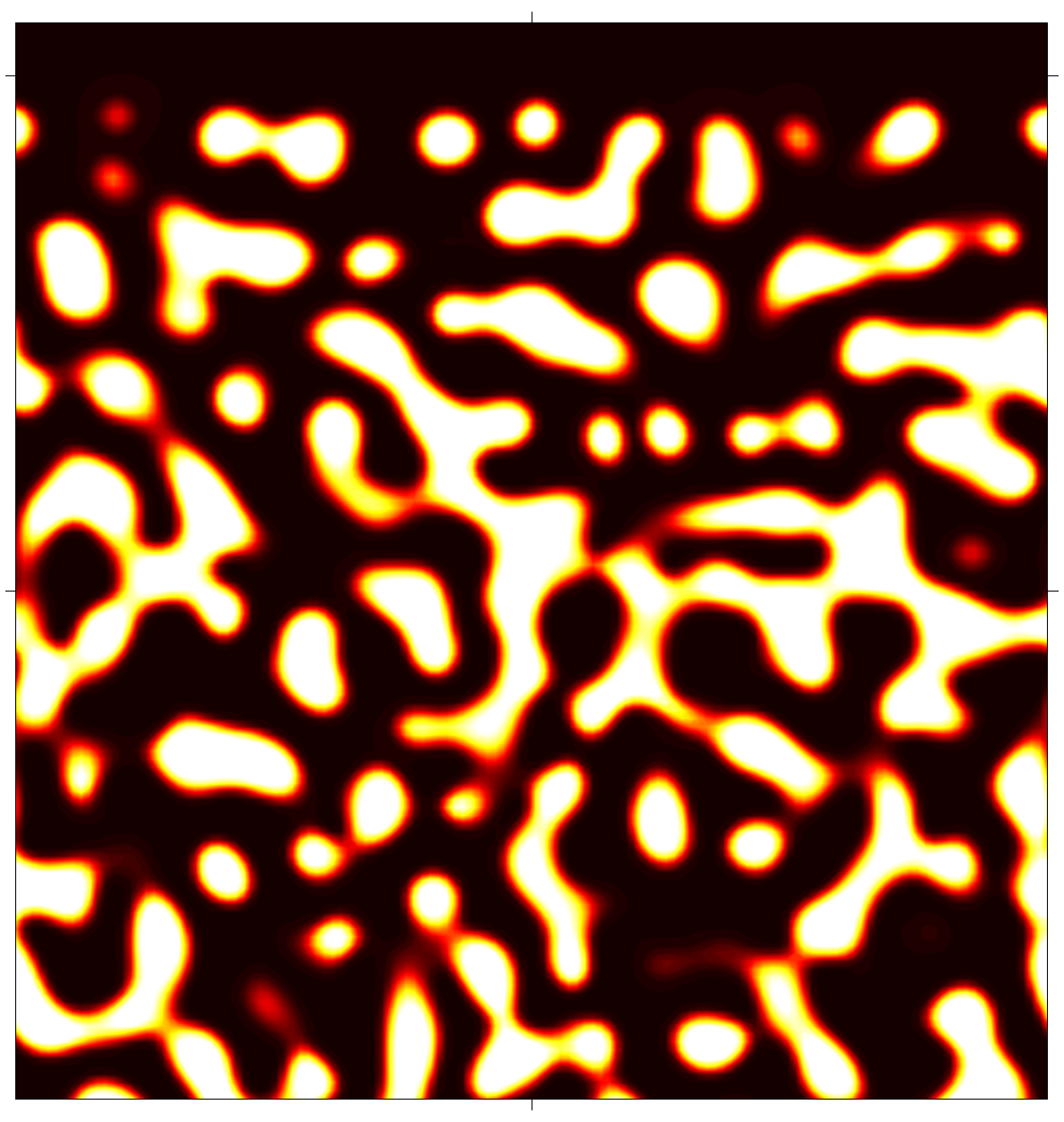} \\
        \hline
        
        \raisebox{1cm}{\textbf{Connectivity}} & 
        \includegraphics[width=0.16\textwidth,trim={3cm 3cm 3cm 4cm},clip]{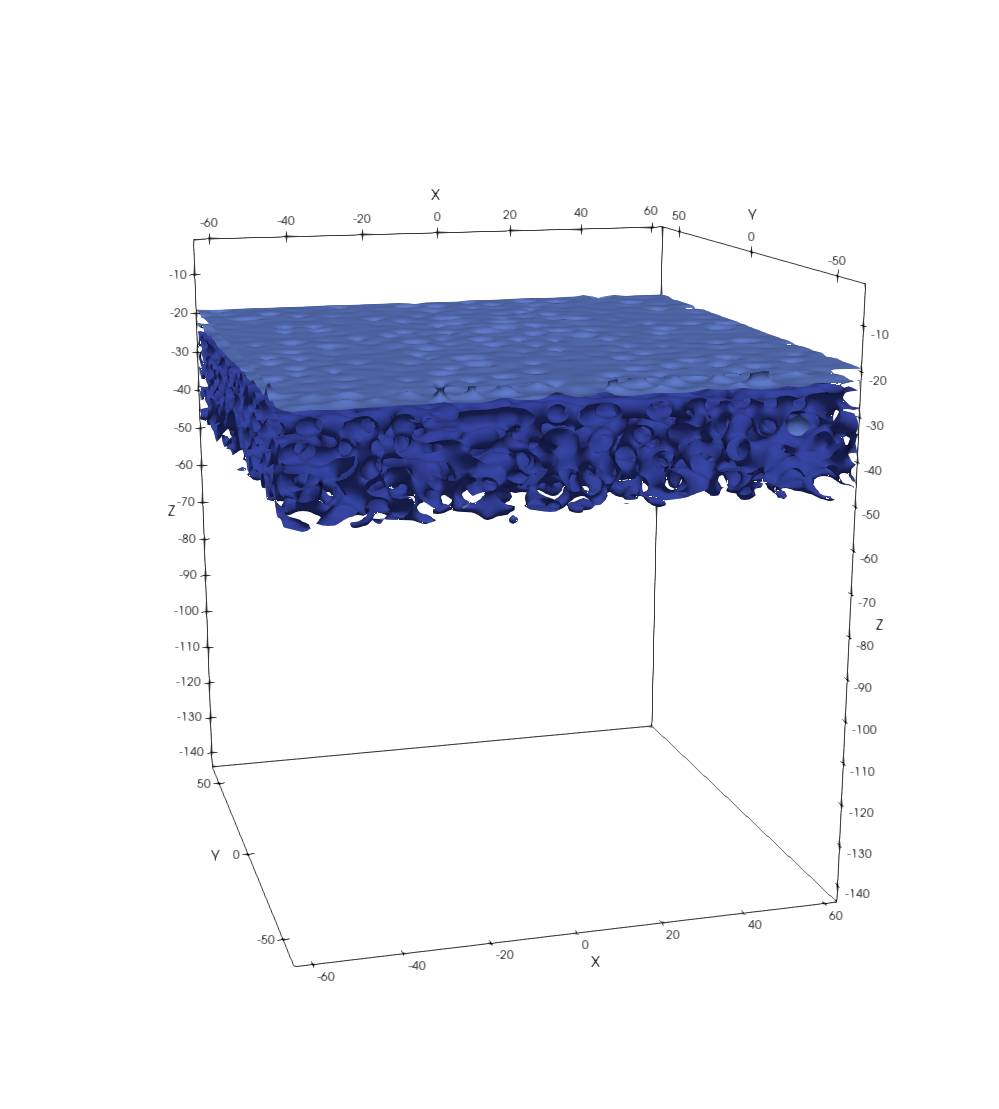} &
        \includegraphics[width=0.16\textwidth,trim={3cm 3cm 3cm 4cm},clip]{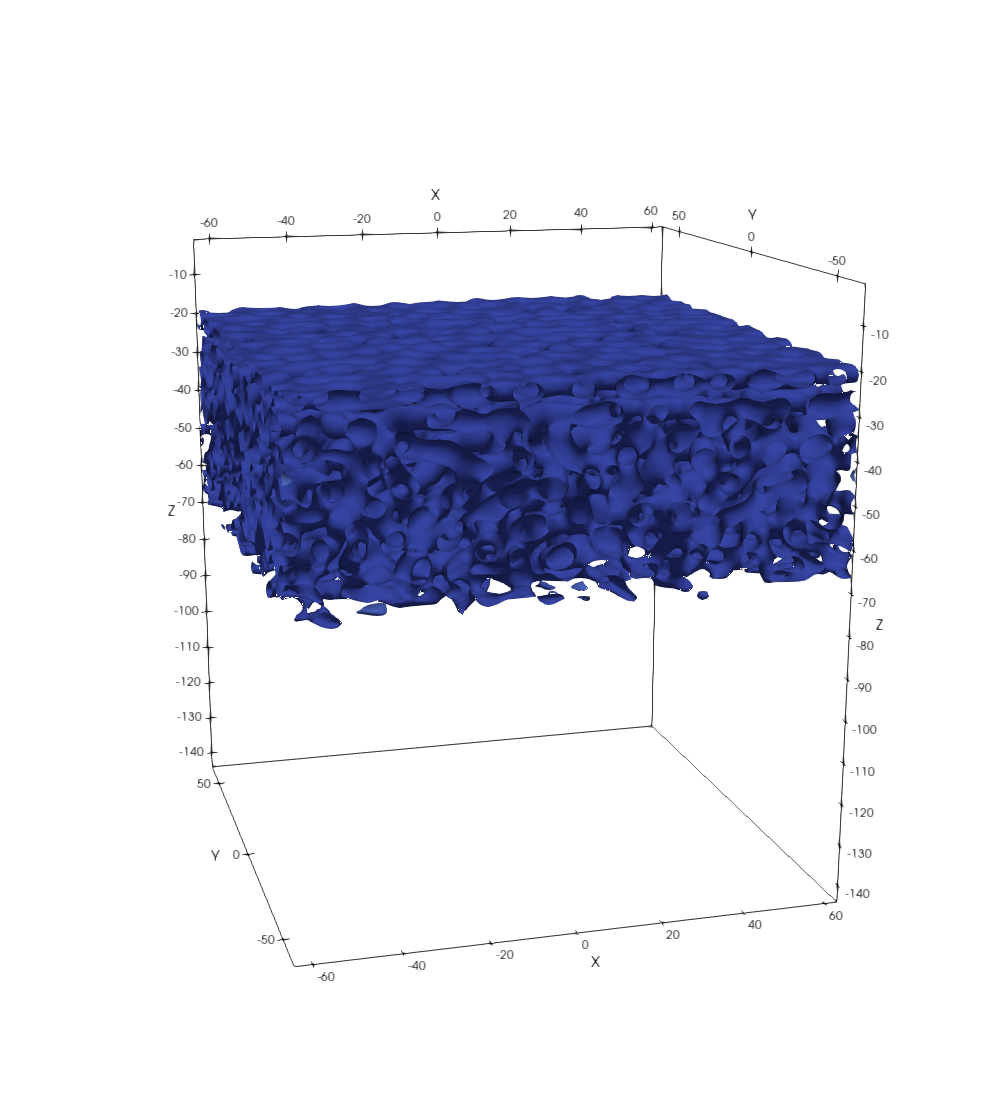} &
        \includegraphics[width=0.16\textwidth,trim={3cm 3cm 3cm 4cm},clip]{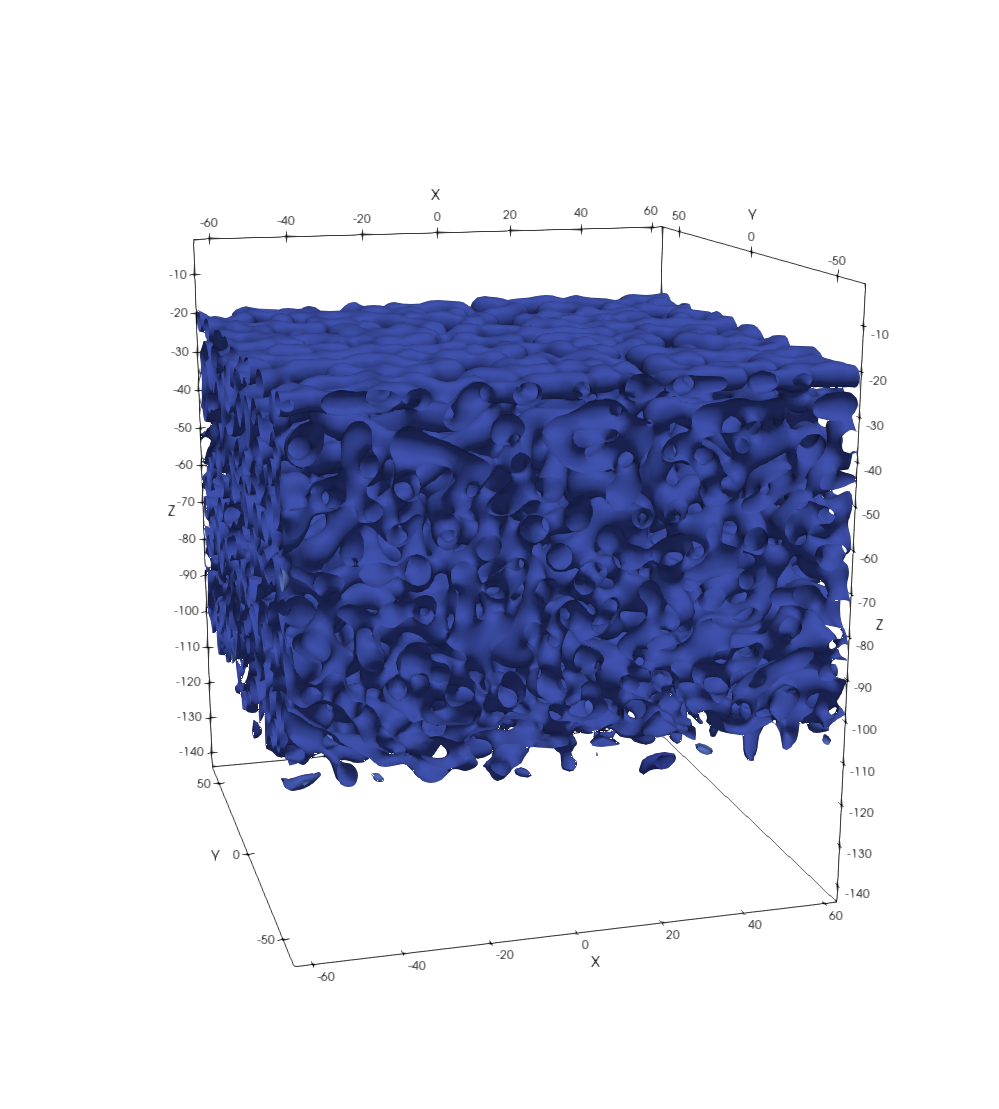} &
        \includegraphics[width=0.16\textwidth,trim={3cm 3cm 3cm 4cm},clip]{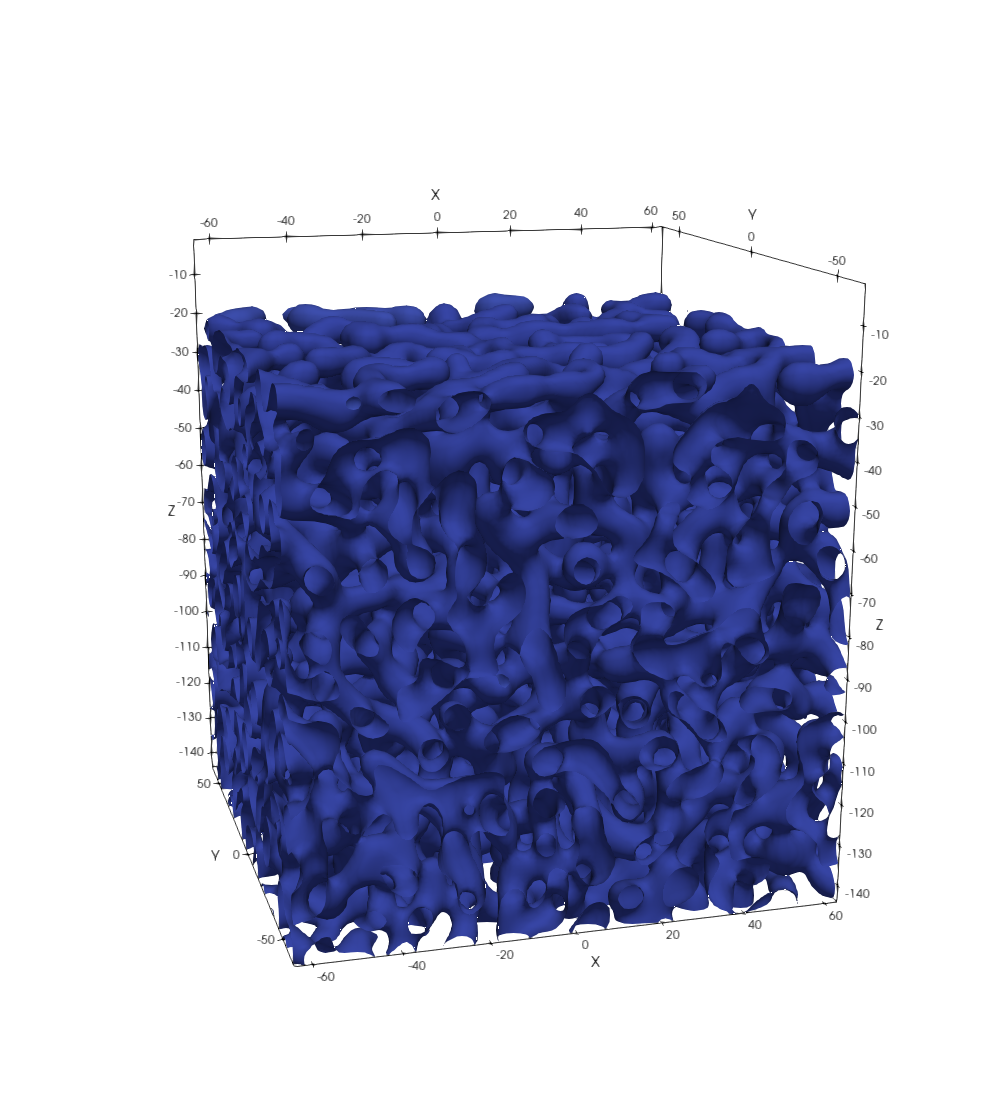} \\
        \hline

    \end{tabular}
    \end{minipage}%
    \hfill
    \raisebox{0.65cm}{
    \begin{minipage}{0.07\textwidth}
        \centering
        \includegraphics[width=\textwidth]{images/colorbar_vert.PNG}
    \end{minipage}}
    
    \caption{ Simulation results for $\phi_p^0 = 0.25$, illustrating the temporal evolution of the microstructure at four time values ($t \times 10^5$). The first row shows 2D simulations, the second row presents a planar slice extracted from the 3D simulations, and the third row displays the connectivity analysis of the microstructure obtained using a ParaView filter. The bath is located at the top of each image.}          

    \label{fig:3D_phi_0.25}
\end{figure}

In Figure \ref{fig:3D_phi_0.25} we illustrate the evolution of a film  that gives rise to a bicontinuous pattern in 3D  as can be seen from the bottom row where the 3D isosurface is plotted:  there is a large polymer rich domain that spans the system from top to bottom. Here we have considered $\phi_p^0=0.25$. The 2D and 3D simulation colorplots (2 top rows) are very similar up to $t=6$ and it is difficult to conclude from such images that the 3D pattern is bicontinuous. On the contrary for $t=12$ there is a clear difference between the colormaps from the 2D and 3D simulations since in the 3D images there are  more complex structure  closer to maze-like patterns. This later point is in contrast with the images obtained for $\phi_p^0=0.15$, where one could only see isolated bubbles (sometimes elongated or deformed).  It must be noted that the early stage pattern that can be seen at $t=6$ is reminiscent of isotropic phase separation. This is due to the effect of the boundary condition at the bottom of the system: both diffusion front collide leading to a geometry similar to the one obtained when considering no flux boundary conditions.\\ 

\begin{figure}
    \centering
    \renewcommand{\arraystretch}{1.5}
    \begin{minipage}{0.85\textwidth}
    \centering
    \begin{tabular}{|c|c|c|c|c|}
        \hline
        t & \textbf{$1.8$} & \textbf{$3$} & \textbf{$4.8$} & \textbf{$12$}  \\
        \hline
        \raisebox{1cm}{\textbf{2D}} & 
        \includegraphics[width=0.13\textwidth]{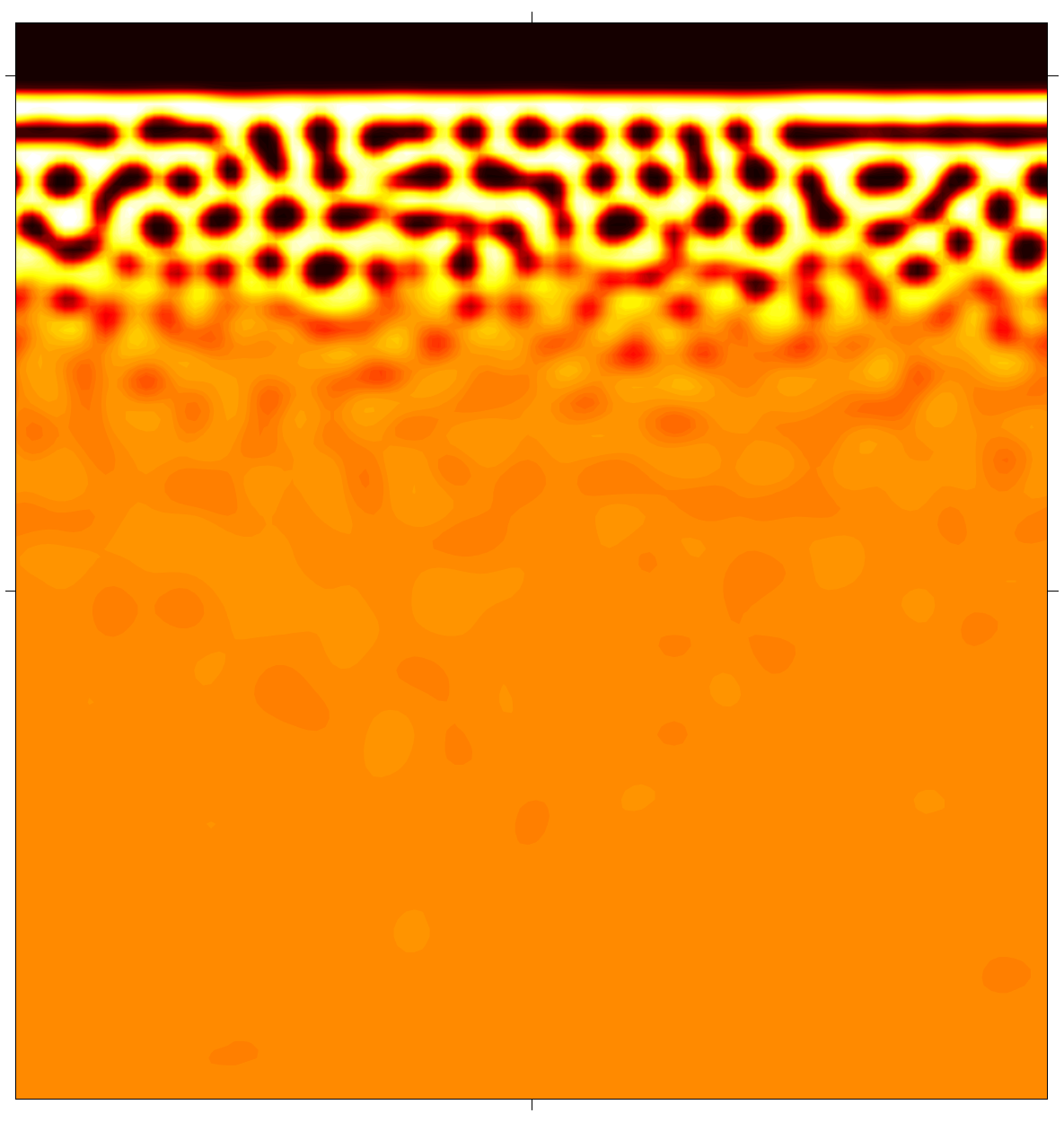} &
        \includegraphics[width=0.13\textwidth]{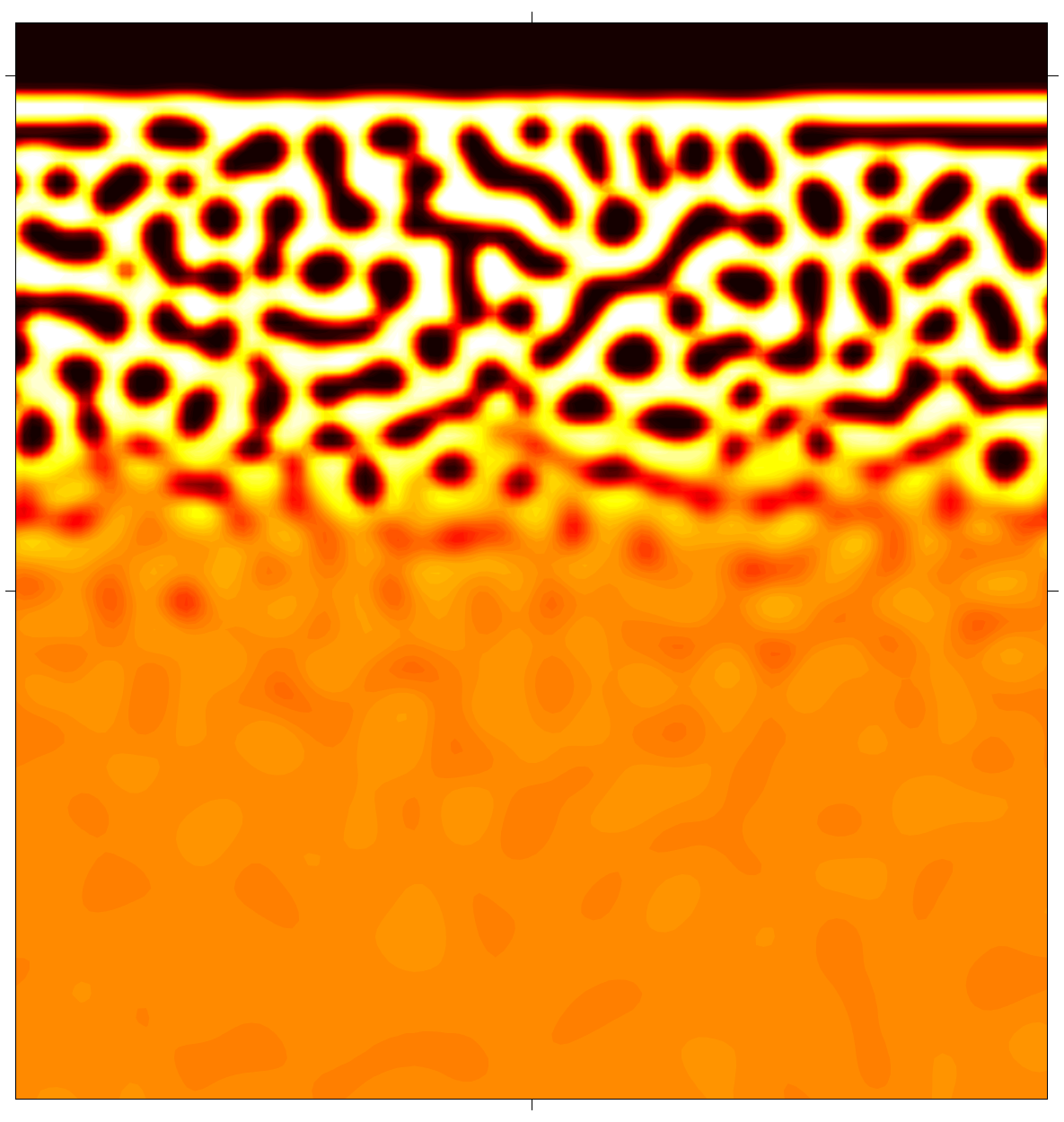} &
        \includegraphics[width=0.13\textwidth]{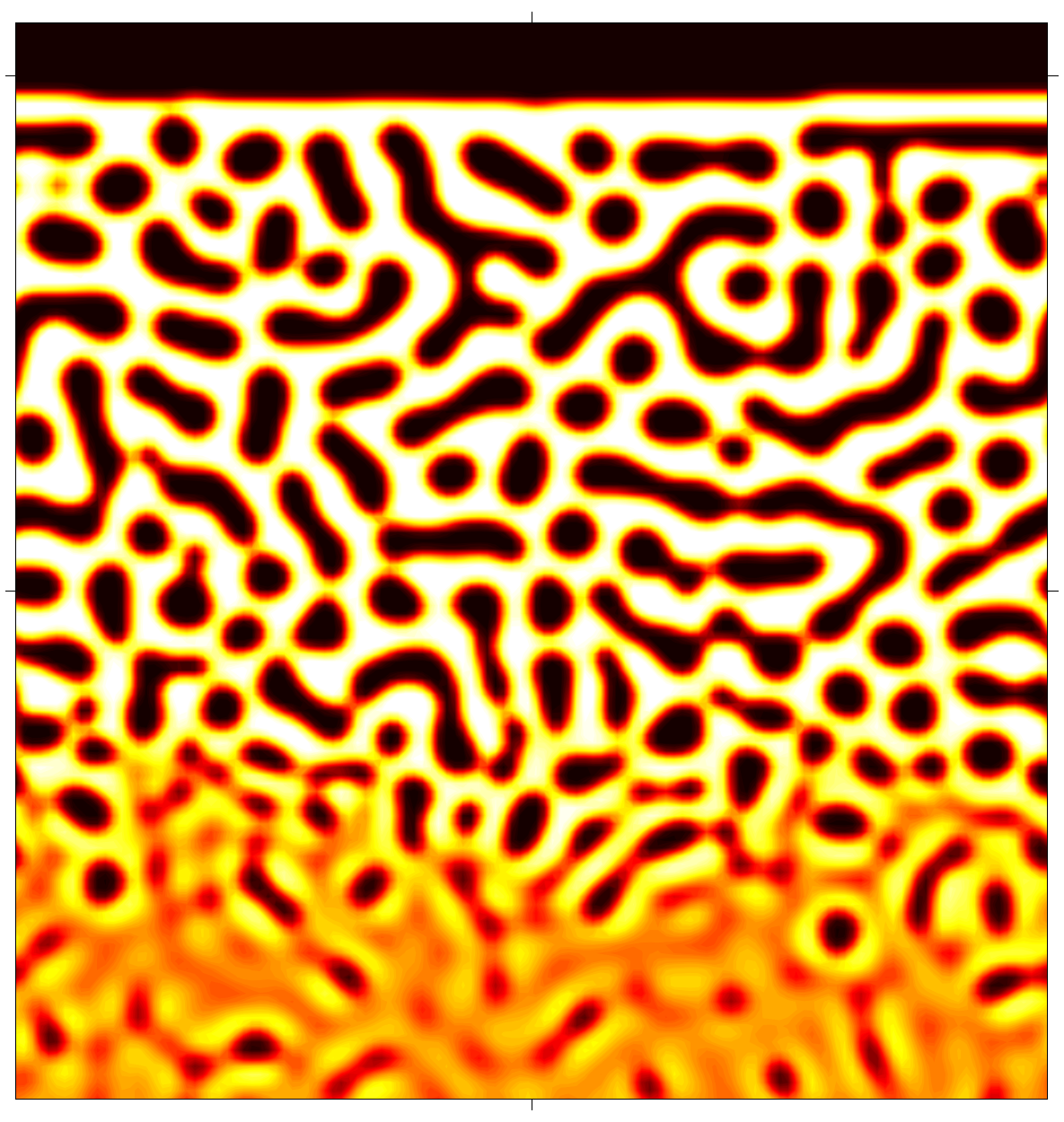} &
        \includegraphics[width=0.13\textwidth]{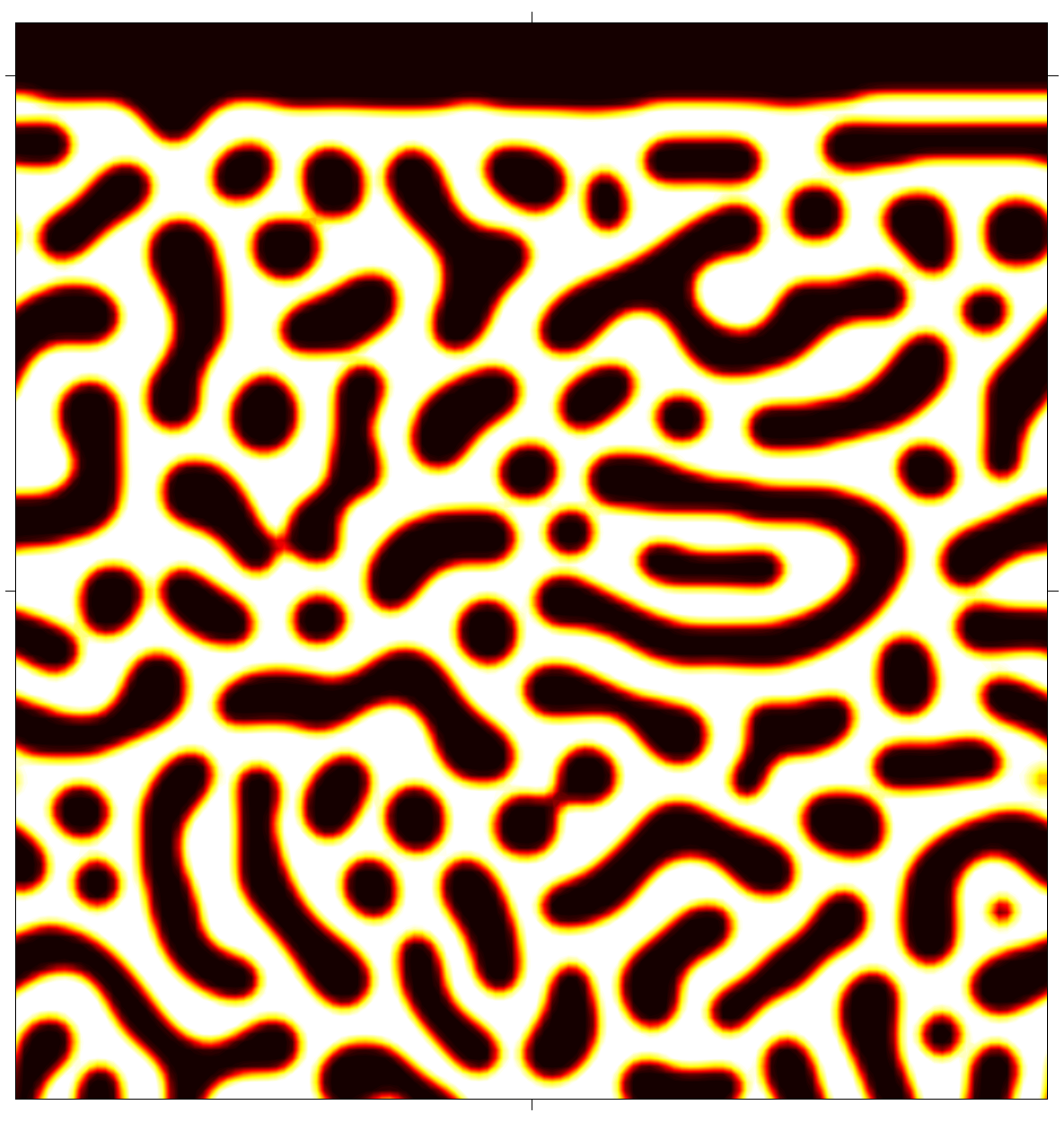} \\
        \hline
        
        \raisebox{1cm}{\textbf{3D slice}} & 
        \includegraphics[width=0.13\textwidth]{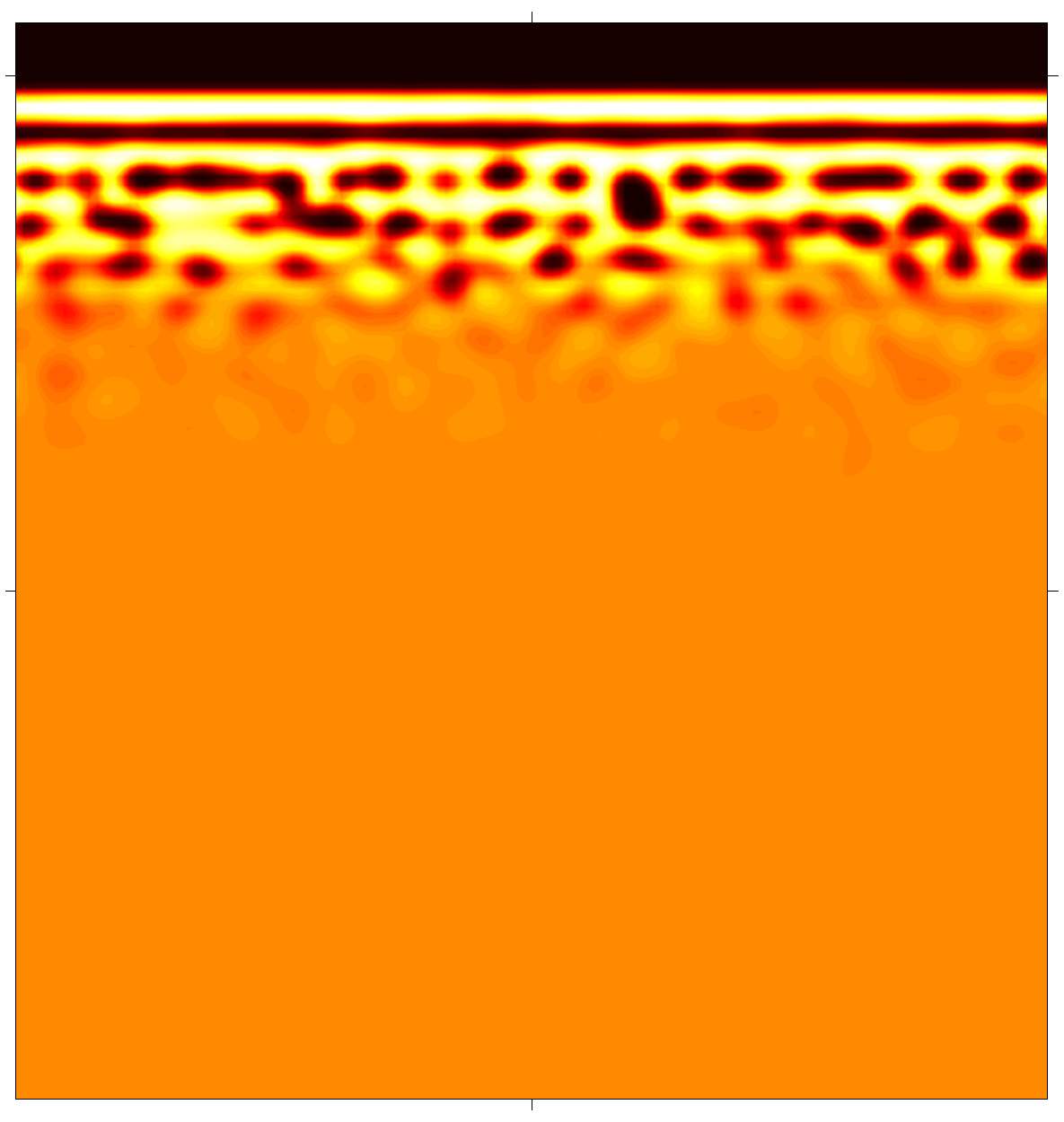} &
        \includegraphics[width=0.13\textwidth]{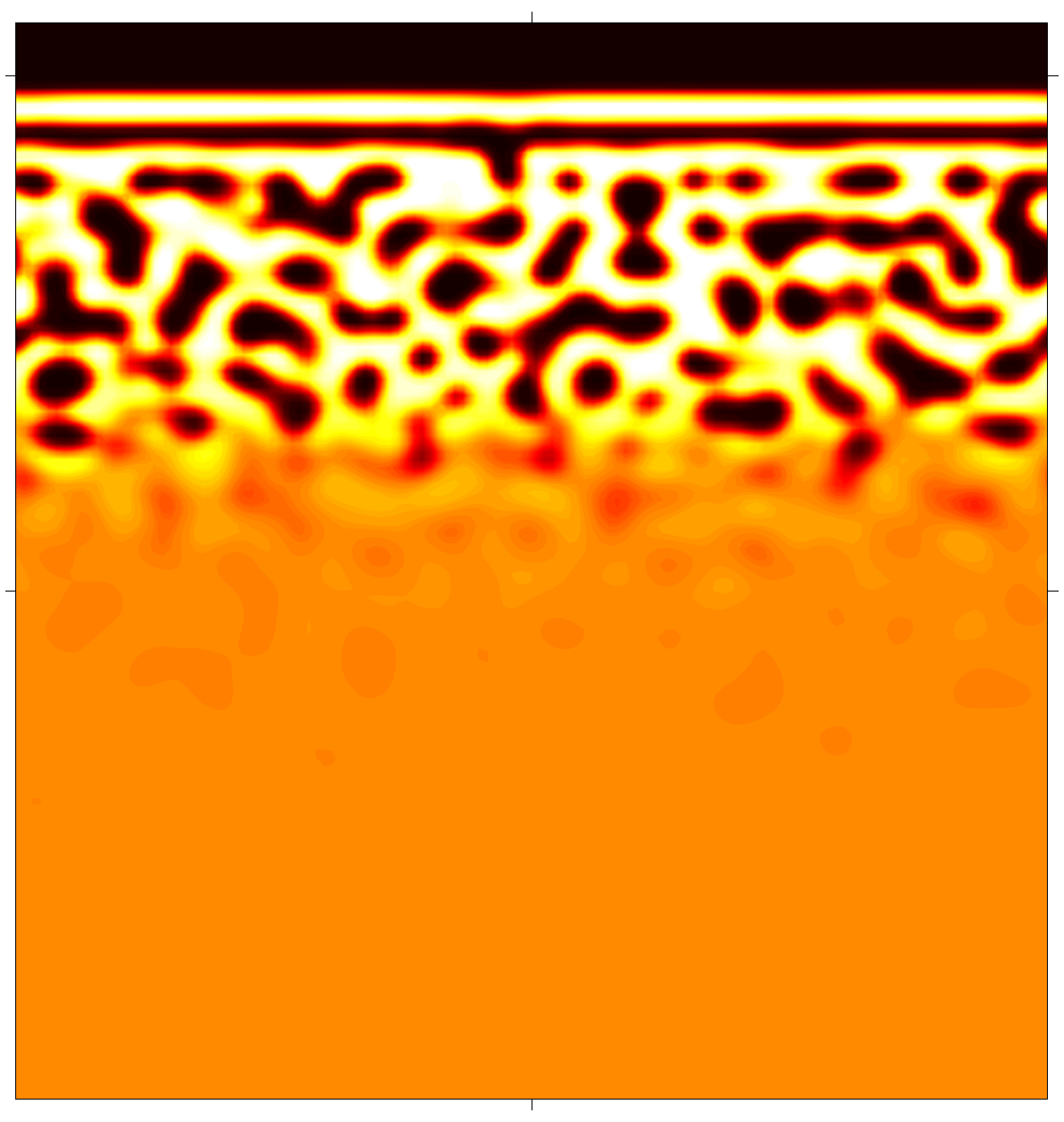} &
        \includegraphics[width=0.13\textwidth]{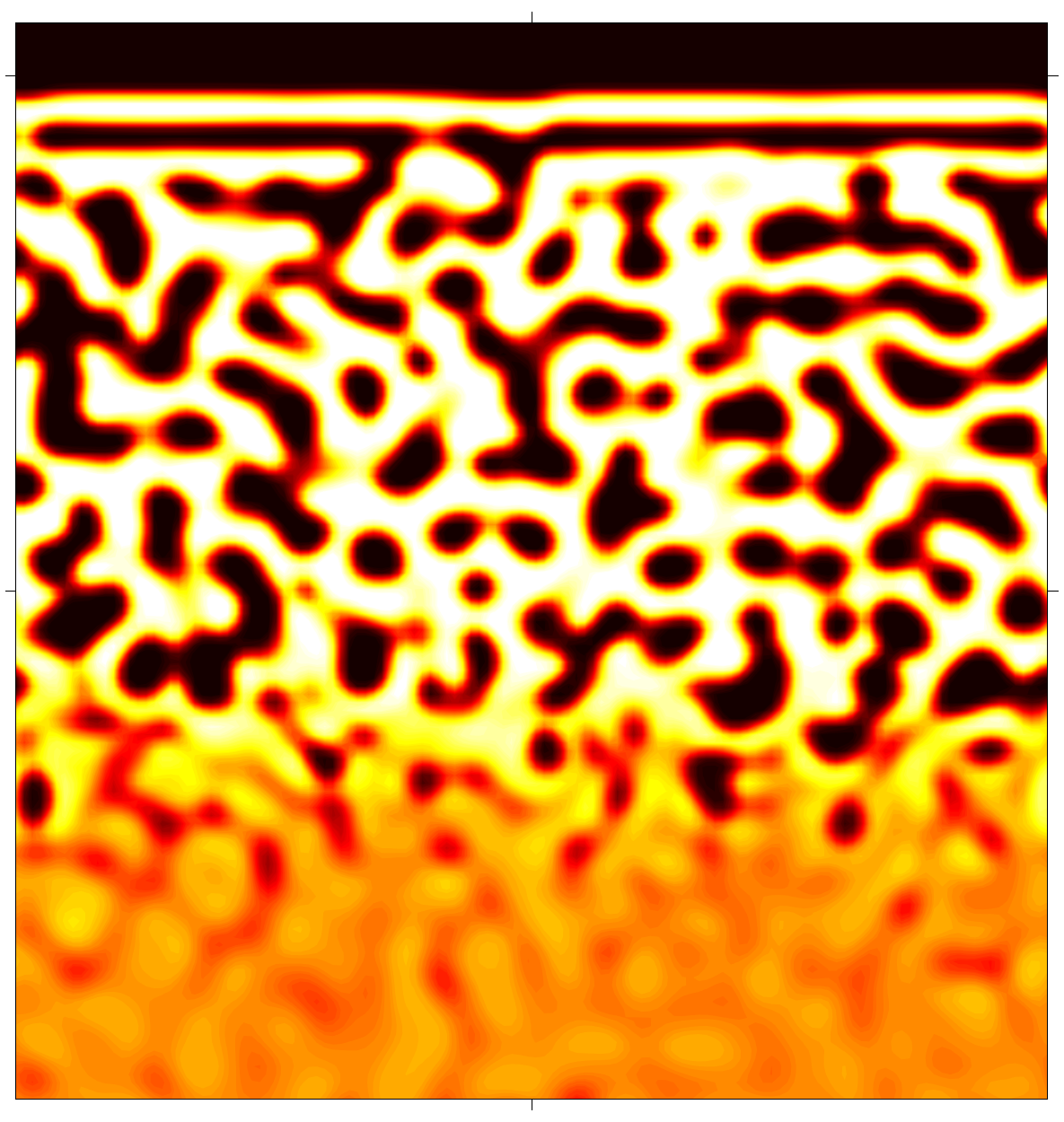} &
        \includegraphics[width=0.13\textwidth]{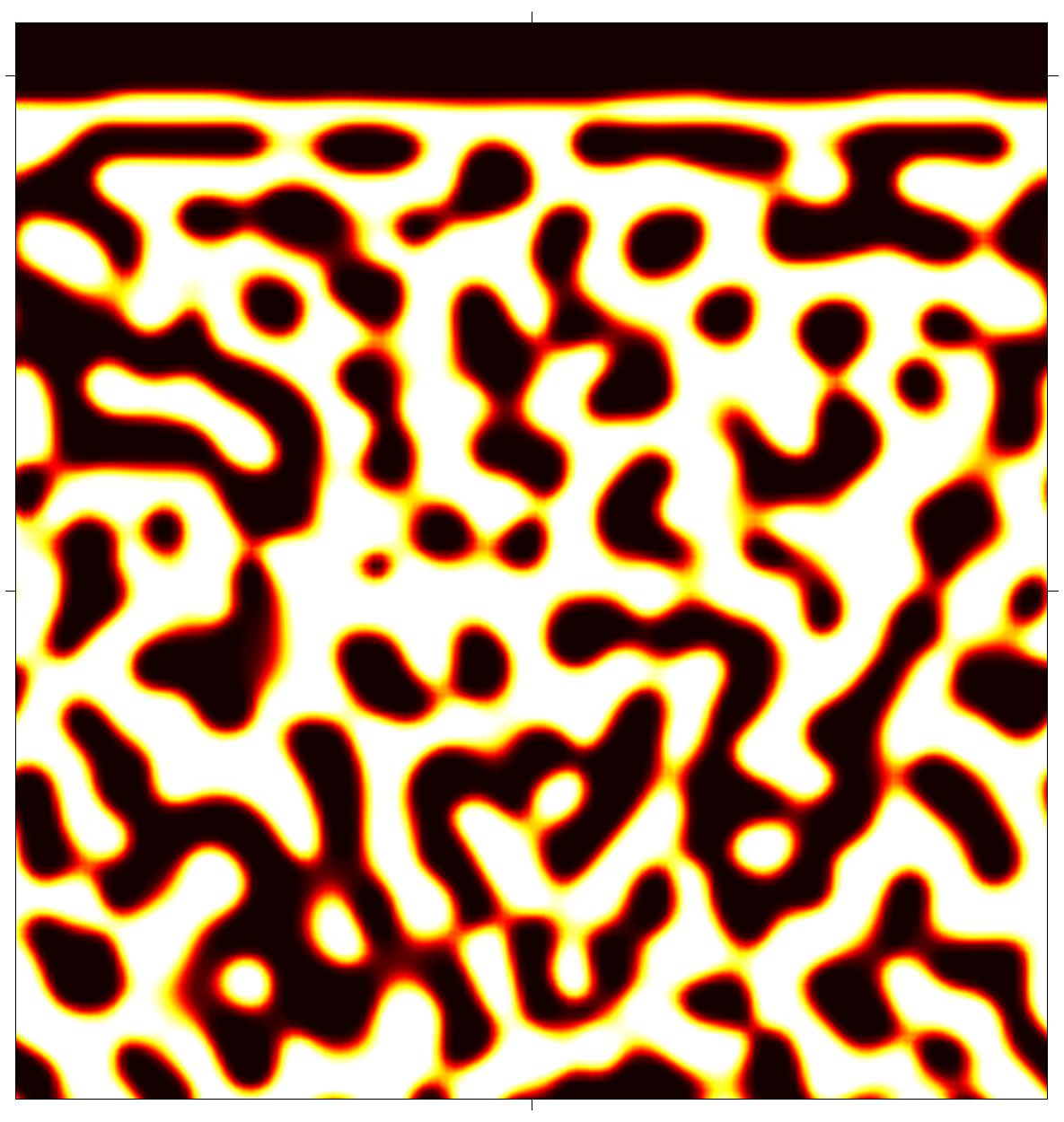} \\
        \hline
        
        \raisebox{1cm}{\textbf{Connectivity}} & 
        \includegraphics[width=0.16\textwidth,trim={3cm 3cm 3cm 4cm},clip]{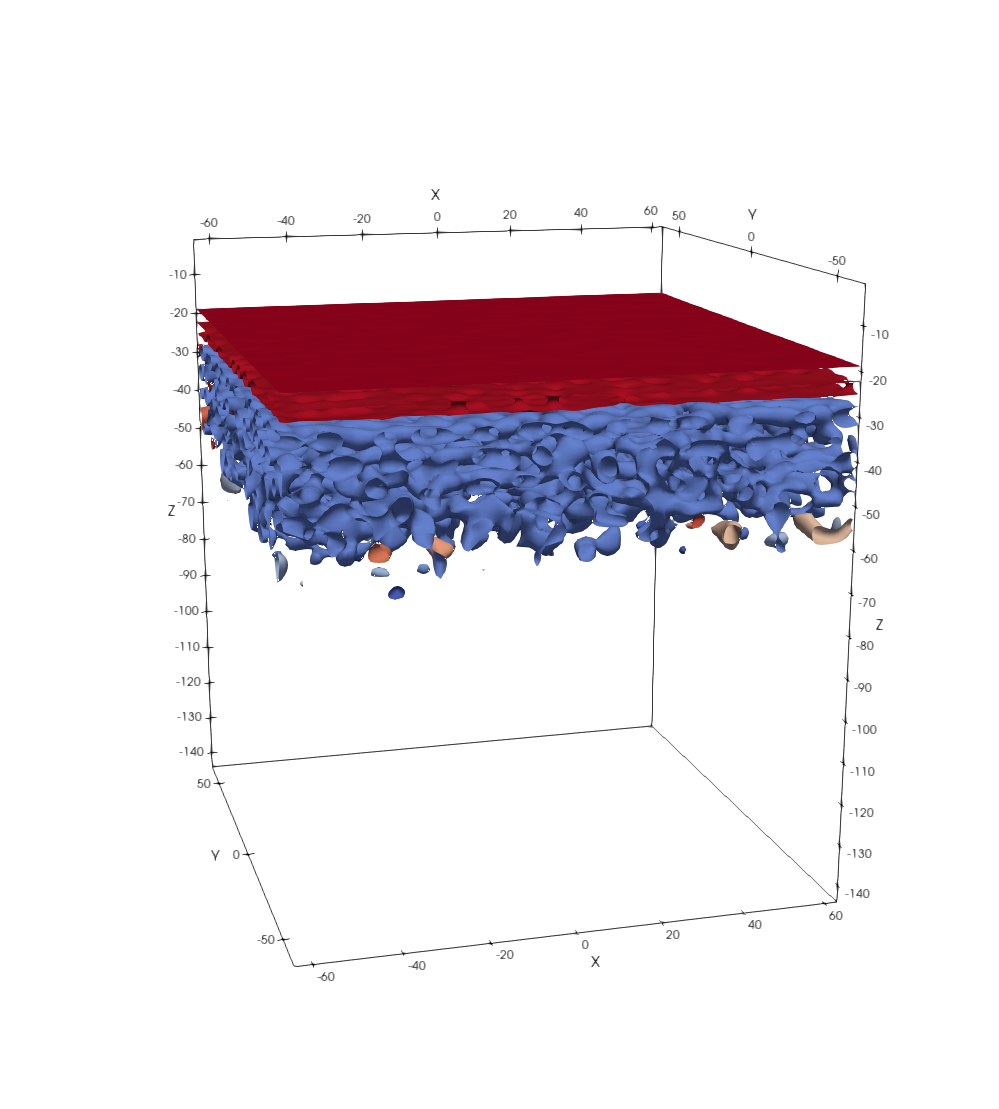} &
        \includegraphics[width=0.16\textwidth,trim={3cm 3cm 3cm 4cm},clip]{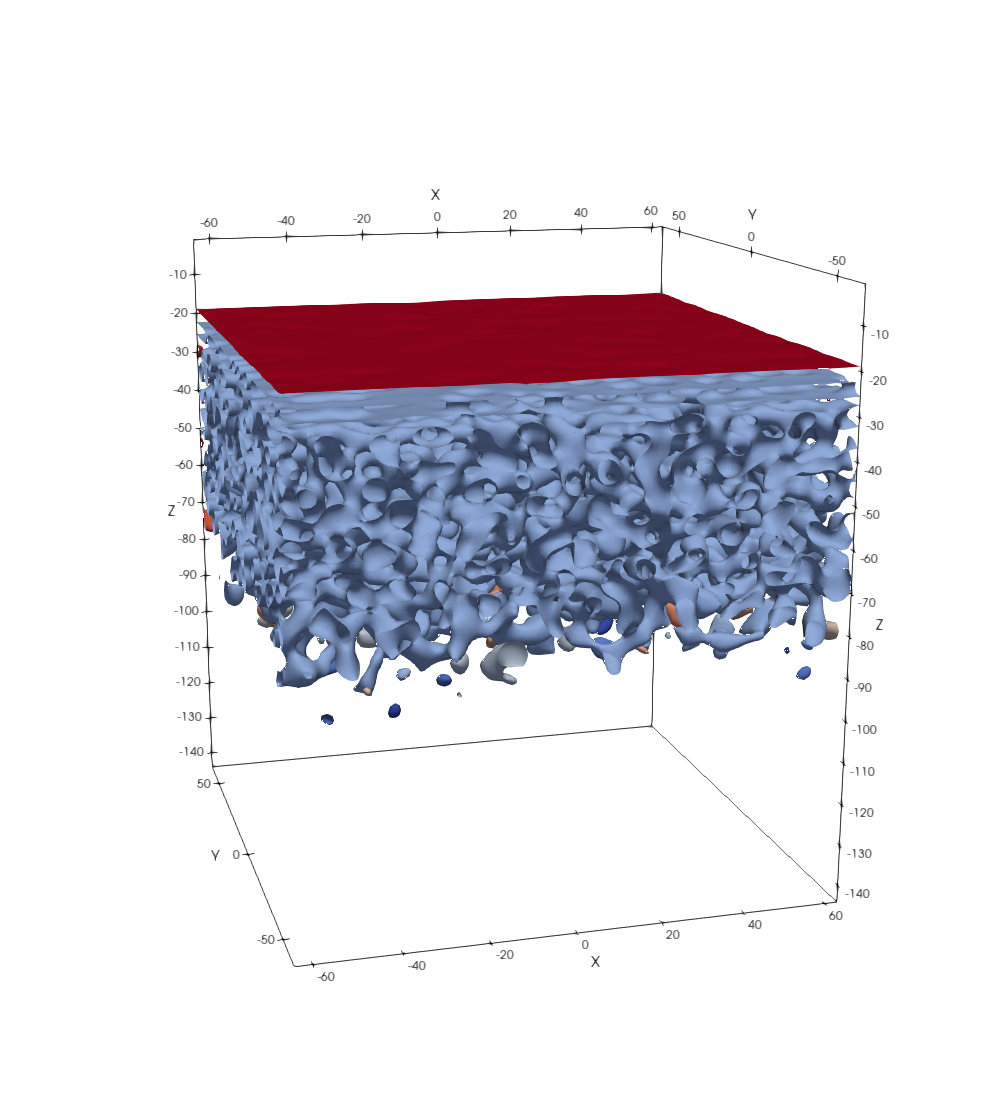} &
        \includegraphics[width=0.16\textwidth,trim={3cm 3cm 3cm 4cm},clip]{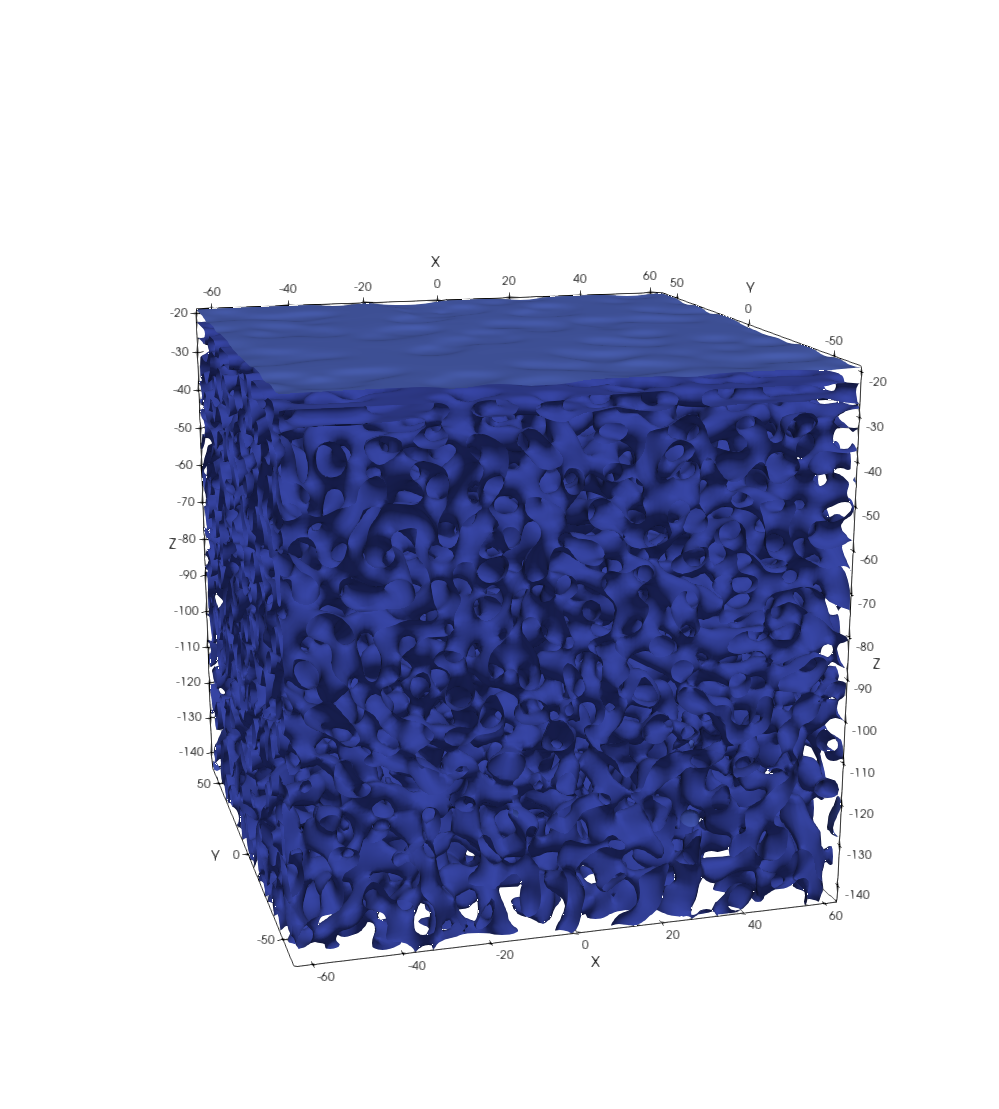} &
        \includegraphics[width=0.16\textwidth,trim={3cm 3cm 3cm 4cm},clip]{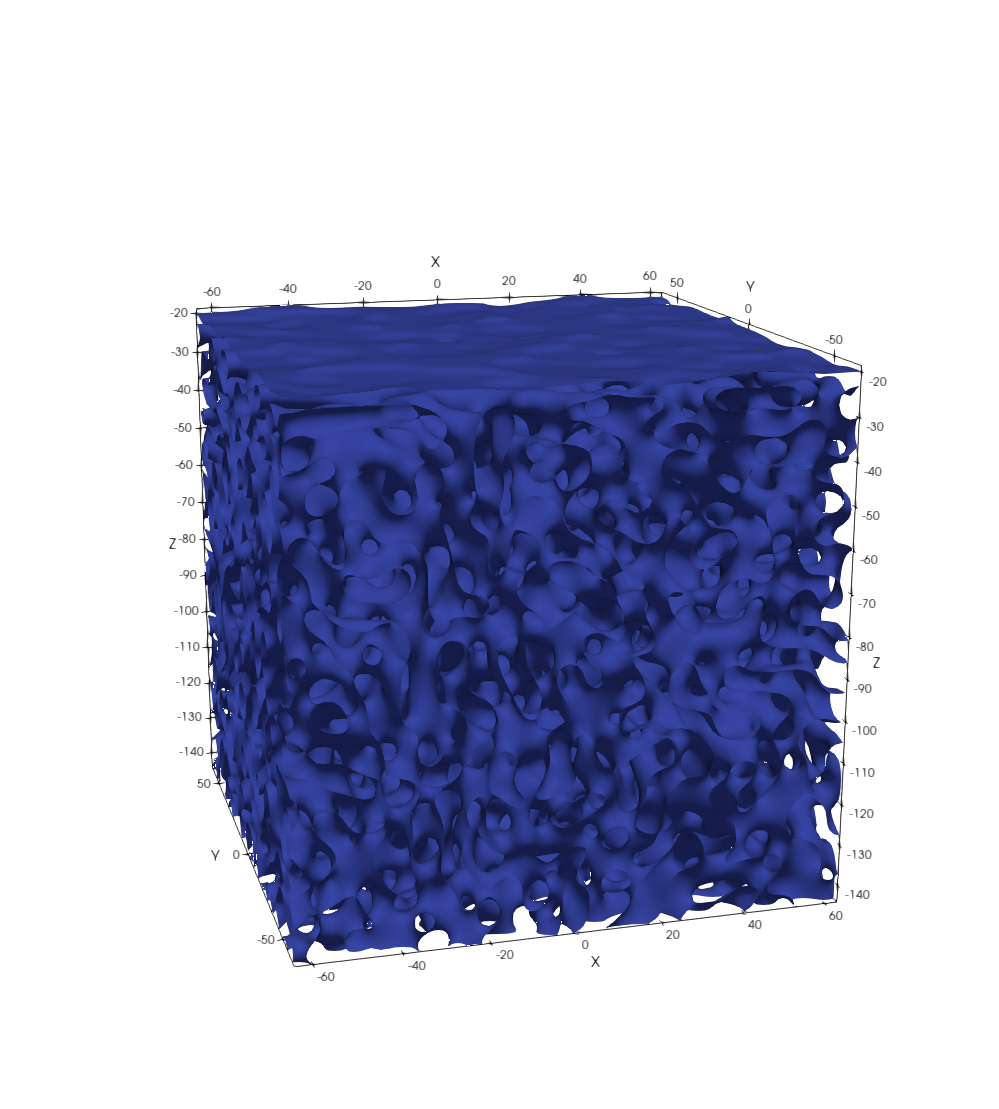} \\
        \hline

    \end{tabular}
    
    \end{minipage}%
    \hfill
    \raisebox{0.65cm}{
    \begin{minipage}{0.07\textwidth}
        \centering
        \includegraphics[width=\textwidth]{images/colorbar_vert.PNG}
    \end{minipage}}

    \caption{ Simulation results for $\phi_p^0 = 0.4$, illustrating the temporal evolution of the microstructure at four time values ($t \times 10^5$). The first row shows 2D simulations, the second row presents a planar slice extracted from the 3D simulations, and the third row displays the connectivity analysis of the microstructure obtained using a ParaView filter. The bath is located at the top of each image.}          

    \label{fig:3D_phi_0.4}
\end{figure}

Simulations using larger values of $\phi_p^0$ lead to similar patterns with less and less disconnected domain until the polymer poor domain loose their connectivity. A typical example of this is illustrated for $\phi_p^0=0.4$ in fig. \ref{fig:3D_phi_0.4}. With this choice of initial composition the final pattern in both 2D and 3D are  reminiscent of a bicontinuous microtructure. \\

From these results it appears that there is a wide range of initial film composition for which bicontinuous microstructure are formed. Obviously the properties  of the phases will depend on the initial film composition, for instance, for $\phi_p^0$ above $0.25$, the polymer phase is likely to be continuous. However, its effective elastic modulo is likely to be much  weaker than for $\phi_p^0=0.4$. And the opposite is expected for the polymer poor region: the ease of having the fluid flow through it is likely to be higher for  $\phi_p^0\approx0.25$. To assess these properties and how they depend on the initial film composition we have chosen to compute the tortuosity  of both  the polymer poor and polymer rich phases. This quantity is a measure of the ease of diffusing through the pattern: that is the effective thickness of the path of the phase that goes from one side of the pattern to the other side. This is a good proxy for both mechanical properties\footnote{The mechanical equilibrium equation and the diffusion equations are both elliptic} and flow properties\footnote{In this later case the characteristic size of the pattern plays also a key role due to the $l^4$ scaling of the poiseuille resistance}. \change{Tortuosity is computed by analogy with electrical conductance, and the method used is presented in the Appendix \ref{conductivity}.}\\


\begin{figure}
    \centering
    \begin{subfigure}[b]{0.49\textwidth}
        \centering
        \includegraphics[width=\textwidth]{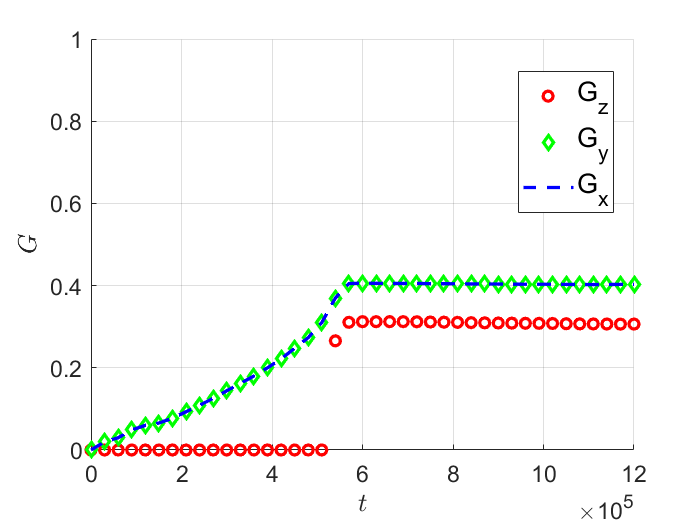}
        \label{fig:cond_t_rich}
    \end{subfigure}
    \begin{subfigure}[b]{0.49\textwidth}
        \centering
        \includegraphics[width=\textwidth]{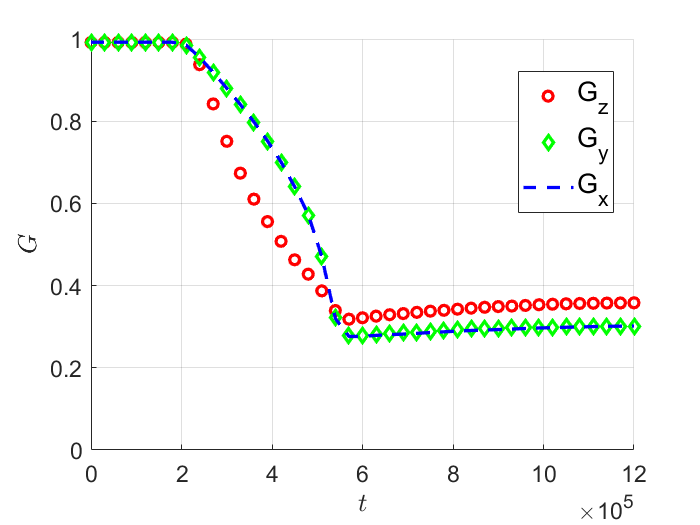}
        \label{fig:cond_t_poor}
    \end{subfigure}
    \caption{Conductances of the \textbf{Left} polymer-rich and \textbf{Right} polymer-poor phases during the coarsening for polymer volume fraction $\phi_p^0=0.4$}
    \label{fig:conductivity_t}
\end{figure}

First in fig. \ref{fig:conductivity_t} we present the evolution of the tortuosity/conductance  of each phase for $\phi_p^0=0.4$ with time in the three directions. For the polymer rich phase along the $x$ and $y$ direction there is a steady increase of the conductance that corresponds to the progressive thickening of the phase separated domain. Along the $z$ direction it remains close to zero until$ t \approx 5.10^5$ and it rapidly increases to reach the same value as along the x and y direction. This indicates that the phase separated region  has reached the bottom of the sample and that the thickness of the polymer poor has reached 0. Thereafter the conductance of the domain in all directions remains constant while the microstructure evolves (one can see  \ref{fig:3D_phi_0.4} that the characteristic size of the microtructure has increased significantly  between $t=4.8 \times 10^5$ and $t=12\times10^5$).  This indicates that a self-similar regime has been reached and that the conductance of both polymer rich and polymer poor phases is a characteristic of the final pattern. This behaviour is also seen for other  values of the initial film composition $\phi_p^0$. Therefore, we use this as descriptor of the final film structure.\\
\begin{figure}
    \centering
    \begin{subfigure}[b]{0.49\textwidth}
        \centering
        \includegraphics[width=\textwidth]{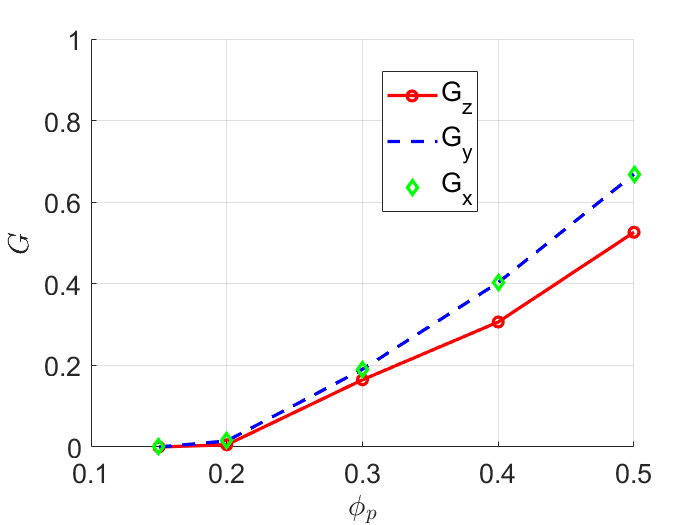}
        \caption{}
        \label{fig:cond_rich}
    \end{subfigure}
    \begin{subfigure}[b]{0.49\textwidth}
        \centering
        \includegraphics[width=\textwidth]{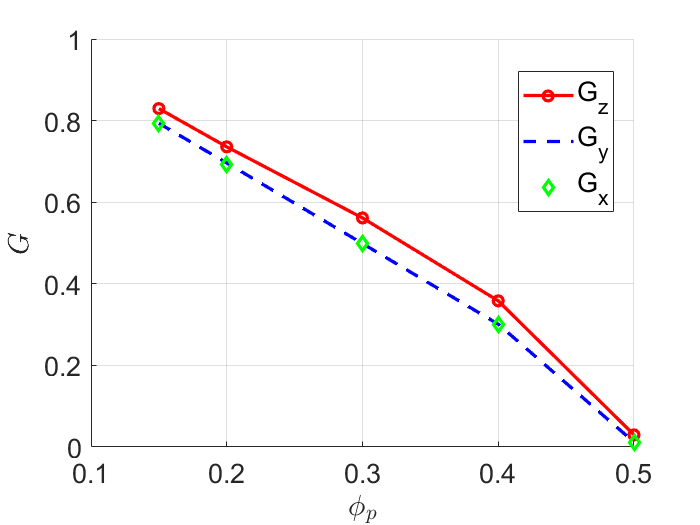}
        \caption{}
        \label{fig:cond_poor}
    \end{subfigure}
    \caption{Conductances of the \textbf{Left} polymer-rich and \textbf{Right} polymer-poor phases as a function of $\phi_p^0$. The values shown here correspond to the conductance reached at long times.}
    \label{fig:conductivity_phi}
\end{figure}

This is illustrated in figure \ref{fig:conductivity_phi} where the  conductance of polymer poor and polymer rich phases  is plotted as a function of $\phi_p^0$ along the three directions $x,\ y$ and $z$.  It must be noted that $\phi_p^0$ is not the volume fraction of the phases because of the asymmetry of the phase diagram. One can see that below a threshold value of $\phi_p^0$ the conductance of he polymer rich region is 0, which corresponds to a disconnected  pattern. Above this threshold, for the considered range (which corresponds to the bicontinuous pattern), the conductance of the polymer rich domain grows linearly. The opposite behaviour is observed for the polymer poor region. This shows that there exist a range of initial film composition for which one can obtain both a significant conductance of the polymer poor region, which will translate in a good permeability if the pore size is sufficient and a good mechanical stability with sufficiently small pore sizes. It must be noted that the  faster drop in the polymer poor region conductance along $z$  between $\phi_p^0=0.4$ and $\phi_p^0=0.5$ is related to the formation of a continuous layer of the polymer rich phase at the top of the film.\\

For $M_p=M/10$, the observed pattern for the parameter regime that leads to phase separation consisted of banded patterns in all studied cases. As in the 2D case we attribute this to the fact that the initial noise has been damped through polymer diffusion. And in 3D it has even been damped more rapidly due to simple dimensional effects.

\section{Conclusion}
In this article, we have studied numerically the influence of the respective mobilities of the polymer and of the solvent and anti-solvent during the NIPS process. Simulations show that for polymeric mobilities comparable to the solvent and anti-solvent mobilities there exists a small parametric domain for which the NIPS process can actually lead to phase separation and the formation of a porous microstructure. On the contrary for low relative polymer mobility there is a large film composition domain for which phase separation can occur. Moreover one can see that in this case, above a relatively small threshold (here a factor of 10), the reduction of the polymer mobility barely affects the initial pattern. As a result from a modeling point of view, capturing extremely high mobility contrast of chemical species is not necessary to capture the initial pattern formation. \\ 

3D simulations, that were limited to a high  value of mobility contrast, have shown  that there exist a wide range of film composition for which the formation of a bicontinuous structure is possible. They have also shown that in this case, despite a process that is anisotropic, the formed pattern is isotropic and invariant along the z axis (with exception of a thick polymer layer at the top of the film for high values of $\phi_p^0$). In addition we have used the tortuosity or the conductance of the polymer poor and polymer rich regions to quantitatively characterize the properties of these two phases.\\

Possible extension of this work are numerous. First of all, the study of the effect of mobility contrast on the initial pattern is clearly of interest. Indeed in the simulations presented here there is no gradient of microstructure along the z direction. However it is clear from preliminary simulations that for intermediate values of mobility contrast some gradient of properties may appear. The consequences of this on membrane properties are still unclear and it is unclear whether in actual membrane system such phenomena can be observed.  Finally and probably most importantly, our work was limited to diffusion, extension toward  hydrodynamic coarsening and the use of more complex rheologies such as viscoelastic flows is necessary.  

\change{\section*{Acknowledgements}
The authors gratefully acknowledge the Agence Nationale de la Recherche
(SIMUMEM Project: ANR-21-CE06-0043) for its financial support. This work was granted access to the HPC resources of
IDRIS under allocation 2024-AD012B07727R2 made by GENCI.}

\newpage
\appendix
\numberwithin{equation}{section}

\change{

\section{Determination of the Gradient Coefficient \(\kappa\) and the Mobility \(M\)\label{Appendice_Mob}}

The total free energy of a binary mixture can be written as
\begin{equation}
F = \int \left[ f_0\,\mathrm{FH}(\phi) + \frac{\kappa}{2}|\nabla \phi|^2 \right] dV,
\end{equation}
where \(f_0\) sets the energy scale of the Flory--Huggins free energy and \(\kappa\) is the gradient coefficient.\\

The corresponding chemical potential is obtained by the variational derivative:
\begin{equation}
\mu = f_0\,\mathrm{FH}'(\phi) - \kappa \nabla^2 \phi
\end{equation}

\subsection*{Linearization for small concentration fluctuations}

For small concentration fluctuations \(\phi = \phi_0 + \delta \phi\) with \(|\delta \phi| \ll 1\), the chemical potential is expanded to first order.  
Neglecting the gradient term in \(\mu\), we obtain:
\begin{equation}
\mu \simeq f_0\,\mathrm{FH}''(\phi_0)\,\delta \phi.
\end{equation}

The linearized Cahn--Hilliard equation then reads:
\begin{equation}
\partial_t \phi = M\,\nabla^2 \mu = M\,f_0\,\mathrm{FH}''(\phi_0)\,\nabla^2 \phi,
\end{equation}
which is equivalent to the standard diffusion equation:
\begin{equation}
\partial_t \phi = D\,\nabla^2 \phi,
\qquad D = M\,f_0\,\mathrm{FH}''(\phi_0),
\end{equation}
where \(D\) is the effective diffusion coefficient. Obviously it depends on the composition of the mixture $\phi_0$ arround which the perturbation analysis is carried.   

\subsection*{Relations between \(\kappa\), \(f_0\), \(\gamma\), and \(\varepsilon\)}

The interfacial tension \(\gamma\) and the interfacial width \(\varepsilon\) scale as:
\begin{equation}
\gamma \sim \sqrt{f_0\,\kappa}, 
\qquad 
\varepsilon \sim \sqrt{\frac{\kappa}{f_0}}.
\end{equation}
Combining both expressions gives:
\begin{equation}
\boxed{\kappa = \gamma\,\varepsilon, \qquad f_0 = \frac{\gamma}{\varepsilon}.}
\end{equation}
These relations provide a direct link between measurable interfacial properties (\(\gamma\), \(\varepsilon\)) and model parameters (\(\kappa\), \(f_0\)).

\subsection*{Choice of the mobility \(M\)}

The mobility \(M\) is chosen such that the diffusion coefficient \(D\) reproduces realistic experimental values:
\begin{equation}
\boxed{M = \frac{D}{f_0\,\mathrm{FH}''(\phi_0)} = \frac{D\,\varepsilon}{\gamma\,\mathrm{FH}''(\phi_0)}.}
\end{equation}
Once \(\gamma\), \(\varepsilon\), and \(\mathrm{FH}''(\phi_0)\) are known, this expression provides a consistent value for \(M\). This approah limited to a binary system can be exteended to a ternary mixture using the mobility mattrix and the Hessian of the free energy density, that is the equivalent of the second derivative. This is explained in detail page 13 of \cite{Tree2017} (equations 26-27).    }
\change{

\section{A simple explanation of the less than naively expected effect of reducing polymer mobility:\label{sublinear}}
 Here, we consider a  toy system model:
 \begin{equation}
   \tau_A \dot{A}= (\alpha t)A 
 \end{equation}
 where  $A$ can be seen as the amplitude of the oscillations of the polymer concentration in the film.  $\tau_A$ is a time constant that can be varied and $\alpha $ is a fixed time constant that corresponds to the rate at which the driving force for the instability grows. $\tau_A$ is a proxy for the polymer mobility and $\alpha t$ corresponds to how deep the film has been driven in  the spinodal domain through solvent/non solvent exchange. $\alpha$ is therefore independent of the polymer mobility.

  For $t<0$, $A=0$  is a stable equilibrium point and for $t>0$ it loses stability. One can easily show that for a given initial amplitude $A_0$ at $t=0$, the solution of this  ordinary differential equation is:
  \begin{equation}
    A(t)=A_0\exp(\frac{\alpha}{2\tau_A}t^2)
  \end{equation}
  As a result the polymer concentration will reach a given amplitude $A_1$ for a time:
  \begin{equation}
    t_1=\sqrt{\frac{\ln(A_1/A_0)\tau_A}{\alpha}}
  \end{equation}
  that does scale as $\sqrt{\tau_A}$.  This sublinear  behaviour  is  similar to the one observed in the actual simulations even though the complexity of the considered system is likely  prevent us from checking the scaling quantitatively.
  }

\change{
	\section{Measure of the conductivity of the system \label{conductivity}}
In order to measure the connectivity of the microstructure (i.e.) of one of the
phases, we compute  the conductivity of the microstructure with one phase
with a conductance of 1 and the other a conductance of $g<<1$. This  has been measured here
by solving the linear PDE for a given microstructure (the size of the
microstrure was set to 1 in all  directions) $c(\mathbf{x})$: 
\begin{equation}
0=\nabla (G(c) \nabla V)\label{eq_laplace}
\end{equation}
whith  $G(c)=1$ if $c(\mbox{resp }(1-c)))>0.5$ and $G(c)=g$ if $c(\mbox{resp
}(1-c)))<0.5$. The boundary conditions are $V(i=L=1)=1 \mbox{ and }
V(i=0)=0$ and periodic at the other faces of the sample where $i$ is either $x$
or $y$ or $z$.  Once $V$ is computed (using a method described below), the flux
along the $i$ direction is:
\begin{equation}
  \Phi_i=\int_{j}\int_{k} g\partial_i V \label{eq_conductance}
\end{equation}
where $j$ and $k$ are the two remaining indices once $i$ is set.
In such a system,  one can easily see that (conducting tubes along the gradient
 of $V$ which is anisotropic) the  maximal conductance (conducting tubes along
 the gradient of $V$ which is anisotropic)   is equal to the volume fraction of
 the conducting phase $\varphi$. Here, we need to take into account the fact
 that the pattern is not invariant along the $z$ axis. To this purpose when
 computing the conductivity  the microstructure was
 \textit{tuned}:
 \begin{itemize}
   \item Conductivity  along $z$: the top layers of the film (close to the
     film/bath) interface were
     replaced by  layers with conductivity 1.
   \item    Conductivity  along $x$ or $y$: the top layers of the film  were
     replaced by layers with conductivity $g<<1$. 
 \end{itemize}

 The solution was computed using a discretized damped wave equation with a
 properly chosen damping $\lambda=0.005$ and varying mass density  to ensure
 fast convergence toward the equilibrium and a constant wave equation in domains
 independently of the phase:
 \begin{equation}
   \partial_{tt} V=\frac{1}{G(c)}(\nabla (G(c) \nabla V))-\lambda\partial_t V
 \end{equation}
 Simulations showed that after $10^4$ iteration, a very good convergence had
 been reached~: the residual were extremely small and the value of $\Phi$ that
 was computed was nearly independent of the position where it is computed. This
 was in stark contrast with results obtained using Gauss Seidel over-relaxation
 for which after a comparable number of iterations, the same value of the
 error(using $L_\infty$ norm) was reached  but where significant  long wavelength variations
 of the flux $\varphi$ were present. An estimate of $\lambda$ in the case of a
 continuous one dimensional system is of the order of magnitude of  $ c/L$ where
 c is the wave-speed and $L$ is the size of the system: considering higher
 values of $\lambda$ would lead to a mode whose amplitude decreases with a rate
 much lower than $\lambda$. 
}


\end{document}